\documentclass[a4paper,12pt,english]{article}

\usepackage{amsfonts,bm,amssymb,euscript,array,babel,cite}
\usepackage{mathtools}
\usepackage{tikz-cd}
\usepackage{amsmath}
\usepackage{hhline}
\usepackage[amsmath]{empheq}
\usepackage{hyperref}
\usepackage{cancel}

\usepackage{mathrsfs}
\usepackage{tikz}
\newcommand*\circled[1]{\tikz[baseline=(char.base)]{
		\node[shape=circle,draw,inner sep=1pt] (char) {#1};}}

\newsavebox{\mysaveboxM}
\newsavebox{\mysaveboxT}
\newcommand*\alphabox[2][A Nice Box]{%
	\sbox{\mysaveboxM}{#2}%
	\sbox{\mysaveboxT}{\fcolorbox{black}{yellow}{#1}}%
	\sbox{\mysaveboxM}{%
		\parbox[b][\ht\mysaveboxM+0.5\ht\mysaveboxT+0.5\dp\mysaveboxT][b]{%
			\wd\mysaveboxM}{#2}%
	}%
	\sbox{\mysaveboxM}{%
		\fcolorbox{black}{white}{%
			\makebox[\linewidth-24.5em]{\usebox{\mysaveboxM}}%
		}%
	}%
	\usebox{\mysaveboxM}%
	\makebox[0pt][r]{%
		\makebox[\wd\mysaveboxM][c]{%
			\raisebox{\ht\mysaveboxM-0.1\ht\mysaveboxT
				+0.5\dp\mysaveboxT-0.5\fboxrule}{\usebox{\mysaveboxT}}%
		}%
	}%
}

\newcommand*\betabox[2][A Nice Box]{%
	\sbox{\mysaveboxM}{#2}%
	\sbox{\mysaveboxT}{\fcolorbox{black}{yellow}{#1}}%
	\sbox{\mysaveboxM}{%
		\parbox[b][\ht\mysaveboxM+0.5\ht\mysaveboxT+0.5\dp\mysaveboxT][b]{%
			\wd\mysaveboxM}{#2}%
	}%
	\sbox{\mysaveboxM}{%
		\fcolorbox{black}{white}{%
			\makebox[\linewidth-27.5em]{\usebox{\mysaveboxM}}%
		}%
	}%
	\usebox{\mysaveboxM}%
	\makebox[0pt][r]{%
		\makebox[\wd\mysaveboxM][c]{%
			\raisebox{\ht\mysaveboxM-0.1\ht\mysaveboxT
				+0.5\dp\mysaveboxT-0.5\fboxrule}{\usebox{\mysaveboxT}}%
		}%
	}%
}

\newcommand*\gammabox[2][A Nice Box]{%
	\sbox{\mysaveboxM}{#2}%
	\sbox{\mysaveboxT}{\fcolorbox{black}{yellow}{#1}}%
	\sbox{\mysaveboxM}{%
		\parbox[b][\ht\mysaveboxM+0.5\ht\mysaveboxT+0.5\dp\mysaveboxT][b]{%
			\wd\mysaveboxM}{#2}%
	}%
	\sbox{\mysaveboxM}{%
		\fcolorbox{black}{white}{%
			\makebox[\linewidth-1.5em]{\usebox{\mysaveboxM}}%
		}%
	}%
	\usebox{\mysaveboxM}%
	\makebox[0pt][r]{%
		\makebox[\wd\mysaveboxM][c]{%
			\raisebox{\ht\mysaveboxM-0.1\ht\mysaveboxT
				+0.5\dp\mysaveboxT-0.5\fboxrule}{\usebox{\mysaveboxT}}%
		}%
	}%
}

\makeatletter

\newcommand{\dd}{\mathrm{d}}

\newcommand{\w}{\wedge}
\newcommand{\bbm}{\left(\begin{matrix}}
\newcommand{\ebm}{\end{matrix}\right)}
\newcommand{\beq}{\begin{eqnarray}}
\newcommand{\eeq}{\end{eqnarray}}

\newcommand{\cbral}{[\![}
\newcommand{\cbrar}{]\!]}

\makeatother

\newtheorem{prop}[equation]{Proposition}
\newtheorem{defn}[equation]{Definition}
\newtheorem{rmk}[equation]{Remark}

\newcommand{\sfrac}[2]{{\textstyle\frac{#1}{#2}}}

\newcommand{\be}{\begin{equation}}
\newcommand{\ee}{\end{equation}}

\newcommand{\beqa}{\begin{eqnarray}}
\newcommand{\eeqa}{\end{eqnarray}} 
\def\nn{\nonumber} \def \bea{\begin{eqnarray}} \def\eea{\end{eqnarray}}

\newcommand{\barr}{\begin{array}}
\newcommand{\earr}{\end{array}}
\numberwithin{equation}{section}

  \def\b{\beta}
  \def\G{\Gamma}
 \def\d{\delta} 
 \def\F{\Phi}   
\def\l{\lambda} \def\L{\Lambda}  
   \def\P{\Pi} 
 \def\S{\Sigma}  

\def\mc{\mathcal}

  \def\cF{{\cal F}} 
 \def\cH{{\cal H}}  
  \def\cL{{\cal L}} 
  \def\cO{{\cal O}} 
\def\cP{{\cal P}}

\def\R{{\mathbb R}} \def\C{{\mathbb C}}  \def\X{\mathbb X} \def \A{\mathbb A}
\def \F{\mathbb F} \def \B{\mathbb B}
\def \L{\mathbb L}
 \def\one{\mbox{1 \kern-.59em {\rm l}}}

\def\sfp{{\mathsf p}}
\def\sfL{{\mathsf L}}

\def\bit{\begin{itemize}} \def\eit{\end{itemize}}

\def\({\left(} \def\){\right)}

 \allowdisplaybreaks[3]

\begin{document}

\makeatother


\parindent=0cm

\renewcommand{\title}[1]{\vspace{10mm}\noindent{\Large{\bf

#1}}\vspace{8mm}} \newcommand{\authors}[1]{\noindent{\large

#1}\vspace{5mm}} \newcommand{\address}[1]{{\itshape #1\vspace{2mm}}}


\begin{titlepage}

\begin{flushright}
\small
EMPG--18--04
\end{flushright}

\begin{center}


\title{ {\Large 
{Double Field Theory and Membrane Sigma-Models}}}


  \authors{ 
\large
    Athanasios
    {Chatzistavrakidis}{$^{\ast,}$}{\footnote{a.chatzistavrakidis@gmail.com}}
    , \ 
    Larisa Jonke{$^{\ast,}$}{\footnote{larisa@irb.hr}} , \\ \vspace{10pt} 
    Fech Scen Khoo{$^{\ast,}$}{\footnote{Fech.Scen.Khoo@irb.hr}} , \ Richard~J. Szabo{$^{\dagger,}$}{\footnote{R.J.Szabo@hw.ac.uk}}}
 
 \vskip 3mm
 
  \address{ $^{\ast}$ Division of Theoretical Physics, Rudjer Bo\v skovi\'c Institute \\ Bijeni\v cka 54, 10000 Zagreb, Croatia \\

 \
 
 $^{\dagger}$ Department of Mathematics, Heriot-Watt University
 \\ Colin Maclaurin Building, Riccarton, Edinburgh EH14 4AS, UK \\
 \vspace{3pt}
 Maxwell Institute for Mathematical Sciences, Edinburgh, UK \\
 \vspace{3pt}
 The Higgs Centre for Theoretical Physics, Edinburgh, UK
 
 }


\begin{abstract}
\noindent
We investigate geometric aspects of double field
theory (DFT) and its formulation as a doubled membrane
sigma-model. Starting from the standard Courant algebroid over the
phase space of an open membrane, we determine a splitting and a projection
to a subbundle that sends the Courant algebroid operations to the corresponding operations in DFT. This describes
precisely how the geometric structure of DFT lies in between two
Courant algebroids and is reconciled with generalized geometry. We construct the membrane sigma-model that
corresponds to DFT, and demonstrate how the standard T-duality orbit
of geometric and non-geometric flux backgrounds is captured by its
action functional in a unified way. This also clarifies the appearence
of noncommutative and nonassociative deformations of geometry in
non-geometric closed string theory. Gauge invariance of the DFT
membrane sigma-model is compatible with the flux formulation of DFT
and its strong constraint, whose geometric origin is explained. Our
approach leads to a new generalization of a Courant algebroid, that we
call a DFT algebroid and relate to other known generalizations, such
as pre-Courant algebroids and symplectic nearly Lie 2-algebroids. We
also describe the construction of a gauge-invariant doubled membrane
sigma-model that does not require imposing the strong constraint.   

\end{abstract}

\end{center}

\vskip 2cm

\end{titlepage}

\setcounter{footnote}{0}
\tableofcontents

\newpage

\section{Introduction and summary}

\subsubsection*{Motivation and goals}

When quantum gravitational effects become important, it is expected
that the geometry of spacetime departs from classical Riemannian
geometry. Such is the case in open string theory, where the endpoints
of open strings ending on D-branes supporting a constant gauge flux
probe a
noncommutative deformation of the worldvolume
geometry~\cite{stncg1,stncg2,stncg3} (see
e.g.~\cite{Douglas:2001ba,Szabo:2001kg} for reviews). However, open
strings are associated to gauge interactions, whereas gravity appears
in the closed string sector. In recent years it was argued that closed
strings propagating in backgrounds with non-geometric fluxes can probe
noncommutative and even nonassociative deformations of the background
geometry~\cite{stnag1,stnag2,stnag3,Mylonas:2012pg} (see
e.g.~\cite{Plauschinn:2012kd,stnag4,Mylonas:2014aga,Blumenhagen:2014sba,Barnes:2016cjm}
for reviews). T-duality plays a prominent role in these developments,
since string backgrounds that are T-dual to each other may correspond
to target spaces with different geometry and topology. It typically reveals the existence of unconventional closed string
geometries where string duality transformations are required as transition
functions, and lead to non-geometric flux backgrounds (see e.g.~\cite{Hull:2009sg} and references therein.)  

Non-geometric backgrounds can be naturally described within the framework
of a doubled formalism for closed
strings~\cite{doubled1,doubled2,doubled3,doubled4,doubled5,Siegel:1993th}. A
double field theory (DFT), where both coordinates conjugate to
momentum modes and dual coordinates conjugate to winding modes of the
closed string are implemented, was constructed
in~\cite{doubled3,Siegel:1993th} and more recently in~\cite{dft1,dft2,dft3,dft4}. In~\cite{Freidel:2015pka} an alternative approach to implementing
  T-duality is given as a linearly realized symmetry. In DFT, the continuous
version of the T-duality group becomes a manifest symmetry of the
action, and as such it has the power to describe different T-dual
backgrounds in a unified way (see e.g.~\cite{dftrev1,dftrev2,dftrev3}
for reviews).

On the other hand, the underlying higher mathematical structures for all these developments appear in the differential geometry of Courant algebroids \cite{courant,liu,dee3,Severa:2017oew} and in generalized geometry \cite{Hitchin:2004ut,Gualtieri:2003dx}. It was realised by~\cite{cg} (based on earlier results of~\cite{Bouwknegt:2003zg}) that T-duality can be understood as an isomorphism of Courant algebroids over two dual manifolds which are principal torus bundles over a common base via projection from a ``large'' structure on the correspondence space.
Relations between Courant algebroids and DFT were already investigated in \cite{dft2}, where it was shown that the C-bracket of DFT is the covariantization of the Courant bracket, in the sense that 
solving the strong constraint of DFT reduces one to the
other. Moreover, precise relations among the two brackets for
different implementations of the strong constraint were proposed in
\cite{Freidel:2017yuv}. However, the geometric \emph{origin} of the
DFT data, such as the C-bracket, the generalized Lie derivative and
the strong constraint, is not clarified within this approach. The
first main goal of the present paper is to establish such a geometric
origin for the structures appearing in DFT and to provide a precise
geometric definition of the corresponding algebroid. Similar goals
were pursued in \cite{Vaisman:2012ke,Deser:2014mxa,Deser:2016qkw,Heller:2016abk} from a different
standpoint, and we shall comment on the similarities and differences with this approach in the main text.

The second main goal of this paper is to use the relations between DFT and Courant algebroids to construct and study a membrane sigma-model, which is a worldvolume formulation of DFT. The starting point for this 
construction is a theorem of Roytenberg stating that there is a one-to-one
correspondence between Courant algebroids and QP2-manifolds
\cite{dee1}. Since the latter are the natural arena for the general
AKSZ construction in three worldvolume dimensions
\cite{Alexandrov:1995kv}, this essentially means that given the data
of a Courant algebroid one can construct, uniquely up to isomorphism,
a membrane sigma-model which is a three-dimensional topological field
theory. This is discussed in detail in \cite{dee2} (see also
\cite{Park:2000au,Ikeda:2002wh,Hofman:2002jz}). This result was
utilized in \cite{Mylonas:2012pg,Aschieri:2015roa} to explain the
origin of nonassociativity in locally non-geometric $R$-flux
backgrounds upon quantization. There it was already argued that the
target space for such models should be a doubled space, in particular
the total space of the cotangent bundle $T^{\ast}M$ of the original target space $M$. This proposal was studied further in \cite{Chatzistavrakidis:2015vka}, where a doubled membrane sigma-model was suggested, albeit without a complete geometric explanation. A similar construction in the language of supermanifolds appears in \cite{Bessho:2015tkk,Heller:2016abk}. 

Let us elaborate on the necessity of open membrane vs. closed string
sigma-models in this context. First, this is natural when non-trivial
fluxes are incorporated. Indeed, the very presence of an NS--NS flux
on a non-trivial background requires the introduction of a Wess-Zumino
term, which already means that one is working with an open membrane
whose worldvolume
boundary is the closed string worldsheet. From a different point of
view, the relationship between supergravity and generalized geometry
\cite{Coimbra:2011nw} indicates that Courant sigma-models,
which require a membrane worldvolume formulation, are the natural
sigma-models to consider. A third argument is related to
quantization. Recall that the deformation quantization of Poisson
manifolds~\cite{Kontsevich:1997vb} is given by the perturbative
expansion of the path integral for open strings in a $B$-field
background, the open Poisson
sigma-model~\cite{Cattaneo:1999fm}. Applying this reasoning in one higher
dimension, the quantization of closed strings with fluxes requires an open membrane sigma-model; further details are found in \cite{Mylonas:2012pg}.  

\subsubsection*{Summary of results and outline}

In order to achieve the goals of this paper that we discussed above,
we begin in Section~\ref{sec2} by considering a doubled spacetime. In this paper we do not
consider global aspects of doubled geometry, and we model the doubled space locally as the cotangent bundle{\footnote{Some progress on the global replacement of this doubled manifold has been reported in \cite{Freidel:2017yuv}.}} $T^{\ast}M$ of the standard target space $M$. The (doubled) local coordinates 
on this space may be identified as the dual momentum and dual winding
coordinates of DFT, or alternatively as phase space coordinates of an
open membrane with configuration space $M$. Since this space itself has the structure of a smooth manifold, one may consider an exact Courant algebroid over it, whose vector bundle is the second order bundle $E=T(T^{\ast}M)\oplus T^{\ast}(T^{\ast}M)$. The sections, symmetric bilinear form, Courant bracket and Dorfman derivative of this `large' Courant algebroid for arbitrary anchor are direct generalizations of the corresponding data of a Courant algebroid over $M$. However, these do not give rise directly to the corresponding DFT data. In order to establish this correspondence, we shall show that a particular splitting $E=L_+\oplus L_-$ should be constructed, accompanied by a projection $\mathsf{p}_+: E\to L_+$. DFT vectors, the constant $O(d,d)$-invariant metric, the C-bracket and the generalized Lie derivative are all obtained by suitably applying the projection map $\mathsf{p}_+$ on the large Courant algebroid data. Combining this with the known result that the DFT data reduce to the structure of a `canonical' Courant algebroid over an undoubled space $M$ when the strong constraint is imposed, our first result is that 
\begin{itemize}
\item The geometric structure of DFT lies in between two Courant
  algebroids, which may be depicted schematically as 
\be \nonumber
\small
\begin{matrix}\text{Large Courant algebroid}\\ \text{over
    $T^{\ast}M$}\end{matrix} \ \xrightarrow{ \ \mathsf{p}_+ \ } \ 
\text{DFT on $L_+$} \ \xrightarrow{ \ \rm{strong} \
} \ \begin{matrix}\text{Canonical Courant algebroid}\\ \text{over $M$}\end{matrix}
\ee
\normalsize
\end{itemize}
We emphasize that (i) projections to subbundles other than $L_+$ would not result in the desired structures, and 
(ii) $L_+$ is \emph{not} an involutive subbundle of $E$ and as such it does not correspond to a Dirac structure. 

Equiped with this result, we then use the one-to-one correspondence
between Courant algebroids and a class of membrane sigma-models to
construct the Courant sigma-model for the large Courant algebroid. As
expected, this does not directly give rise to the field content of
DFT, but instead the projection $\mathsf{p}_+$ should be used once more. This task results in a topological doubled sigma-model. A connection to the dynamics of string sigma-models can be reached by adding a symmetric boundary term. Then our second result is that 
 \begin{itemize}
 	\item The $O(d,d)$-invariant membrane sigma-model for DFT is given by the action functional
 	\bea  
 	S[\X , A,F]&=&\int_{\S_3}\,\big(F_I\w \dd\X^I+\eta_{IJ}\,A^I\w\dd A^J-(\rho_+)^{I}{}_{J}\,A^J\w F_I\big) \nn\\ && +\,\int_{\S_3}\,\sfrac 16\, T_{IJK}\,A^I\,\w A^J\w A^K+\int_{\partial\S_3}\,\sfrac 12\, {g}_{ I J}(\X)\, A^{ I}\w\ast A^{ J}~,\label{dftmsmintro}
 	\eea
 	where $\X=(\X^I): \S_3\to T^{\ast}M~, I=1,\dots,2d$, are maps
        from the membrane worldvolume $\Sigma_3$ to the doubled target
        space (pullbacks of the DFT coordinates), $A^I$ is a
        worldvolume 1-form (pullback of a DFT vector), $F_I$ is a worldvolume 2-form, and the rest of the quantities are explained in Section~\ref{sec24}, where a coordinate-independent formulation of the action is also given.    
 \end{itemize}

One direct test for the proposed DFT membrane sigma-model is whether it describes simultaneously all entries of the standard T-duality chain relating geometric and non-geometric flux configurations \cite{Shelton:2005cf}
\bea
H_{ijk}\stackrel{{\sf T}_k}{\longleftrightarrow}
f_{ij}{}^k\stackrel{{\sf T}_j}{\longleftrightarrow}
Q_i{}^{jk}\stackrel{{\sf T}_i}{\longleftrightarrow} R^{ijk} \ ,
\eea
where ${\sf T}_i$ denotes a T-duality transformation along $x^i\in
M$. In Section~\ref{sec3} we shall demonstrate that
\begin{itemize}
	\item All four T-dual backgrounds with $H$-, $f$-, $Q$- and $R$-flux are captured by \eqref{dftmsmintro}.
\end{itemize}
In particular, we shall explain how the T-fold is obtained in this
framework and provide a precise explanation of its relation to closed
string noncommutativity, thus filling a gap in the analysis of
\cite{Mylonas:2012pg}. Furthermore, we shall revisit the locally
non-geometric $R$-flux frame and confirm the previously obtained
result of \cite{Mylonas:2012pg} on the appearance of closed string
nonassociativity in this case; as expected, these noncommutative and
nonassociative polarizations violate the strong constraint of DFT. We comment on different types of $R$-flux, including a comparison with the Poisson $R$-flux sigma-model considered in \cite{Bessho:2015tkk}.

In Section~\ref{sec4} we investigate the relation of our membrane sigma-model to the flux formulation of DFT \cite{dftflux1,dftflux2,dftflux3,dftflux4}. Recall that invariance of the Courant sigma-model under gauge transformations is guaranteed by a set of conditions that may be identified as the local coordinate expressions of the Courant algebroid axioms \cite{Ikeda:2002wh}. From a different point of view, these expressions give the fluxes of generalized geometry and their Bianchi identities \cite{Halmagyi}. We shall show that gauge invariance of the DFT membrane sigma-model leads to the local coordinate expressions for the DFT fluxes (interpreted here as generalized Wess-Zumino terms) and Bianchi identities, as they appear e.g. in \cite{dftflux3}. One additional requirement for gauge invariance is identified as the analog of the strong constraint in this context, as expected. Thus the main result of Section~\ref{sec4} is that 
\begin{itemize}
	\item Gauge invariance of the membrane sigma-model \eqref{dftmsmintro} is compatible with the flux formulation of DFT.
	\end{itemize}   
Along the way we also find that, for special choices of structure
maps, the DFT membrane sigma-models reduce on their boundaries to the usual
worldsheet sigma-models for doubled target space geometries, as studied in e.g.~\cite{Hull:2009sg,Bakas:2016nxt},
and our worldvolume framework provides an alternative to the gauging
procedures for obtaining T-dual background configurations, such as
those discussed in Section~\ref{sec3}.

In Section~\ref{sec5} we exploit the structural similarity of the
expressions appearing in the flux formulation of DFT to the local
coordinate expressions for the axioms of a Courant algebroid to
reverse-engineer a precise geometric definition for the DFT structures. Our strategy is to replace the Courant algebroid data with the corresponding DFT data and examine which of the axioms of a Courant algebroid are obstructed. In this process, the origin of the strong constraint acquires a clear geometric explanation. We shall find that two of the Courant algebroid axioms, the Leibniz rule and the compatibility condition, are unobstructed and use them to define the structure of a DFT algebroid:
\begin{itemize}
	\item A DFT algebroid is the structure given by Definition~\ref{dftalg1}.
	\end{itemize}
We also demonstrate precisely how this definition reduces to a
canonical Courant algebroid upon solving the strong constraint, which
amounts to a choice of polarization, as in the explicit examples of Section~\ref{sec3}, and how $O(d,d)$-transformations corresponding to changes of polarization naturally give rise to isomorphisms of Courant algebroids, similarly to~\cite{cg}.

It is useful pointing out that the five Courant algebroid axioms of
\cite{liu}, which we recall in Appendix~\ref{seca}, are not a minimal
set, since two of them (the homomorphism property of the anchor and
the image of the derivation lying in the kernel of the anchor) follow
from the rest, as shown for example in \cite{Uchino2002}. This is no
longer the case when the Jacobi identity is relaxed, as in the notion
of a pre-Courant algebroid \cite{preca} or Courant algebroid twisted
by a 4-form~\cite{Hansen:2009zd}. In such cases, the two additional
properties should be included in the set of axioms. However, one may
consider relaxing these properties as well, and moreover in an independent
way. As we discuss in Appendix~\ref{seca}, two additional geometric
structures may be defined in this fashion, which we call ante-Courant
algebroid (where only the homomorphism property is relaxed) and
pre-DFT algebroid (where both additional properties are relaxed). The 
latter is a metric algebroid in the terminology of~\cite{Vaisman:2012ke}; 
it has a corresponding realization in the language of graded
geometry and is called a symplectic nearly Lie 2-algebroid
\cite{preca2}. Our results imply that a DFT algebroid is a special
case of a pre-DFT algebroid in which imposing that the image of the
derivation is in the kernel of the anchor reduces it directly to a
Courant algebroid, without passing through the intermediate structures
of ante-Courant and pre-Courant algebroids. All cases may be
characterized in terms of an underlying $L_\infty$-algebra
structure~\cite{prehequiv,Deser:2016qkw}. In Appendix~\ref{seca4} we
provide examples highlighting the features of each of these structures. 

The structure of a pre-DFT algebroid suggests a natural geometric
weakening of the strong constraint of DFT. The final problem we address in this paper is whether a generalized
doubled membrane sigma-model can be constructed whose gauge invariance
does not rely on the strong constraint. A key element in our approach
to this problem is relaxing the assumption that the fiber metric of
the underlying algebroid is constant. This indicates a
departure from DFT, where the $O(d,d)$-invariant metric is
constant. However, non-constant fiber metrics were considered before,
for example in \cite{Freidel:2017yuv,Hansen:2009zd}. We shall show
that this new ingredient in principle allows us to dispense with the
strong constraint, as long as a certain partial differential equation
for the fiber metric is satisfied. This appears in Section~\ref{sec6}, where we also discuss the closure of the algebra of sigma-model gauge transformations for both constant and non-constant fiber metric. 

\paragraph{Note added.} After completion of this work, the paper \cite{Svoboda:2018rci} appeared, where global aspects of DFT in 
the framework of para-Hermitian manifolds are discussed with some overlapping similarities. 

\section{From doubled membrane sigma-models to DFT}
\label{sec2}

In this section we will derive the $O(d,d)$-invariant open membrane
sigma-model associated to DFT, whose boundary dynamics will govern the
motion of
closed strings in backgrounds with both geometric and non-geometric fluxes in a manifestly T-duality invariant way.

\subsection{Courant algebroids and membrane sigma-models}
\label{sec21}

We consider as starting point a theorem of Roytenberg stating that there is a one-to-one
correspondence between Courant algebroids and QP2-manifolds
\cite{dee1}.\footnote{See Appendix~\ref{seca} for relevant details
  about Courant algebroids, including their definition and properties
  (together with local coordinate expressions), and some examples.}
Since the latter are the natural arena for the general
AKSZ construction in three dimensions, this essentially means that from a Courant algebroid one can construct uniquely, up to isomorphism, a membrane sigma-model which is a three-dimensional topological field theory. This is discussed in detail in \cite{dee2}.

The full BV action, including ghosts, antifields and ghosts-for-ghosts, is constructed in \cite{dee2}, but in this paper we shall focus only on the
classical ``bosonic'' action obtained by setting all of the latter
fields to zero. We are exclusively interested in exact Courant algebroids (with a Lagrangian splitting), whose underlying vector bundle over a manifold $M$ of $d$ dimensions is $E=TM\oplus T^\ast M$.
This defines a standard membrane sigma-model with action
\beqa \label{scasm}
S_{0}[X,A,F]&=&\int_{\S_{3}}\, \big(F_i\w\dd X^i+\sfrac 12\, \eta_{IJ}\,
  A^I\w\dd A^J-\rho^i{}_{I}(X)\, A^I\w F_i \nonumber \\ && \qquad
  \qquad +\, \sfrac 16\, T_{IJK}(X)\, A^I\w A^J\w A^K\big)~,
\label{csm0}\eeqa
where $\S_3$ is the membrane worldvolume, $X=(X^i):\S_3\to M$ is the mapping of the worldvolume to the target space $M$, 
$A\in \Omega^1(\S_3,X^{\ast}E)$ is a worldvolume 1-form valued in $E$, 
and $F\in \Omega^2(\S_3,X^{\ast}T^{\ast}M)$ is an auxiliary 
worldvolume 2-form with values in the cotangent bundle of $M$. 
The index ranges are $i=1,\dots,d$ (target space) and $I=1,\dots,2d$ (algebroid).
The tensor
\be \label{eta}
\eta=(\eta_{IJ})=\begin{pmatrix}
	0 & 1_d \\ 
	1_d & 0
\end{pmatrix}
\ee
is the matrix of the symmetric bilinear form of the Courant algebroid, which defines an $O(d,d)$-invariant metric,\footnote{In the following capital Latin indices $I,J,\dots$ are raised and lowered with this metric.} and 
$\rho$ and $T$ are the anchor and twist of the Courant algebroid,
respectively, with the latter generating a generalized Wess-Zumino term. 
This shows that given the data of a Courant algebroid over $M$, i.e. a quadruple $(E,[\, \cdot \,,\, \cdot \,]_E,\langle\, \cdot \,,\, \cdot \,\rangle_{E},\rho)$ (see Appendix \ref{seca}), 
one can uniquely reconstruct the action \eqref{csm0}, which is thereby called a Courant sigma-model. 
This becomes particularly transparent if we write the action in
basis-independent form as 
\be \label{csmif}
S_{0}[X,A,F]=\int_{\S_{3}}\, \big(\langle F,\dd X\rangle+ \langle A,\dd A\rangle_E-\langle F,\rho(A)\rangle+
\sfrac 13\, \langle A,[A,A]_{E}\rangle_E\big)~,
\ee
where the bilinear form $\langle\, \cdot \,,\, \cdot \,\rangle$ (without subscript) is
the canonical dual pairing between the tangent and cotangent bundles; in the case of exact Courant algebroids, the two pairings are essentially identical.
This action indeed contains just the anchor, the bracket and the bilinear form of $E$. The bracket is the 
Courant bracket twisted by a generalized 3-form $T$. Denoting $A=A_V+A_F\in \G(E)$ where $A_V\in\G(TM)$ and
$A_F\in\G(T^{\ast}M)$, it is given as{\footnote{Here $\dd$ and
    $\dd_{\ast}$ are exterior differentials increasing the $p$-form and $p$-vector degree by one, respectively.}}
\bea 
[A,B]_E&=&[A_V,B_V]+{\cal L}_{A_F}B_V-{\cal L}_{B_F}A_V+\sfrac 12\, \dd_{\ast} (\iota_{A_V}B_F-\iota_{B_V}A_F)
\nn\\
&+&[A_F,B_F]+{\cal L}_{A_V}B_F-{\cal L}_{B_V}A_F-\sfrac 12\, \dd (\iota_{A_V}B_F-\iota_{B_V}A_F)+T(A,B)~.\label{generalcourant}
\eea  
It is precisely its last term, the twist, that yields the generalized
Wess-Zumino term in \eqref{csm0} from the last term in \eqref{csmif},
since all the other terms in $[A,A]_E$ are trivially zero. (The factor
of 2 difference is due to the fact that the non-degenerate bilinear
form is defined as $\langle A,B\rangle_E=\sfrac 12\, \eta_{IJ}\, A^I\,B^J$.) A special case of this bracket is the more familiar twisted Courant bracket that corresponds to the standard Courant algebroid, where the anchor is chosen to be the projection to the tangent bundle, which reads as 
\bea 
[A,B]_{{\rm s}E}=[A_V,B_V]+{\cal L}_{A_V}B_F-{\cal L}_{B_V}A_F-\sfrac 12\, \dd (\iota_{A_V}B_F-\iota_{B_V}A_F)+T(A_V,B_V)~.
\eea  
Courant algebroids with arbitrary anchor are not however compatible with this choice of bracket, but only with the general bracket \eqref{generalcourant}.

We can summarize the present discussion as
\begin{itemize}
\item	Given the data of a Courant algebroid one can write a unique membrane sigma-model, whose action is given in \eqref{csm0}.
\end{itemize}

\subsection{Doubling the target space}
\label{sec22}

In order to make contact with DFT, we would like to double the target space of the membrane sigma-model; 
for the purposes of this paper, we therefore take the target space to be the cotangent bundle $T^{\ast}M$ 
instead of the original manifold $M$. This is possible because the
total space of $T^{\ast}M$ has itself the structure of a smooth manifold. 
Such an approach has been advocated previously in \cite{Mylonas:2012pg,Freidel:2015pka,Aschieri:2015roa,Chatzistavrakidis:2015vka,Blumenhagen:2016vpb,Deser:2016qkw,Aschieri:2017sug}.
As most of our considerations in the following will be local, we can
assume $M$ is contractible and thus identify\footnote{This
  identification holds more generally when $M$ is only required to be
  parallelizable, which will be the case for some of the examples we
  discuss in Section~\ref{sec3}.} $T^{\ast}M=M\times (\R^d)^{\ast}$, which we equip with local coordinates $(x,p)$ where $x=(x^i)$ are local coordinates on the base manifold $M$ and $p=(p_i)$ are local fiber coordinates. This provides a local model for the doubled spacetime of DFT, with $p$ playing the role of winding coordinates which are T-dual to $x$. Alternatively, we may wish to regard $T^\ast M$ as the kinematical phase space of the membrane configuration space $M$, with $p$ the dual momentum coordinates to $x$ with respect to the canonical symplectic form. The relations between these two perspectives are discussed in~\cite{Freidel:2015pka,Aschieri:2015roa,Blumenhagen:2016vpb,Aschieri:2017sug}, and we shall refer to both points of view interchangeably in what follows. 

To write down the open membrane sigma-model, we consider a map 
\be 
\X:\Sigma_3\longrightarrow T^{\ast}M~.
\ee
The components of this map are denoted
\be 
\X=(\X^I)=(\X^i,\X_i)=:(X^i,\widetilde X_i)~. 
\ee
The fields $X^i$ and $\widetilde X_i$ are thus identified with the pullbacks of the coordinate functions, i.e.~$X^i=\X^{\ast}(x^i)$ and $\widetilde X_i=\X^{\ast}(p_i)$.

We take the vector bundle $E=\mathbb T(T^{\ast}M):=T(T^{\ast}M)\oplus T^{\ast}(T^{\ast}M)$, which is a second-order bundle over $M$, being the generalized tangent 
bundle of the cotangent bundle of $M$.
We introduce a worldvolume 1-form{\footnote{In what follows we use blackboard bold typeface style for quantities in the Courant algebroid $E$, and we reserve ordinary typeface style for DFT quantities to be encountered later in this section, e.g. $\A$ vs. $A$.}} $\A\in
\Omega^1(\S_3,\X^{\ast}\mathbb T(T^{\ast}M))$ and an auxiliary 
worldvolume 2-form $\F\in \Omega^2(\S_3,\X^{\ast}T^{\ast}(T^{\ast}M))$. The Courant sigma-model is given by the 
coordinate-free action functional
\be \label{csm1}
S[\X,\A,\F]=\int_{\S_{3}}\, \big(\langle \F,\dd \X\rangle+ \langle \A,\dd \A\rangle_E-\langle \F,\rho(\A)\rangle+
\sfrac 13\,\langle \A,[\A,\A]_E\rangle_E\big)~.
\ee
 This action is formally the same as \eqref{csm0} with $M$ substituted by its cotangent bundle 
 and the various fields living over the corresponding bundles. 
In local coordinate form, the action functional may be written as 
\be \label{aksz}
S[\X,\A,\F]=\int_{\S_{3}}\, \big(\F_I\w\dd \X^I+ \sfrac 12\,\eta_{\hat I\hat J}\,\A^{\hat I}\w\dd \A^{\hat J}-
\rho^I{}_{\hat I}(\X)\,\A^{\hat I}\w \F_I+
\sfrac 16\,T_{\hat I\hat J\hat K}(\X)\,\A^{\hat I}\w \A^{\hat J}\w \A^{\hat K}\big)
\ee
where $I=1,\dots, 2d$ and the extended Courant algebroid index is now $\hat I=1,\dots, 4d$.
Finally, we add a general symmetric term to the action on the boundary of the membrane which is given by
\be \label{symboundary}
S_{\text{sym}}[\X,\A]=\int_{\partial\S_3}\,\|\A\|_{{g}}:= \int_{\partial\S_3}\,\sfrac 12\, {g}_{\hat I\hat J}(\X)\, \A^{\hat I}\w\ast \A^{\hat J}~,
\ee
where $g\in\Gamma(\mathbb{T}(T^{\ast}M) \otimes
\mathbb{T}(T^{\ast}M))$ is a (possibly degenerate) symmetric
generalized $(2,0)$-tensor and $\ast$ is the Hodge duality operator
with respect to a chosen Riemannian metric on the worldsheet
$\partial\S_3$; this term breaks the topological symmetry of the
Courant sigma-model on the boundary. To completely define the action,
one should of course also specify suitable boundary conditions on
$\partial\S_3$; we shall address this point later.

So far we have not achieved much. We merely wrote the membrane sigma-model for a doubled target space. 
It is clear that this cannot be directly associated to DFT. The reason is that by doubling both the target space and the bundle over it, we slightly ``overdoubled''. For instance, the fields $\A^{\hat{I}}$ have too many components to be associated with DFT vectors. In other words, the membrane sigma-model over $M$ carries less 
information than DFT, while the one over $T^{\ast}M$ carries too much information. Clearly, we should be looking for 
something in between, and below we shall construct a suitable DFT membrane sigma-model as a restriction of the Courant sigma-model on the doubled space.

\subsection{Projecting the large Courant algebroid to DFT}
\label{sec23a}

Recall that the data needed to define a Courant algebroid and the
corresponding Courant sigma-model are a vector bundle $E$ over a
manifold $M$, together with a skew-symmetric bracket and a symmetric
bilinear form on its sections, and a map $\rho$ from $E$ to the
tangent bundle $TM$, as discussed before. Here we take the vector
bundle
\be
E=\mathbb T(T^{\ast}M)=T(T^{\ast}M)\oplus T^{\ast}(T^{\ast}M) \ , 
\ee
the generalized tangent bundle of the cotangent bundle of $M$, with sections $(\A^{\hat I})=(\A^I,\widetilde \A_I)=(\A^i,\A_i,\widetilde \A_i,\widetilde \A^i)$ and
\be
\A=\A_{V}+\A_{F}:=\A^I\,\partial_I+\widetilde \A_I\,\dd\X^I~,
\ee
where we defined basis vectors and forms on $T^\ast M$ as
$(\dd\X^I):=(\dd X^i,\dd \widetilde X_i)$ and
$(\partial_I)=(\partial/\partial X^i,\partial/\partial \widetilde
X_i)=:(\partial_i,\tilde\partial^i)$. By the \emph{large} Courant
  algebroid we mean the vector bundle $E$ over $T^*M$ with the symmetric bilinear form constructed using the usual contraction of vectors and 1-forms, 
\be
\langle \A,\B\rangle_E=\sfrac 12\, \big(\A^I\,\widetilde \B_I+\widetilde \A_I\, \B^I\big)=\sfrac 12\, \eta_{\hat{I}\hat{J}}\,\A^{\hat{I}}\,\B^{\hat{J}}~,
\ee
and the bracket on $\Gamma(E)$ given by the twisted Courant bracket
with twist{\footnote{Note that this is an ``$H$-type'' twist from the
    perspective of the large Courant algebroid. In other words, in
    this subsection we take the large Courant algebroid $E$ over the
    doubled space to be the standard one, though we do not indicate it
    explicitly in the notation for the Courant bracket, as our results immediately generalize
    to any Courant algebroid as we discuss later on.}} $T$,
\bea\label{cbrac}
[\A,\B]_E=
[\A_V,\B_V]+{\cal L}_{\A_V}\B_{F}-{\cal L}_{\B_V}\A_{F}-\sfrac 12\, \dd (\iota_{\A_V}\B_F-\iota_{\B_V}\A_F)+\iota_{\B_V}\iota_{\A_V}T ~.
\eea
Here we have introduced the standard Lie derivative along a vector on $T^{\ast} M$ acting on forms, 
\be
{\cal L}_{\A_V}=\dd\circ \iota_{\A_V}+\iota_{\A_V}\circ\dd~,
\ee
where $\iota_{\A_V}=\iota_{\A^I\,\partial_I}$ denotes contraction along the vector, and the exterior derivative is expanded as $\dd f=\partial_I f\,\dd\X^I=\partial_if\,\dd X^i+\tilde\partial^if\,\dd\widetilde X_i$ for a function $f(\X)$.
Written in components the Courant bracket \eqref{cbrac} becomes
\bea\label{cbracomp}
[\A,\B]_E&=&\big(\A^I\,\partial_I\B^J-\B^I\,\partial_I\A^J\big)\,\partial_J \nn \\
&+&\big(\A^I\,\partial_I\widetilde \B_J-\B^I\,\partial_I\widetilde \A_J-\sfrac 12\,(\A^I\,\partial_J\widetilde \B_I-\widetilde \B_I\,\partial_J\A^I-\B^I\,\partial_J\widetilde \A_I+\widetilde \A_I\,\partial_J\B^I)\big)\,\dd\X^J \nn\\ &+& T_{IJK}\,\A^I\,\B^J\,\dd \X^K~.
\eea
Our aim now is to extract the various fields and geometric operations of DFT from this large Courant algebroid structure.

\subsubsection*{DFT vectors}

It is convenient to introduce the notation
\be\label{polarise}
\A_{\pm}^I  =\sfrac 12\,\big( \A^I\pm\eta^{IJ}\,\widetilde \A_J\big)~, 
\ee
and rewrite everything in terms of $\A_\pm$ using the inverse relations 
\bea 
\A^I=\A_{+}^I+\A_{-}^I \qquad \text{and} \qquad \widetilde{\A}_I=\eta_{IJ}\,\big(\A_+^J-\A_-^J\big)~.\label{plusminusinv}
\eea
A crucial point in this discussion is that the metric $\eta_{IJ}$ appearing here is the $O(d,d)$-invariant metric and \emph{not} the metric $\eta_{\hat{I}\hat{J}}$ of the Courant algebroid structure on $E=\mathbb T(T^{\ast}M)$. Thus, although our starting point is the large Courant algebroid $E$ over $T^{\ast}M$, here some information of 
a `small' algebroid over $M$ enters. However, for the time being we do not even consider the latter structure; we simply use the fixed tensor \eqref{eta} to rotate the components of a generalized vector
of $E$. In other words, this structure is already present in the large Courant algebroid as becomes manifest from
\be
\langle \A,\B\rangle_E=\sfrac 12\, \eta_{\hat{I}\hat{J}}\,\A^{\hat{I}}\,\B^{\hat{J}}=\eta_{IJ}\,\big(\A^I_+\,\B^J_+-\A^I_-\,\B^J_-\big)~.
\ee
The generalized vector is then given as
\be
\A=\A^I\,\partial_I+\widetilde \A_I\,\dd\X^I=\A_+^I\,e_I^++\A_-^I\,e_I^-~,\ee
where we defined 
\be\label{eq:eIpm}
e^{\pm}_I=\partial_I\pm \eta_{IJ}\,\dd\X^J~.
\ee  
One then notices that taking the components $\A_-^I=0$ and renaming $\A_+^I=A^I$ leads to a special generalized vector of $E$ given by
\be\label{dftvector}
A=A_i\,\big(\dd X^i+\tilde\partial^i\big)+A^i\,\big(\dd \widetilde X_i+\partial_i\big)~.
\ee
This is precisely a DFT vector, as written e.g. in \cite{Deser:2016qkw}. 

However, setting some components of the vector $\A$ to zero is not a good operation, since it is not invariant. Alternatively, we note that the local frame \eqref{eq:eIpm} defines a decomposition of the generalized tangent bundle as 
\be\label{eq:EL+L-}
E=\mathbb{T}(T^\ast M) = L_+\oplus L_- \ , 
\ee
where $L_\pm$ is the bundle whose space of sections is spanned locally by $e_I^\pm$. Then 
the same special set of sections \eqref{dftvector} may be reached by projection to the subbundle $L_+$ of $E$ by introducing the bundle map 
\bea \label{pplus}
\sfp_+: E \longrightarrow L_+ \ , \quad
(\A_V,\A_F) \longmapsto \A_+:=A~.
\eea 
This indeed gives 
\be 
\sfp_+(\A)=\A_+=\A_+^I\,e_I^+=\sfrac 12\,\big(\A_i+\widetilde{\A}_i\big)\,\big(\dd X^i+\tilde{\partial}^i\big)+\sfrac 12\, \big(\A^i+\widetilde{\A}^i\big)\,\big(\dd \widetilde X_i+\partial_i\big)~,
\ee
which is identical to \eqref{dftvector} upon identifying $A_i=\sfrac 12\,\big(\A_i+\widetilde{\A}_i\big)$ and $A^i=\sfrac 12\, \big(\A^i+\widetilde{\A}^i\big)$. 
 The pairing of two such vectors $A=\sfp_+(\A)$ and $B=\sfp_+(\B)$, called DFT vectors from now on, is 
\be 
\langle A,B\rangle_{L_+} = A_i\,B^i+A^i\,B_i=\eta_{IJ}\,A^I\,B^J~,
\ee 
as expected in DFT. Retrospectively, we observe why the introduction of the splitting \eqref{eq:EL+L-} is necessary: Had we attempted to project to $T(T^{\ast}M)$ or $T^{\ast}(T^{\ast}M)$, we would have not been able to derive the $O(d,d)$-structure from the large Courant algebroid in this way. The same is true of the C-bracket and the generalized Lie derivative, as we show below.

\subsubsection*{C-bracket}

Let us now reconsider the Courant bracket \eqref{cbracomp} of $E$ in
light of the above result: Is the projection $\sfp_+$ sufficient to
reduce the large Courant bracket to the C-bracket of DFT? For this,
let us rewrite the Courant bracket \eqref{cbracomp} in terms of
$\A^I_\pm$, setting the twist $T$ to zero for the moment. We find
\bea
[\A,\B]_E&=&\eta_{IK}\,\big( (\A^K_+\,\partial^I\B^L_+-\B^K_+\,\partial^I\A_+^L+\A^K_-\,\partial^I\B^L_+-\B^K_-\,\partial^I\A_+^L)\,e^+_L \nn\\
 && \quad \ +\,(\A^K_+\,\partial^I\B^L_--\B^K_+\,\partial^I\A_-^L+\A^K_-\,\partial^I\B^L_--\B^K_-\,\partial^I\A_-^L)\,e_L^- \\
&& \quad \ -\, (\A^K_+\,\partial^L\B^I_++\B^K_-\,\partial^L\A_-^I-\A^K_-\,\partial^L\B^I_--\B^K_+\,\partial^L\A_+^I)\,\eta_{LM}\, \dd \X^M\big)~. \nn
\eea
We can rewrite the last term using $\eta_{LM}\,\dd\X^M=\sfrac 12\, (e_L^+-e^-_L)$ to obtain
\bea
[\A,\B]_E&=& \eta_{IK}\,\big(\A^K_+\,\partial^I\B^L_+ +\A^K_-\,\partial^I\B^L_+-\sfrac 12\, (\A^K_+\,\partial^L\B^I_+ -\A^K_-\,\partial^L\B^I_- )-\{\A\leftrightarrow \B\}\big)\,e^+_L \nn\\
 &+&\eta_{IK}\,\big(\A^K_+\,\partial^I\B^L_- +\A^K_-\,\partial^I\B^L_- +\sfrac 12\,(\A^K_+\,\partial^L\B^I_+ -\A^K_-\,\partial^L\B^I_-)-\{\A\leftrightarrow \B\}\big)\,e_L^-~.\nn\\\label{courantdft}
\eea
 This form of the Courant bracket should be compared with the C-bracket of DFT vectors, which reads as{\footnote{ We denote the C-bracket by double brackets, as in e.g. \cite{Freidel:2017yuv}.}} 
\bea\label{cbra}
\cbral A,B \cbrar_{L_+}^J= A^K\,\partial_K B^J-\sfrac 12\, A^K\,\partial^JB_K - \{A\leftrightarrow B\}~.\eea
Clearly, projecting with the map $\sfp_+$, i.e. taking the Courant bracket of DFT vectors $[\sfp_+(\A),\sfp_+(\B)]_E$, eliminates  the components $\A^I_-$ and $\B^I_-$ from the right-hand side of \eqref{courantdft}.  However, this is not sufficient in order to reduce to the C-bracket. This happens because the Courant bracket on $L_+$ is not a closed operation, or in other words $L_+$ is not an involutive subbundle of $E$ (and thus neither a Dirac structure), in contrast to the subbundles $T(T^*M)$ and $T^*(T^*M)$ which themselves become Lie algebroids under the respective restrictions of the Courant bracket and
anchor of $E$. Additionally, a further 
projection of the $E$-section which is the result of the operation $[\sfp_+(\A),\sfp_+(\B)]_E$ is necessary. 
More precisely, this may be expressed as a relation between the Courant bracket on $E$ and the C-bracket on $L_+$ given by
\begin{empheq}
[box = {\alphabox[C-bracket on $L_+$ vs. Courant bracket on $E$]}]
{equation}
\cbral A,B\cbrar_{L_+}=\sfp_+\big([\sfp_+(\A),\sfp_+(\B)]_E\big) \label{couranttoc}
\end{empheq}
Note that this differs from the result of~\cite{Freidel:2017yuv}
where the C-bracket is related to the Courant bracket of a `small'
Courant algebroid, whereas our relation involves the 
Courant bracket on the \emph{large} Courant algebroid $E$.
This indicates that for each extra operation in DFT, one has to
perform anew a projection from the Courant algebroid structure on
$E$. 

\subsubsection*{Generalized Lie derivative}

Let us also examine the reduction of the Dorfman derivative to the generalized Lie derivative of DFT. The Dorfman derivative for the standard Courant algebroid is defined as 
\bea
\L_{\A}{\B}=[\A_V,\B_V]+{\cal L}_{\A_V}\B_F-\iota_{\B_V}\dd \A_F ~,
\eea 
and its antisymmetrization yields the Courant bracket
\be 
[\A,\B]_E=\L_{\A}\B-\L_{\B}\A~.
\ee 
Rewritten in terms of the redefined components $\A^\pm$, the Dorfman derivative takes the form 
\bea
\L_{\A}{\B}&=&\eta_{IK}\,\big(\A^K_+\,\partial^I\B^L_+-\B^K_+\,\partial^I\A_+^L+\A^K_-\,\partial^I\B^L_+-\B^K_-\,\partial^I\A_+^L+ \B^K_+\,\partial^L\A_+^I-\B^K_-\,\partial^L\A_-^I\big)\,e^+_L \nn\\
 &+&   \eta_{IK}\,\big(\A^K_+\,\partial^I\B^L_--\B^K_+\,\partial^I\A_-^L+\A^K_-\,\partial^I\B^L_--\B^K_-\,\partial^I\A_-^L-\B^K_+\,\partial^L\A_+^I+\B^K_-\,\partial^L\A_-^I\big)\,e^-_L~.\nn\\\label{dorfcomp}
\eea
Then it is evident that taking the Dorfman derivative of
$\sfp_+$-projected vectors, which effectively amounts to setting
$\A_-^I=\B_-^I=0$, and recalling that $\A_+^I=A^I$ and $\B_+^I=B^I$, we obtain
\bea
\L_AB=\eta_{IK}\,\big(A^K\,\partial^IB^L-B^K\,\partial^IA^L+B^K\,\partial^LA^I \big)\,e_L^+-\eta_{IK}\,B^K\,\partial^LA^I\, e_L^- \ .
\eea
 When restricted to $L_+$ via the map $\sfp_+$, this expression corresponds to the standard one for the generalized Lie derivative in DFT given by
\be 
(\sfL_AB)^J=A^I\,\partial_IB^J-B^I\,\partial_IA^J+B_I\,\partial^JA^I~. 
\ee
Equivalently, the relation between the two derivatives may be expressed in the form 
\begin{empheq}
[box = {\betabox[DFT generalized Lie derivative on $L_+$ vs. Dorfman derivative on $E$]}]
{equation}
\sfL_{A}B=\sfp_+\big(\L_{\sfp_+(\A)}\sfp_+(\B)\big) 
\end{empheq}

The discussion of the generalized Lie derivative of DFT raises one
more question: What is the role of the strong constraint of DFT here?
The Dorfman derivative \eqref{dorfcomp} over the large Courant
algebroid $E$ automatically satisfies the closure identity 
\be 
[\L_\A,\L_\C]=\L_{[\A,\C]_E}~.
\ee
 However, what happens when we calculate this expression for $\sfp_+$-projected derivatives, i.e. for generalized Lie derivatives of DFT?
  Although the result is well-known, let us for completeness repeat the argument here.
We have
\bea
\sfL_C\sfL_A B &=&\eta_{IK}\,\eta_{JM}\,\big(C^K\,\partial^I(A^J\,\partial^MB^L-B^J\,\partial^MA^L+B^J\,\partial^LA^M) \nn\\
&& +\, (A^J\,\partial^MB^K-B^J\,\partial^MA^K+B^J\,\partial^KA^M)\,(\partial^LC^I-\partial^IC^L)\big)\,e_L^+~,
\eea
and
\bea
\sfL_{\cbral C,A\cbrar_{L_+}}B&=&\eta_{IK}\,\eta_{JM}\,\big(\partial^I B^L\,(C^M\,\partial^JA^K-A^M\,\partial^JC^K-\sfrac 12\, C^M\,\partial^KA^J+\sfrac 12\, A^M\,\partial^K C^J) \nn\\
&& -\, B^K\,\partial^I(C^M\,\partial^JA^L-A^M\,\partial^JC^L-\sfrac 12\, C^M\,\partial^LA^J+\sfrac 12\, A^M\,\partial^L C^J) \nn\\
&& +\, B^K\,\partial^L(C^M\,\partial^JA^I-A^M\,\partial^JC^I-\sfrac 12\, C^M\,\partial^IA^J+\sfrac 12\, A^M\,\partial^I C^J)\big)\,e_L^+~,
\eea
giving altogether
\bea
\big([\sfL_C,\sfL_A]-\sfL_{\cbral
  C,A\cbrar_{L_+}}\big) B&=&\eta_{IK}\,\eta_{JM}\,\big(B^J\,\partial^KC^M\, \partial^IA^L-B^J\,\partial^KA^M\,\partial^IC^L
\nn \\
&& +\, \sfrac 12\, C^M\,\partial^KA^J\,\partial^I B^L-\sfrac 12\,
A^M\,\partial^KC^J\,\partial^I B^L\big)\,e_L^+~. \label{constraint}
\eea
The right-hand side of \eqref{constraint} corresponds to the result obtained in \cite[eq.~(3.24)]{dft4}, giving the strong constraint 
\be \label{eq:strongconstraint}
\eta^{IJ}\,\partial_I f\,\partial_Jg=0~,
\ee
for all fields $f,g$ of DFT.
The situation is summarized schematically in the diagram
\be
\begin{tikzcd}
\big(E\,,\,\L_{\, \cdot \,}\, \cdot \,\,,\,\rho\,,\,\langle\, \cdot \,\,,\,\, \cdot \,\rangle_E\big) \arrow[dashed]{r}{} \arrow[swap]{d}{\sfp_+} & {[}\L,\L]-\L_{[\, \cdot \,,\, \cdot \,]_E}=0 \arrow{d}{\sfp_+} \\%
\big(L_+\,,\,\sfL_{\, \cdot \,}\, \cdot \,\,,\,\rho_+\,,\,\langle\, \cdot \,\,,\,\, \cdot \,\rangle_{L_+}\big) \arrow[swap,dashed]{r}{\text{strong}}& 
{[}\sfL,\sfL]-\sfL_{\cbral\, \cdot \,,\, \cdot \,\cbrar_{L_+}}=0
\end{tikzcd}
\ee
The horizontal arrows here are not maps, but implications of the
structure maps from the left. In the upper-left corner we encounter the
large Courant algebroid over $T^{\ast}M$. This is projected to the
corresponding structure of DFT, appearing in the lower-left corner. As we will discuss momentarily, the latter structure does not constitute a Courant algebroid. 
The upper-right corner contains the closure identity
for Dorfman derivatives of $E$. This is trivially projected to the
corresponding closure identity for 
generalized Lie derivatives of DFT. However, in order to reach this
lower-right corner at the level of the DFT structure, the strong
constraint is required. Thus we have shown that starting from the
large Courant algebroid over the doubled target space we can obtain
known DFT structures by choosing a suitable projection map $\sfp_+$ in
\eqref{pplus}.  Furthermore, in Section 5 we shall utilize the
structure of large Courant algebroid to obtain a geometric
interpretation of the strong constraint itself and DFT data in general. 

\subsection{Projecting to the DFT membrane sigma-model}
\label{sec24}

We are now ready for our original goal, which is to find the $O(d,d)$-invariant membrane sigma-model that corresponds to DFT. The way to do this is to rewrite the Courant sigma-model over 
$E$ in terms of $\A^I_{\pm}$ and $e^{\pm}_I$, and then impose the projection we found above. Focusing first on the topological sector, the action 
\eqref{aksz} may be written as
\bea 
S&=&\int_{\S_3}\, \Big(\F_I\w \dd \X^I+\eta_{IJ}\,\big(\A_+^I\w\dd \A_+^J-\A_-^I\w\dd \A_-^J\big)-\big((\rho_+)^{I}{}_{K}\,\A^K_{+}+(\rho_-)^I{}_K\,\A_-^K\big)\w \F_I \nn\\ && \qquad \qquad +\, \sfrac 16\, T_{IJK}\,\A^I_+\w \A^J_+\w \A^K_++\sfrac 12\, T'_{IJK}\,\A^I_-\w \A^J_+\w \A^K_+ \nn\\ && \qquad \qquad +\,\sfrac 12\, T''_{IJK}\,\A^I_+\w \A^J_-\w \A^K_-+\sfrac 16\, T'''_{IJK}\,\A^I_-\w \A^J_-\w \A^K_-\Big)~,
\eea 
where, with respect to the anchor $\rho^I{}_{\hat{J}}=(\rho^I{}_J,\widetilde\rho^{\,IJ})$ of $E$, we defined 
\bea
(\rho_{\pm})^{I}{}_{J}=\rho^{I}{}_{J}\pm \eta_{JK}\,\widetilde \rho^{\,IK}~,
\eea 
which are maps from $L_\pm$ to the tangent bundle $T(T^*M)=TM\oplus T(\R^d)^\ast$ on the doubled space. 
The components of $T,T',T'',T'''$ are combinations of the twist
components
\be
T_{\hat{I}\hat{J}\hat{K}}:=\begin{pmatrix}A_{IJK}&B_{IJ}{}^K\\
  C_{I}{}^{JK}&D^{IJK} \end{pmatrix} \ .
\ee 
Their explicit expressions are not important for our purposes, apart from the first one, which is equal to 
\be 
T_{IJK}=A_{IJK}+3\,B_{[IJ}{}^L\,\eta_{K]L}+3\,C_{[I}{}^{LM}\,\eta_{J\underline{L}}\,\eta_{K]M}+D^{LMN}\,\eta_{[I\underline{L}}\,\eta_{J\underline{M}}\,\eta_{K]N}~,
\ee 
where the underlined indices are not antisymmetrized.
Now we project with the map $\sfp_+$, i.e. we impose $\A_-^I=0$, and identify $\A^I_+= A^I$ and $\F_I=F_I$. 
 The resulting action is{\footnote{A remark is in order regarding the
     generalized Wess-Zumino term here. When the twisted Courant
     bracket is considered, the projected twisted C-bracket is
     obtained with a twist $\sfrac 12\, T$. In more precise terms,
     taking the twisted brackets $[\A,\B]_{{\rm t}
       E}:=[\A,\B]_E+T(\A,\B)$ and $\cbral A,B\cbrar_{{\rm t}
       L_+}:=\cbral A,B\cbrar_{L_+}+{\hat T}(A,B)$, their relation is found to be 
 \be 
\sfp_+\big([\sfp_+(\A),\sfp_+(\B)]_{{\rm t} E}\big)=\cbral
A,B\cbrar_{L_+}+\sfp_+\big(T(A,B)\big)=\cbral A,B\cbrar_{L_+}+\sfrac
12\, T(A,B) \nn
 \ee 
giving $\hat{T}(A,B)=\sfrac 12\, T(A,B)$. Thus the Wess-Zumino term can also be written as $\sfrac 13\, \hat{T}_{IJK}\,A^I\w A^J\w A^K$ in terms of the C-bracket twist.\label{footnote}}}
\begin{empheq}
[box = {\gammabox[DFT membrane sigma-model]}]
{equation}\label{dftsm}
S[\X , A,F]=\int_{\S_3}\,\big(F_I\w \dd\X^I+\eta_{IJ}\,A^I\w\dd A^J-(\rho_+)^{I}{}_{J}\,A^J\w F_I+\sfrac 16\, T_{IJK}\,A^I\,\w A^J\w A^K\big)
\end{empheq} 
or in coordinate-free form
\bea
S[\X , A,F] = \int_{\S_{3}}\, \big(\langle F,\dd \X\rangle+ \langle A,\dd A\rangle_{L_+}-\langle F,\rho_+(A)\rangle+
\sfrac 13\, \langle A,\cbral A,A\cbrar_{L_+}\rangle_{L_+}\big) \ .
\eea

We conclude that this is the topological sector of the membrane sigma-model that corresponds to DFT. It is 
satisfying to observe that this very action was essentially proposed
already in \cite{Chatzistavrakidis:2015vka}, albeit without
explanation. Here we cover this gap by providing precise argumentation
for that action. However, in
\cite{Chatzistavrakidis:2015vka} it was implicitly assumed that the
resulting sigma-model still corresponds to the Courant algebroid
$E$. Here it becomes clear that no Courant algebroid is associated to
the action \eqref{dftsm}. In particular, the action \eqref{dftsm} does not define a Courant sigma-model and so its gauge invariance is not immediate. We shall analyse this point in detail in Section~\ref{sec4}. The Courant algebroid structure is broken on
the way from the large Courant algebroid $E$ to DFT and it is recovered, as is well-known, once the strong constraint is solved and the dual coordinates are eliminated; then the Courant algebroid over $M$ becomes the relevant structure. Thus we see that DFT lies in between 
the Courant algebroid over $M$ and the large Courant algebroid over the doubled space. We shall further quantify this observation in Section \ref{sec5}.

Regarding the remaining symmetric term, which is  necessary in order to reach any connection with the dynamics of string sigma-models, we follow the same procedure of $\sfp_+$-projecting the corresponding term in \eqref{symboundary}. This leads to 
\be 
S_{\text{sym}}[\X,A]=\int_{\partial\S_3}\, \| A\|_g = \int_{\partial\S_3}\,\sfrac 12\, {g}_{ I J}(\X)\, A^{ I}\w\ast A^{ J}~.
\ee
Then the full action we consider from now on is 
\be 
S_{\text{DFT}}=S+S_{\text{sym}}~.
\ee
To completely specify the sigma-model, the bulk action $S$ should be
supplemented with suitable Dirichlet and Neumann boundary conditions
for the fields on $\partial\S_3$. For the Courant sigma-model this is
discussed in detail
in~\cite{Mylonas:2012pg,Hofman:2002jz,Ikeda:2012pv}, whereby suitable
boundary conditions are imposed to ensure BV gauge invariance of the
induced boundary worldsheet sigma-models. In the following we shall
treat gauge invariance of our sigma-models from a different
perspective of DFT in Section~\ref{sec4}, and hence we will only assume implicitly that
suitable boundary conditions are defined, whose details are not
important for the boundary reductions which follow. Moreover, the
breaking of topological symmetry by the explicit boundary term $S_{\rm
  sym}$ furthermore ensures consistency of the bulk theory in the
presence of non-geometric flux
deformations, as discussed in~\cite{Mylonas:2012pg,Halmagyi}.

In writing the DFT membrane sigma-model we started with a
general Courant algebroid and its corresponding Courant sigma-model,
in contrast to Section~\ref{sec23a} where we started with the Courant bracket \eqref{cbrac} of the standard Courant algebroid. In this case, it is useful to also write down the C-bracket obtained via the double projection prescription \eqref{couranttoc} on the general form of the Courant bracket \eqref{generalcourant}. 
This leads to the general C-bracket
\be \label{cbragen}
\cbral A,B\cbrar^J_{L_+} = (\rho_+)^L{}_I\,\big(A^I\,\partial_L
B^J-\sfrac 12\, \eta^{IJ}\,A^K\,\partial_LB_K - \{A\leftrightarrow
B\}\big)+\sfrac 12\, T_{IK}{}^{J}\,A^I\,B^K~.
\ee
This bracket is to be used whenever the initial large Courant
algebroid is not the standard one (see e.g.~\cite{dft2}). It will also assist in determining a set of axioms and properties for the higher geometric structure associated to DFT in Section \ref{sec5}. 

\section{Examples}
\label{sec3}

In order to corroborate our proposal that \eqref{dftsm} is the DFT
membrane sigma-model, let us test it on some simple yet illustrative
cases. In the following we consider the worldsheet theories for the
four T-dual closed string backgrounds associated to the 3-torus $M$
with constant $H$-, $f$-, $Q$- and $R$-fluxes, as found e.g. in~\cite{Hull:2009sg}, and show that they are
all contained in the single 
action~\eqref{dftsm}. 

Let us introduce the following notation. 
 The components of $\rho_+$ are generally given as 
\be \label{rho1}
(\rho_+)^I{}_{ J}=\begin{pmatrix}\rho^i{}_{j}&\rho^{ij}\\
  \rho_{ij}&\rho_i{}^{j} \end{pmatrix}~,
\ee
while the components of a DFT vector $A$ and of the twist $T$ are
written respectively as\footnote{In this section, $p_i$ are always
  worldvolume 1-forms and should not be confused with the local fiber coordinates 
of Section 2.2.} 
\be 
A^I=(q^i,p_i)\qquad \text{and} \qquad
T_{IJK}= \begin{pmatrix}H_{ijk}&f_{ij}{}^k\\ Q_{i}{}^{jk}&R^{ijk}\end{pmatrix}~.
\ee
The symmetric term has components{\footnote{The components of $\rho$
    and $g$ with different positionings of indices on the right-hand
    sides of \eqref{rho1} and \eqref{g1} are in general unrelated. In
    particular, $g^{ij}$ is not generally the inverse of $g_{ij}$. }}
\be \label{g1}
g_{IJ}=\begin{pmatrix}g_{ij}&g_i{}^{j}\\ g^i{}_j&g^{ij}\end{pmatrix}~.
\ee
Our main goal here is to describe the standard T-duality chain relating
geometric and non-geometric flux configurations schematically through~\cite{Shelton:2005cf}
\bea
H_{ijk}\stackrel{{\sf T}_k}{\longleftrightarrow}
f_{ij}{}^k\stackrel{{\sf T}_j}{\longleftrightarrow}
Q_i{}^{jk}\stackrel{{\sf T}_i}{\longleftrightarrow} R^{ijk} \ ,
\eea
where ${\sf T}_i$ denotes a T-duality transformation along $x^i$.
We shall derive the corresponding $O(d,d)$ transformations among the structure
maps above, and demonstrate how the DFT membrane sigma-model
correctly captures the anticipated geometric and non-geometric
descriptions in each T-duality frame.

\subsection{NS--NS flux and the Heisenberg nilmanifold}
\label{sec31}

Let us start with the supergravity frames. In order to describe the geometric $H$-flux frame on the 3-torus $M$, we choose the data\footnote{The choices are not unique.}
\bea 
(\rho_+)^I{}_J=\begin{pmatrix}\d^i{}_j&0\\ 0&0\end{pmatrix}~,
\qquad  T_{IJK}=\begin{pmatrix}H_{ijk}&0\\ 0&0\end{pmatrix}
\qquad \mbox{and} \qquad g_{IJ}=\begin{pmatrix}0&0\\ 0&g^{ij}\end{pmatrix}~,
\eea
where here and below $g^{ij}$ denotes a constant metric on the dual space with inverse $g_{ij}$.
Then the membrane action becomes
\bea \label{H0}
S_{\text{DFT}} &=& \int_{\S_{3}}\, \big(F_I\w\dd \X^I+q^i\w\dd p_i+p_i\w\dd q^i-
q^{i}\w F_i+
\sfrac 16\,H_{ijk}\,q^{i}\w q^{j}\w q^{k}\big) \nn \\ && +\,
\int_{\partial\S_3}\,\sfrac 12\, g^{ij}\,p_i\w\ast p_j~.
\eea
We are interested in the on-shell membrane
theory. The equation of motion for $F_I$ yields two relations, one from
$F_i$ and the other from $F^i$, giving
\be \label{eomH}
q^i=\dd X^i \qquad \text{and} \qquad \dd \widetilde X_i=0~.
\ee
The action now takes the form
\bea 
 \int_{\partial\S_3}\, \big(p_i\w\dd X^i+\sfrac 12\, g^{ij}\,p_i\w\ast p_j\big)+\int_{\S_3}\,\sfrac 16\, H_{ijk}\,\dd X^i\w\dd X^j\w\dd X^k~,
\eea
which, after integrating out $p_i$ using $\ast^2=1$, takes precisely the desired form 
\be
S_H[X]:=\int_{\partial{\S_3}}\,\sfrac 12\, g_{ij}\,\dd X^i\w\ast \dd X^j+\int_{\S_3}\,\sfrac 16\, H_{ijk}\,\dd X^i\w\dd X^j\w\dd X^k
\ee
for the closed string sigma-model on $\partial\S_3$ with 3-torus target space and NS--NS flux.
 We obtained this action in a rather unnecessarily complicated fashion, however the advantage is that 
exactly the same steps may be followed for any other T-duality frame without the need for major adjustments. 

The T-dual of the above configuration corresponds to a twisted 3-torus
$N$ that has a purely metric flux
(torsion). It can be constructed as the quotient of the three-dimensional
non-compact Heisenberg group by a cocompact discrete subgroup, and in particular $N$ is parallelizable. The simplest way
to describe it in our formalism is to introduce a globally defined left-invariant (inverse) vielbein as a
component of the anchor map and choose the
data{\footnote{Topologically, the tangent bundles $TN$ and
    $TM$ are (non-canonically) isomorphic, and the components of $\rho_+$ in
    \eqref{fdata} correspond to a chosen isomorphism from
    $TN$ to $TM$. Since this is relevant only in the simple case discussed here, we shall not delve into further details.}}
\bea \label{fdata}
(\rho_+)^M{}_J=\begin{pmatrix}E^\mu{}_j&0\\ 0&0\end{pmatrix}~,
\qquad T_{IJK}=\begin{pmatrix}0&2\,f_{ij}{}^k\\ 0&0\end{pmatrix}
\qquad \mbox{and} \qquad g_{IJ}=\begin{pmatrix}0&0\\ 0&g^{ij}\end{pmatrix}~,
\eea
where here we use the convention that Greek indices $\mu,\nu,\dots$ label local coordinates while Latin indices $i,j,\dots$ label frames, and $f_{ij}{}^k=-2\,E^\mu{}_{[i}\, E^\nu{}_{j]}\, \partial_\mu E^k{}_\nu$ are structure constants of the three-dimensional Heisenberg algebra.  
Then the membrane action becomes
\bea \label{f0}
S_{\rm DFT} &=& \int_{\S_{3}}\, \big(F_\mu\w\dd X^\mu{+}\widetilde F^\mu\w\dd\widetilde X_\mu{+}q^i\w\dd p_i{+}p_i\w\dd q^i{-}E^\mu{}_j\,
q^{j}\w F_\mu{+}f_{ij}{}^k\,q^{i}\w q^{j}\w p_{k}\big) \nn\\ 
&&  +\,
\int_{\partial\S_3}\,\sfrac 12\, g^{ij}\,p_i\w\ast p_j~.
\eea
The equations of motion for $F_M=(F_\mu,\widetilde F^\mu)$ yield two relations
\be \label{eomf}
q^i=E^i:= E^i{}_\mu\,\dd X^\mu\qquad \text{and} \qquad \dd \widetilde X_\mu=0~.
\ee
Using the Maurer-Cartan structure equations
\bea
\dd E^i=-\sfrac 12\, f_{jk}{}^i\,E^j\w E^k 
\eea
we obtain 
\bea 
\int_{\partial\S_3}\, \big(p_i\w E^i+\sfrac 12\, g^{ij}\,p_i\w\ast p_j\big) ~,
\eea
which, after integrating out $p_i$, takes precisely the desired form 
\be
S_f[X]:=\int_{\partial{\S_3}}\,\sfrac 12\, g_{ij}\,E^i\w\ast E^j
\ee
for the closed string sigma-model with target the geometric T-dual of the 3-torus with NS--NS flux.

\subsection{The T-fold and noncommutativity}
\label{sec32}

To describe the globally non-geometric $Q$-flux frame corresponding to
a parabolic monodromy from this point of view, we choose 
\bea \label{Qdata}
(\rho_+)^I{}_J=\begin{pmatrix}\delta^i{}_j& \beta^{ij}(X) \\
  0&-\d_i{}^j\end{pmatrix} \qquad \mbox{and} \qquad
T_{IJK}=\begin{pmatrix}0&0\\ Q_{i}{}^{jk}&0\end{pmatrix}~,
\eea
where
\bea\label{eq:betaQ}
\beta^{ij}(X)=-Q_k{}^{ij}\, X^k
\eea
defines a local bivector
$\beta=\sfrac 12\,\beta^{ij}(x)\,\partial_i\wedge\partial_j$ on $M$
which is ``T-dual'' to the Kalb-Ramond field $B_{ij}(X)=H_{ijk}\, X^k$ of the supergravity frame~\cite{Andriot:2012wx}.
We take the only non-vanishing components of the constant $Q$-flux to be $Q_3{}^{12}=-Q=-Q_3{}^{21}$, and
\be
g_{IJ}=\begin{pmatrix}0&\delta_{3}{}^{j}\\ 0&{g}^{ij}\end{pmatrix} \qquad \mbox{with} \quad  g^{ij}=\text{diag}(1,1,0) \ .
\ee
With this choice the topological part of membrane action \eqref{dftsm}
is
\bea
S&=&\int_{\S_3}\,\big( F_I\w \dd\X^I+q^i\w\dd p_i+p_i\w\dd q^i-q^i\w F_i+p_i\w F^i \nn\\ && \qquad\qquad -\,Q\,X^3\,p_2\,\w F_1+Q\,X^3\,p_1\w F_2-
 Q\,p_1\w p_2\w q^3\big)~.
\eea
By integrating out the auxiliary fields $F_I$ we obtain
\bea
q^m=\dd X^m-Q_3{}^{mn}\,X^3\,p_n \qquad \mbox{for} \quad m,n=1,2 \qquad \mbox{and} \qquad 
q^3=\dd X^3 \ , 
\eea
and
\bea
p_i=-\dd\widetilde X_i \qquad \mbox{for} \quad i=1,2,3~. 
\eea
Using these field equations, the three-dimensional membrane action drops to the boundary, and adding the symmetric term we get
\bea
\int_{\partial\S_3}\,\big( \dd\widetilde
X_m\w\dd X^m+Q\, X^3\, \dd\widetilde X_1\w\dd\widetilde X_2 + \sfrac 12\, \dd X^3\w\ast\dd X^3+\sfrac
12\, \dd\widetilde X_m\w\ast\dd\widetilde
X_m\big)~. \label{Qaction}
\eea
The first term plays an important role here. For smooth worldvolume
manifolds with boundary, i.e.~when the boundary of $\partial\Sigma_3$
is empty, one may naively just drop the first term and obtain a
two-dimensional action corresponding to the T-duality exchange of fields
$X^m \leftrightarrow \widetilde{X}_m$, for $m=1,2,$ from the
sigma-model for the 3-torus with $H$-flux. 
However, for the 3-torus the coordinate fields are not globally
defined, so the first term cannot be ignored and the situation is different. Using $\ast^2=1$, integrating out $\widetilde X_m$ yields
\bea
\dd\widetilde X_m=-\frac{1}{1+(Q\,X^3)^2}\,\big(\ast\dd
X^m-Q_3{}^{mn}\,X^3\,\dd X^n\big)~, 
\eea
and the resulting action 
\bea\label{eq:SQX}
S_Q[X]:=\int_{\partial{\S_3}}\, \big(\sfrac 12\,\dd X^3\w\ast\dd
X^3+\sfrac  {1}{2(1+(Q\,X^3)^2)}\,\dd X^m\w\ast\dd X^m- \sfrac{Q\,X^3}{1+(Q\,X^3)^2}\,\dd X^1\w\dd X^2
\big)
\eea
is the anticipated worldsheet action associated to the T-fold
which is the globally non-geometric T-dual of the 3-torus with NS--NS
flux.

An alternative perspective on this global non-geometry is the proposal
of~\cite{stnag2} that
closed strings which wind in the $Q$-flux background probe a noncommutative
deformation of the background geometry. This effect cannot be observed
in the membrane sigma-model by viewing the closed strings as boundary
modes of open membranes, as we have done until now, but instead we
should regard them as wrapping modes of closed membranes. For this, we
view the target space as $M=M_2\times S^1$, with $M_2$ the 2-torus and
$X^3$ the coordinate on $S^1$, and take the membrane worldvolume to be
a product space $\Sigma_3=\Sigma_2\times S^1$,  with $\sigma^3$
denoting the worldvolume coordinate on $S^1$. We wrap the membrane on the target
$S^1$ by making a partial gauge-fixing
\bea
X^3(\sigma)=w^3\,\sigma^3
\eeq
of the worldvolume diffeomorphism symmetry, where
$w^3$ is the winding number of the worldvolume circle around the
target space circle. The symmetric part of the action is now defined
over the closed string worldsheet $\Sigma_2$. Dimensional reduction of
the topological action \eqref{dftsm} proceeds by restricting all
membrane fields $X^m(\sigma)$ and $\widetilde X_i(\sigma)$ to
configurations which are independent of $\sigma^3$. Proceeding as
above, integration over the worldvolume $S^1$ then yields the worldsheet action
\bea
S_{Q,w}[X,\widetilde X\,] := \int_{\S_2}\, \big(\sfrac 12\, \dd\widetilde X_m\w\ast\dd\widetilde
X_m + \dd\widetilde
X_m\w\dd X^m+ \sfrac 12\, Q_3{}^{mn}\, w^3\,\dd\widetilde X_m\wedge\dd\widetilde X_n\big) \ .
\eea
The inverse of the $B$-field appearing in the topological term here defines a bivector
$\theta=\frac12\,\theta^{mn}\, \partial_m\wedge\partial_n+\partial_m\wedge\tilde\partial^m$,
showing that the closed string coordinates have noncommutative phase
space Poisson brackets
\bea\label{eq:NCwinding}
\{X^m,X^n\}_\theta = \theta^{mn} = Q_3{}^{mn}\, w^3 \ , \qquad
\{X^m,\widetilde X_n\}_\theta=\delta^m{}_n \qquad \mbox{and} \qquad
\{\widetilde X_m,\widetilde X_n\}_\theta=0
\eea
in the approach of \cite{Mylonas:2012pg} whereby $\Sigma_2$ is effectively an open string worldsheet, which confirms the
expectations of~\cite{stnag2,Condeescu:2012sp}. Note that even for vanishing $Q$-flux the
coordinates and their duals do not commute, which agrees with the recent
suggestion of intrinsic closed string noncommutativity
\cite{Freidel:2017wst}. In fact, as the noncommutativity parameter $\theta^{mn}$
is induced entirely by the generalized Wess-Zumino term from above,
dimensional reduction of our membrane sigma-model corroborates and
clarifies the proposal~\cite{Andriot:2012an} that the general relation
between the globally non-geometric flux and closed string
noncommutativity is provided by a Wilson line of the $Q$-flux through
\bea
\theta^{ij}=\oint_{C_k}\, Q_k{}^{ij}\, \dd X^k \ ,
\eea
where $C_k=S^1$ are the 1-cycles of $M$.

The metric and $B$-field in the worldsheet action \eqref{eq:SQX} are
locally defined but are not single-valued under periodic shifts of the
circle coordinate $X^3$. Within the framework of the DFT membrane
sigma-model, this global non-geometry is due to the fact that the
anchor $\rho_+$ in \eqref{Qdata} is not globally defined. The
correct global parameterization of the non-geometric space is
defined by the
open-closed field redefinition~\cite{Grana:2008yw,dft1,Andriot:2012an}
\bea\label{eq:openclosed}
\tilde g^{-1}+\beta = (g+B)^{-1}
\eea
which maps the closed string metric and $B$-field $(g,B)$ appearing in
\eqref{eq:SQX} to the open string bivector $\beta$ in \eqref{eq:betaQ} and globally defined
metric $\tilde g={\rm diag}(1,1,1)$. The relation \eqref{eq:openclosed} is just a particular T-duality transformation~\cite{Andriot:2012an}, and in this non-geometric
parameterization the anchor of \eqref{Qdata} is modifed to 
\bea\label{eq:anchorQ}
(\rho_+)^I{}_J=\begin{pmatrix}\delta^3{}_j&0\\ 0&\rho_i{}^j\end{pmatrix} \qquad \mbox{with} \quad \rho_i{}^j={\rm diag}(1,1,0) \ ,
\eea
which is now globally defined; the remaining structure maps are as
above. By following the same steps as before, we arrive at the
worldsheet sigma-model action
\bea
S_Q[X,\widetilde{X}\,] &:=& \int_{\partial\S_3}\, \big( \sfrac 12\, \dd X^3\w\ast\dd X^3+\sfrac
12\, \dd\widetilde X_m\w\ast\dd\widetilde
X_m\big) \nn\\ && +\, \int_{\S_3}\, \sfrac 12\, Q_3{}^{mn}\, \dd
X^3\wedge\dd\widetilde X_m\wedge\dd\widetilde X_n \ ,
\label{eq:SQdual}\eea 
which is now indeed the naive T-dual of the sigma-model with $H$-flux.

\subsection{Locally non-geometric flux and nonassociativity}
\label{sec33}

The corresponding locally non-geometric $R$-flux frame, which has no
conventional target space description on $M$, is described within our framework by
choosing the anchor $\rho_+$ to be 
\be\label{Ranchor} 
(\rho_+)^I{}_{J}=\begin{pmatrix}\delta^i{}_j&\beta^{ij}(\widetilde
X\,)\\ 0&-\delta_i{}^j\end{pmatrix}~, \ee 
where
\bea
\beta^{ij}(\widetilde
X\,)=R^{ijk}\,\widetilde X_k 
\eea
is the T-dual image of the bivector \eqref{eq:betaQ}~\cite{Andriot:2012wx}. 
The bracket twist $T$ and the symmetric form $g$ are chosen as 
\be 
T_{IJK}=\begin{pmatrix}0&0\\ 0&R^{ijk}\end{pmatrix} \qquad \mbox{and}
\qquad g_{IJ}=\begin{pmatrix}0&0\\ 0&g^{ij}\end{pmatrix}~.
\ee 
The topological part of the membrane action \eqref{dftsm} becomes 
\bea \label{R1action1}
S&=&\int_{\S_{3}}\, \big(F_I\w\dd \X^I+q^i\w\dd p_i+p_i\w\dd q^i-q^i\w
F_i+p_i\w F^i \nn\\ && \qquad \qquad -\,R^{ijk}\,\widetilde X_k\,p_{j}\w F_i+
\sfrac 16\,R^{ijk}\,p_{i}\w p_{j}\w p_{k}\big) 
~.
\eea
Integrating out the auxiliary fields $F_I$ gives
\be \label{feom12}
q^i=\dd X^i- R^{ijk}\,\widetilde X_k\,p_j \qquad \mbox{and} \qquad p_i=-\dd\widetilde X_i~.
\ee 
The second equation implies $\dd p_i=0$, so for constant $R^{ijk}$ all
the rest of the terms drop to the two-dimensional boundary giving
\bea
\int_{\partial\S_3}\, \big( -q^i\w \dd\widetilde X_i- \sfrac 12\,
R^{ijk}\,\widetilde X_k\,\dd\widetilde X_i\w \dd\widetilde X_j+\sfrac
12\, g^{ij}\,p_i\w\ast p_j \big)~.
\eea 
Restricting \eqref{feom12} to the boundary we obtain
\bea\label{naR}
S_R[X,\widetilde X\,]:=\int_{\partial\S_3}\,\big( \sfrac 12\,
g^{ij}\,\dd\widetilde X_i\w\ast \dd\widetilde X_j + \dd\widetilde X_i\w\dd X^i+ \sfrac 12\, R^{ijk}\,\widetilde X_k\,\dd\widetilde X_i\w \dd\widetilde X_j\big)~,
\eea 
which, in contrast to the case of the T-fold, cannot even be locally
expressed in terms of fields on the target space $M$.
This is precisely the membrane sigma-model proposed in
\cite{Mylonas:2012pg} which captures the nonassociative phase
space structure of the $R$-flux background that is formally T-dual to the associative algebra \eqref{eq:NCwinding}~\cite{stnag2}; here we have shown that it
is also included in the DFT membrane sigma-model \eqref{dftsm}. 
Following \cite{Mylonas:2012pg}, membranes propagating in the locally
non-geometric target space do not have smooth worldvolumes, but rather
$\Sigma_3$ should now be regarded as a manifold with corners of codimension two, as
suggested by the open-closed string duality of the $R$-flux background
which implies that $\partial\S_3$ has non-empty boundary. Thus in this
parameterization, the inverse of the $B$-field appearing in the
doubled space sigma-model action
\eqref{naR} defines a bivector $\Theta=\sfrac
12\,\Theta^{IJ}\,\partial_I\w\partial_J$ on phase space $T^*M$ with
\bea 
\Theta^{IJ}=\begin{pmatrix}R^{ijk}\,\widetilde X_k & \delta^i{}_j \\ -\delta_i{}^j &0
	\end{pmatrix}~.
\eea
It induces a twisted Poisson bracket given by 
\bea
\{\X^I, \X^J\}_\Theta=\Theta^{IJ}~,
\eea 
which reads explicitly as{\footnote{We emphasize that due to the additional twisted Poisson structure, $\widetilde{X}_i$ are regarded here as canonically conjugate momenta to $X^i$ and not as T-dual winding coordinates.}} 
\bea 
\{X^i,X^j\}_{\Theta}=R^{ijk}\,\widetilde X_k~, \qquad
\{X^i,\widetilde{X}_j\}_{\Theta}=\d^i{}{}_j \qquad \mbox{and} \qquad \{\widetilde{X}_i,\widetilde{X}_j\}_{\Theta}=0~. 
\eea
This leads to the non-vanishing Jacobiator
\be
\{X^i, X^j, X^k\}_\Theta:=\sfrac 13\,
\{\{X^i,X^j\}_\Theta,X^k\}_{\Theta} + \mbox{cyclic} = -R^{ijk}~. 
\ee
Deformation quantization of this twisted Poisson structure was carried
out in~\cite{Mylonas:2012pg}
via perturbative quantization of the sigma-model in the formalism
of~\cite{Kontsevich:1997vb,Cattaneo:1999fm}, and reproduced in various
other contexts in~\cite{Bakas:2013jwa,Mylonas:2013jha,Barnes:2014ksa,Bakas:2015gia,Kupriyanov:2015dda}.

Alternatively, we may choose to work in a suitable global reparameterization of the
locally non-geometric space, analogously to the global non-geometry of
the $Q$-flux frame. For this, we modify the anchor \eqref{Ranchor} to
the globally defined map 
\bea \label{Ranchor2}
(\rho_+)^I{}_J=\begin{pmatrix}0&0\\ 0&\d_i{}^j\end{pmatrix}~.
\eea
Following the same steps as above, the resulting worldsheet action is 
\be\label{rp}
S_R[\widetilde X\,]:=\int_{\partial{\S_3}}\,\sfrac 12\, g^{ij}\,\dd \widetilde X_i\w\ast \dd \widetilde X_j+\int_{\S_3}\,\sfrac 16\, R^{ijk}\,\dd \widetilde X_i\w\dd \widetilde X_j\w\dd\widetilde X_k~,
\ee
which is the same as the sigma-model action with $H$-flux under
the naive T-duality exchanges of all fields $X^i$ with $\widetilde{X}_i$.{\footnote{In this case there is no 
		(twisted) Poisson structure and $\widetilde{X}_i$ are interpreted as T-dual winding coordinates to $X^i$.}} The
difference between the two membrane sigma-models is that the choice of
anchor \eqref{Ranchor} violates the strong constraint of DFT, while
\eqref{Ranchor2} does not. This agrees with the observation~\cite{Blumenhagen:2013zpa} that the
nonassociative deformation of the closed string background is not compatible with the
strong constraint between the background $R^{ijk}$ and fluctuations around it. 

These results clarify the appearance of noncommutativity and
nonassociativity in closed string theory. It is known that the
$H$-flux frame can also be described simply by a Courant sigma-model. Recall that the action \eqref{H0} is \emph{not} a Courant sigma-model action, as already explained generally in Section~\ref{sec24}. However, imposing solely the second of the field equations \eqref{eomH} would lead to an 
action which \emph{is} a Courant sigma-model action, and in particular the one associated to the \emph{standard} Courant algebroid over the target space $M$~\cite{Mylonas:2012pg}. 
The same is true for the other three cases under the exchange of $X^i$
with $\widetilde{X}_i$, whose final worldsheet action results from a Courant sigma-model
corresponding to the standard Courant algebroid, albeit not over $M$
but over other slices of the doubled target space, as in
e.g.~\eqref{eq:SQdual} and \eqref{rp}. 
Thus in terms of the doubled space of DFT, the four T-dual backgrounds
with $H$-, $f$-, $Q$- and $R$-flux all correspond to the {standard
  Courant algebroid} over different polarizations of the doubled
space. However, this does not include the noncommutative and
nonassociative backgrounds discussed above, which violate the strong
constraint of DFT and therefore do not correspond to Courant
sigma-models; as such, the corresponding membrane sigma-models do not
possess the usual (higher) BV gauge symmetries. Later on we shall
describe how the strong constraint can be weakened and how gauge
invariance of the membrane sigma-model is reconciled in this case.
In more complicated cases, for instance when fluxes coexist, this
picture gets suitably modified.

\subsection{$R$-flux with Poisson structure}
\label{sec34}

There is another simple yet interesting example involving the $R$-flux, wherein the target space $M$ is a Poisson manifold with non-degenerate Poisson bivector $\Pi=\frac12\,\Pi^{ij}(x)\, \partial_i\wedge\partial_j$.
We choose the anchor $\rho_+$ to be 
 \be 
 (\rho_+)^I{}_{J}=\begin{pmatrix}0&-\Pi^{ij}\\ 0&0\end{pmatrix}~, 
 \ee 
and the bracket twist $T$ and the symmetric form $g$ are chosen as 
 \be 
 T_{IJK}=\begin{pmatrix}0&0\\ 2\,\partial_k\Pi^{ij}&R^{ijk}\end{pmatrix} \qquad \mbox{and}
\qquad g_{IJ}=\begin{pmatrix}g_{ij}&0\\ 0&0\end{pmatrix}~.
 \ee 
 The topological part of the membrane action \eqref{dftsm} becomes 
 \bea 
S&=&\int_{\S_{3}}\, \big(F_I\w\dd \X^I+q^i\w\dd p_i+p_i\w\dd q^i+
\Pi^{ij}\,p_{j}\w F_i\nn\\&& \qquad \qquad +\, \partial_k\Pi^{ij}\,q^k\w p_i\w p_j+
\sfrac 16\,R^{ijk}\,p_{i}\w p_{j}\w p_{k}\big) 
~. \label{R1action}
\eea
Taking the $F_I$ equations of motion,
\be 
\dd X^i=-\Pi^{ij}\,p_j \qquad \mbox{and} \qquad \dd\widetilde X_i=0~,
\ee 
the non-degeneracy assumption on the bivector allows us to invert the first equation and write 
\be 
p_i=-\Pi^{-1}_{ij}\,\dd X^j~.
\ee
Since $\Pi$ is a Poisson bivector and thus its Schouten bracket with itself vanishes, $[\Pi,\Pi]_{\text{S}}=0$, or in local coordinates
\be 
\P^{l[i}\,\partial_l\Pi^{jk]}=0~,
\ee
the topological part of the action takes the form
 \bea \label{R1action2} 
-\int_{\partial\S_3}\,\Pi^{-1}_{ij}\,q^i\w\dd X^j-\int_{\S_3}\,\sfrac 16\, R^{lmn}\,\Pi^{-1}_{li}\,\Pi^{-1}_{mj}\,\Pi^{-1}_{nk}\,\dd X^i\w\dd X^j\w\dd X^k~.
\eea
Concerning the kinetic part of the boundary action, it is convenient to add an additional term
\be \label{Rkin}
\int_{\partial\S_{3}}\,\big(\sfrac 12\, g_{ij}\,q^i\w\ast q^j + \sfrac 12\, g_{ij}\,\dd X^i\w\ast \dd X^j\big) \ , 
\ee
and after taking the equation of motion for $q^i$ into account, we obtain the worldsheet sigma-model action
\bea
S_{R,\Pi}[X]&:=&\int_{\partial\S_3}\,\sfrac 12\, \big(g_{ij}-\Pi^{-1}_{ik}\,g^{kl}\,\Pi^{-1}_{lj}\big)\,\dd X^i\w\ast\dd X^j \nn\\ && -\,\int_{\S_3}\,\sfrac 16\,R^{lmn}\,\Pi^{-1}_{li}\,\Pi^{-1}_{mj}\,\Pi^{-1}_{nk}\,\dd X^i\w\dd X^j\w\dd X^k~. 
\label{rx}\eea
An alternative option would be to take only the second term in \eqref{Rkin}; then, adding also an extra 2-form topological term $\sfrac 12\, B_{ij}\,q^i\w q^j$ and choosing appropriate boundary conditions, one obtains the $R$-flux sigma-model considered in \cite{Bessho:2015tkk}. In the case where the twist $R$ and the Poisson bivector considered here are constant (as is the case, for instance, for a toroidal target in Darboux coordinates), the topological term falls 
locally on the boundary as $\sfrac 12\, R^{lmn}\,\Pi^{-1}_{li}\Pi^{-1}_{mj}\,\Pi^{-1}_{nk}\,X^k\,\dd X^i\w\dd X^j$.

There is an important difference between the $R$-flux models with
actions \eqref{naR} and \eqref{rx}. The former is a
 sigma-model on the doubled space, while the latter is a
Courant sigma-model on $M$. (Note also that the metrics in the two
actions are not generally related, as we are slightly abusing notation
here.) Only the former one should be properly understood as a
sigma-model for non-geometric $R$-flux in the sense that it can be
obtained from a generalized T-duality transformation of a geometric
background. The second $R$-flux is itself a geometric
flux.\footnote{The precise relation between the two models is
  clarified in~\cite{Kokenyesi2018}, where it is shown that the
  degenerate limit $\Pi=0$ of the Courant sigma-model of~\cite{Bessho:2015tkk}
with a particular BV gauge-fixing coincides exactly with
the $R$-twisted membrane sigma-model of~\cite{Mylonas:2012pg}.}
Comparing the actions \eqref{rp} and \eqref{rx}, we note that the
reason for the existence of both models is that there are two distinct
Courant algebroids, one being the standard Courant algebroid on the
dual winding space, and the other the non-standard Courant algebroid on $M$ with its anchor given by a Poisson bivector (see Appendix~\ref{seca4}).
  
\section{DFT fluxes from the membrane sigma-model}
\label{sec4}

In this section we discuss how the membrane sigma-model \eqref{dftsm} captures the flux formulation of DFT, in particular the role of the generalized Wess-Zumino term in formulating the geometric and non-geometric fluxes, and the manner in which the standard Bianchi identities for DFT fluxes are generated by the gauge symmetries of the action.

\subsection{Three roads to DFT fluxes}
\label{sec41}

In DFT, the potential expressions for the four types of fluxes 
$(H,f,Q,R)$ are modified with respect to the ones of generalized geometry, receiving additional contributions due to the dual coordinate dependences of fields. In a holonomic frame they read as \cite{dftflux1,dftflux2,dftflux3,dftflux4}
\bea 
H_{ijk}&=&3\,\partial_{[i}B_{jk]}
+3\,B_{[i\underline{l}}\,\tilde\partial^{l}B_{jk]}~,
\label{H}\\[4pt]
f_{ij}{}^k&=&\tilde{\partial}^kB_{ij}+\beta^{kl}\,H_{lij}~,
\label{F}\\[4pt]
Q_{k}{}^{ij}&=&\partial_{k}\beta^{ij}
+B_{kl}\,\tilde{\partial}^l\beta^{ij}
+2\,\beta^{l[i}\,\tilde{\partial}^{j]}B_{lk}+\beta^{il}\,\beta^{jm}\,H_{lmk}~,
\label{Q}\\[4pt]
R^{ijk}&=&3\,\tilde{\partial}^{[i}\beta^{jk]}
+3\,\beta^{[i\underline{l}}\,\partial_l\beta^{jk]} \nn\\ &&
+\,3\,B_{lm}\,\beta^{[i\underline{l}}\,\tilde{\partial}^m\beta^{jk]}
+3\,\beta^{[i\underline{l}}\,\beta^{j\underline{m}}\,\tilde{\partial}^{k]}B_{lm}
+\beta^{il}\,\beta^{jm}\,\beta^{kn}\,H_{lmn}~, 
\label{R}\eea
where $B$ is the Kalb-Ramond 2-form field and the bivector field $\beta$ its ``T-dual'' in DFT.
The fluxes in generalized geometry are simply the ones with
$\tilde{\partial}^i=0$ \cite{Halmagyi}, which is a solution of the
strong constraint \eqref{eq:strongconstraint}. These expressions, and their counterparts in an arbitrary non-holonomic frame, may be obtained in the following ways.

\subsubsection*{Generalized vielbein}

In \cite{dftflux3} a generalized vielbein formulation of DFT is considered. Starting from the $d$-dimensional Minkowski metric{\footnote{Hereby indices $a,b,c,\dots$ refer to flat quantities and indices $i,j,k,\dots$ to curved quantities. The corresponding capitalized indices are doubled.}} $s_{ab}$, and introducing the $O(1,d-1)\times O(1,d-1)$-invariant metric $S_{AB}=\text{diag}(s^{ab},s_{ab})$, the covariant generalized metric $\cal H$ is written as 
\be 
{\cal H}_{IJ}={\cal E}^{A}{}_I\,S_{AB}\,{\cal E}^{B}{}_J~,
\ee
where ${\cal E}^{A}{}_I$ is a generalized vielbein. 
One also introduces a flat derivative 
\be
{\cal D}_A={\cal E}_A{}^{I}\,\partial_I
\ee
and the generalized Weitzenb\"ock connection 
\be 
\Omega_{ABC}={\cal D}_A{\cal E}_{B}{}^I\,{\cal E}_{CI}~.
\ee
It is shown in~\cite{dftflux3} that the DFT fluxes{\footnote{Note that we identify the DFT fluxes with the twist $\hat T$ of the C-bracket rather than the twist $T$ of the large Courant bracket; the two twists are related as explained in footnote \ref{footnote}.}} $\hat T_{ABC}$ are given as 
\be \label{fluxes1}
\hat T_{ABC}=3\,\Omega_{[ABC]}~,
\ee
which agrees with the expanded formulas upon the choice of parametrization for the generalized vielbein given by
\be \label{genviel}
{\cal E}^A{}_{I}=\begin{pmatrix} e_{a}{}^i & e_a{}^j\,B_{ji} \\ e^a{}_j\,\beta^{ji} & e^a{}_i+e^a{}_j\,\beta^{jk}\,B_{ki}\end{pmatrix}~,
\ee
where $e$ is a standard vielbein. As usual, when the vielbein $e$ is the
identity and we identify
\be\label{eq:hatTIJKHfQR}
\hat
T_{IJK}=\begin{pmatrix}\hat T_{ijk}&\hat T_{ij}{}^k\\
  \hat T_i{}^{jk}&\hat T^{ijk}\end{pmatrix}=:\begin{pmatrix}H_{ijk}&f_{ij}{}^k\\
  Q_{i}{}^{jk}&R^{ijk}\end{pmatrix} \ , 
\ee
these formulas reproduce the ones appearing in
\eqref{H}--\eqref{R}. These expressions
are not unique as a different parametrization of the generalized
vielbein would yield different expressions, essentially the equivalent
ones in a different $O(d,d)$ frame.

\subsubsection*{C-bracket}

Alternatively, the fluxes may be obtained directly from the
C-bracket. For this, first recall that in generalized geometry
one can consider the Roytenberg bracket~\cite{dee1,Halmagyi}, which is the Courant bracket
with an arbitrary generalized 3-form twist. One way to obtain explicit
expressions for the fluxes is to act with the twist operator
$e^{B}\,e^{\beta}$ on the basis $\partial_i$ and $\dd x^i$ to get
\bea \label{dxe1}
\partial_i &\overset{e^B\,e^{\beta}}\longrightarrow&
e_i:=\partial_i+B_{ij}\,\dd x^j~,\\[4pt]
\dd x^i &\overset{e^B\,e^{\beta}}\longrightarrow& e^i:=\dd x^i+\beta^{ij}\,\partial_j+\beta^{ij}\,B_{jk}\,\dd x^k=\dd x^i+\beta^{ij}\,e_j
~.\label{dxe2}
\eea
Then computing the \emph{untwisted} Courant brackets of the new basis, one obtains 
\bea 
[e_i,e_j]_E&=&H_{ijk}\,e^k+f_{ij}{}^ke_k~,\nn\\[4pt]
{[}e_i,e^j]_E&=&f_{ik}{}^j\,e^k+Q_i{}^{jk}\,e_k~,\nn\\[4pt]
{[}e^i,e^j]_E&=&Q_{k}{}^{ij}\,e^k+R^{ijk}\,e_k~,
\eea
where the generalized structure functions appearing on the right-hand
side are precisely given by the expressions \eqref{H}--\eqref{R} upon
setting $\tilde{\partial}^i=0$. Once again, these expressions are
not unique, since they depend on the way one twists the
basis. Different operators, e.g.~$e^{\beta}\,e^B$, would give the
fluxes in a different $O(d,d)$ frame \cite{Chatzistavrakidis:2013wra}. Now in the DFT case, we choose the components of the anchor $\rho_+$ to
be given by
\be \label{rhoplus}
(\rho_+)^I{}_J=\begin{pmatrix}\d^i{}_j&\beta^{ij}\\ B_{ij}&\d_i{}^j+\beta^{jk}\,B_{ki}\end{pmatrix}~,
\ee
in close relation to the generalized vielbein \eqref{genviel} in a
holonomic frame; this is
similar (up to signs) to what we chose in \eqref{Qdata} and \eqref{Ranchor} in the case
of the 3-torus with purely non-geometric $Q$-flux and $R$-flux, respectively. 
We then consider 
\be 
\hat e^+_J=(\rho_+)^{I}{}_{J}\,e^+_I
\ee 
as the analog of \eqref{dxe1} and \eqref{dxe2}. Then a straightforward computation of the untwisted C-bracket establishes that
\bea\label{eq:Cbracketflux}
 \cbral \hat e^+_M,\hat e^+_N\cbrar_{L_+} =
 3\,\eta_{IK}\,
 (\rho_+)^{K}{}_{[M}\,\partial^I(\rho_+)^{L}{}_{N}\,(\rho_+)_{\underline{L}P]}\,
 \eta^{PQ}\, \hat e^+_Q~,
\eea
which on comparing \eqref{fluxes1}, \eqref{genviel} and
\eqref{rhoplus} is seen to be the desired result $\cbral \hat
e^+_M,\hat e^+_N\cbrar_{L_+} = \hat T_{MN}{}^Q\,\hat e^+_Q$. This last
computation also appears in \cite{dftflux3}, wherein $\rho_+$ is a
duality twist. (An
alternative derivation, based on the commutator algebra of two differential
operators, appears in \cite{Blumenhagen1}.)

\subsubsection*{Generalized Wess-Zumino term}

In the spirit of our approach, the expressions for the DFT fluxes may be derived from the DFT membrane sigma-model \eqref{dftsm}. One can confirm this in two alternative ways. First, let us recall that the Wess-Zumino term in the Courant sigma-model is obtained in the basis-independent formulation from the term 
\be
\langle \A,[\A,\A]_{E}\rangle_{E}~.
\ee
This term is zero for the untwisted Courant bracket; the generalized Wess-Zumino term is obtained from the twist of the bracket. Similarly, in the DFT membrane sigma-model, one may write the generalized Wess-Zumino term as 
\be
\langle A,\cbral A,A\cbrar_{L_+}\rangle_{L_+}~.
\ee
The bracket is now the C-bracket of DFT and $A$ is a DFT vector; the
term is trivially zero when it is untwisted but non-zero when
twisted. Recall now that the background field local expressions for
the fluxes are obtained from the \emph{untwisted} bracket. Thus, in
order to derive these expressions in our approach, we consider the
untwisted DFT membrane sigma-model, namely \eqref{dftsm} \emph{without} the last term. 
As in all other approaches, the precise expressions depend on the
parametrization. With the goal of obtaining the result in a
holonomic frame, we take the components of the anchor $\rho_+$ to again
be given by \eqref{rhoplus}.
The DFT membrane sigma-model becomes
\bea
S_{\rm DFT}=\int_{\S_3}\,\big(F_I\w(\dd \X^I-(\rho_+)^I{}_J\,A^J)+\eta_{IJ}\,A^I\w\dd A^J\big)+\int_{\partial\S_3}\,\sfrac 12\, g_{IJ}\,A^I\w\ast A^J~.
\eea
Taking the equation of motion for the worldvolume 2-form $F_I$ in three
dimensions, we obtain $\dd \X^I=(\rho_+)^I{}_J\,A^J$ which implies
\be 
A^I=(\rho_+)_J{}^I\,\dd \X^J~,
 \ee
 where we used the fact that the particular anchor $\rho_+$ of
 \eqref{rhoplus} is invertible with inverse
\be
(\rho_+)_I{}^J=\begin{pmatrix}\d_i{}^j+B_{ik}\,\beta^{kj}&B_{ij}\\
  \beta^{ij}&\d^i{}_j\end{pmatrix} \ .
\ee
Eliminating $F_I$, the action takes the form 
 \bea
&& \int_{\partial\S_3}\,\big(\sfrac 12\, g_{IJ}\,A^I\w\ast
A^J+\eta_{IJ}\,(\rho_+)_K{}^I\,A^J\w\dd \X^K\big) \nn\\ && \qquad
\qquad +\,\frac 13\,\int_{\S_3}\, 3\,\eta_{IM}\,(\rho_+)^L{}_K\,(\rho_+)_{N}{}^M\,\partial_L(\rho_+)^N{}_J\,A^I\w A^J\w A^K~.
 \eea
Comparing with \eqref{eq:Cbracketflux}, it is observed that the three-dimensional term in this action indeed encodes the correct DFT fluxes $\hat T$.
Moreover, the kinetic term may be written in the second order
formalism, and the resulting action describes the motion of a closed
string with worldsheet $\partial\S_3$ in the doubled target space
$T^*M$ as a standard non-linear sigma-model (see e.g.~\cite{Hull:2009sg})
\be 
S_{\cH,\cF}[\X]:= \int_{\partial\S_3}\, \sfrac 12\, {\cal H}_{IJ}\,\dd \X^I\w\ast\dd \X^J + \int_{\S_3}\, \sfrac 13\, \cF_{IJK}\, \dd\X^I\w\dd\X^J\w\dd\X^K \ ,
\ee
where 
\be
{\cal H}_{IJ}:=(\rho_+)_I{}^K\,g_{KL}\,(\rho_+)_J{}^L\qquad \mbox{and} \qquad \cF_{IJK}:= (\rho_+)_I{}^L\, (\rho_+)_J{}^M\, (\rho_+)_K{}^N\, \hat T_{LMN} \ .
\ee
We can identify $\cH_{IJ}(\X)$ with the covariant generalized metric on $T^*M$, provided we take
a diagonal symmetric form $g_{IJ}$, i.e. $g_i{}^j=g^i{}_j=0$.
Indeed, substituting the components of $\rho_+$ from \eqref{rhoplus}, we find that ${\cal
  H}_{IJ}$ is then given by 
\be 
\begin{pmatrix} g_{ij}{-}B_{ik}\,g^{kl}\,B_{lj} & -B_{ik}\,g^{kj}{+}g_{ik}\,\beta^{kj}{-}B_{ik}\,g^{kl}\,B_{lm}\,\beta^{mj} \\ 
g^{ik}\,B_{kj}{-}\beta^{ik}\,g_{kj}+\beta^{im}\,B_{mk}\,g^{kl}\,B_{lj} & g^{ij}{-}\beta^{ik}\,g_{kl}\,\beta^{lj}
{+}2\,g^{(i\underline{l}}\,B_{ln}\,\beta^{\underline{n}\,j)}{+}\beta^{im}\,B_{mk}\,g^{kl}\,B_{ln}\,\beta^{nj}
\end{pmatrix}~.
\ee
As expected, when $\beta=0$ one obtains the familiar geometric parameterization
\be 
{\cal H}_{IJ}=\begin{pmatrix} g_{ij}-B_{ik}\,g^{kl}\,B_{lj} & -B_{ik}\,g^{kj} \\ 
g^{ik}\,B_{kj} & g^{ij}
\end{pmatrix}~,
\ee
while dually for $B=0$ one obtains the non-geometric parameterization implied
by the open-closed background field redefinition \eqref{eq:openclosed} as
\be
\widetilde{\cal H}_{IJ} = \begin{pmatrix}
g_{ij} & g_{ik}\,\beta^{kj} \\
-\beta^{ik}\,g_{kj} & g^{ij}-\beta^{ik}\,g_{kl}\,\beta^{lj}
\end{pmatrix} \ .
\ee

\subsection{Gauge invariance and Bianchi identities}
\label{sec42}

A systematic way to derive the Bianchi identities for the fluxes is to
examine the gauge invariance of the DFT membrane sigma-model action \eqref{dftsm}. 
For this, we consider the infinitesimal gauge
transformations\footnote{In this subsection we simplify the notation
  for the components of the map $\rho_+$ by denoting $(\rho_+)^{I}{}_{J}$ as~$\rho^{I}{}_{J}$.}
\bea\label{gt1}
\delta_\epsilon \X^I&=&\rho^{I}{}_{J}(\X)\,\epsilon^J~,\\[4pt] \label{gt2}
\delta_\epsilon A^I&=&\dd\epsilon^I+\eta^{IJ}\,{\hat T}_{JKL}(\X)\,A^K\,\epsilon^L~,
\eea
where $\epsilon$ is a gauge parameter which is a function only of the
worldvolume coordinates on $\S_3$.
To test the invariance of the action \eqref{dftsm} under these transformations, first we introduce the worldvolume  derivative
 \be 
D\X^I=\dd\X^I-\rho^{I}{}_{J}(\X)\,A^J~,
\ee
 which accompanies the auxiliary fields $F_I$ in the sigma-model action. 
  It transforms under \eqref{gt1} and \eqref{gt2} as
\bea\label{deltadx1}
\delta_\epsilon D\X^I=\epsilon^J\,\partial_K\rho^{I}{}_{J}\,D\X^K+
\big(\rho^{K}{}_{L}\,\partial_K\rho^{I}{}_{M}-\rho^{K}{}_{M}\, \partial_K\rho^{I}{}_{L}
-\rho^{I}{}_{J}\,\eta^{JK}\,
\hat T_{KLM}\big)\,A^L\,\epsilon^M~.\eea
Had we required that this derivative transforms covariantly,  
as would have been the case for a Courant sigma-model, the second term would have to vanish. However, one can easily verify that it does not. 
Indeed, the DFT fluxes $\hat T$ satisfy 
\be\label{c1}
\rho^{K}{}_{L}\,\partial_K\rho^{I}{}_{M}-\rho^{K}{}_{M}\, \partial_K\rho^{I}{}_{L}-\sfrac
12\, \rho_{KL}\,\partial^I\rho^K{}_M+\sfrac 12\,
\rho_{KM}\,\partial^I\rho^K{}_L=\rho^{I}{}_{J}\,\eta^{JK}\,\hat
T_{KLM}~,
\ee
since they are obtained via the C-bracket. 
This implies that 
\bea
\delta_\epsilon D\X^I=\epsilon^J\,\partial_K\rho^{I}{}_{J}\,D\X^K+
\rho_{K[L}\,\partial^{I}\rho^K{}_{M]}\,A^L\,\epsilon^M~.
\eea
Later we will prove that the last term does not contribute to the
gauge variation of the action when the strong constraint
\eqref{eq:strongconstraint} is
satisfied. Moreover, in Section \ref{sec6} we will suggest a way of
eliminating this term altogether.{\footnote{It appears as if it is
    possible to get rid of this term already by allowing the
    transformation \eqref{gt2} to contain an extra term such that the
    combination \eqref{c1} appears as such in
    \eqref{deltadx1}. However, in that case new terms of the form
    $A\w\dd A$ would arise in the gauge variation of the action, whose interpretation is unclear.}}

Equipped with this relation, we proceed with the evaluation of the
gauge variation of the action \eqref{dftsm} to get
\bea
\delta_\epsilon S &=& \int_{\Sigma_3}\,\big(\eta_{IJ}\,\dd\epsilon^I\w
\dd A^J+\rho_{K[L}\,\partial^I\rho^K{}_{M]}\,\epsilon^M\, F_I\w A^L \nn\\
&& \qquad +\,\d_{\epsilon}F_K\w
D\X^K+\epsilon^J\,(\partial_K\rho^{I}{}_{J}\,F_I-\partial_K \hat
T_{ILJ}\,A^I\w A^L)\w D\X^K \label{gvS}\\
&& \qquad +\, \epsilon^L\,(\eta^{MN}\,\hat T_{MJK}\,\hat
T_{ILN}+\rho^{M}{}_{I}\,\partial_M \hat T_{KJL}+\sfrac 13\,
\rho^{M}{}_{L}\,\partial_M\hat T_{IJK})\,A^I\w A^J\w A^K\big)~. \nn
\eea
The first term is a total derivative, while the second line vanishes upon postulating that the gauge variation of the auxiliary 2-form $F_I$ is 
\be 
\d_{\epsilon}F_K=-\epsilon^J\,\big(\partial_K\rho^{I}{}_{J}\,F_I-\partial_K \hat T_{ILJ}\,A^I\w A^L\big)~.
\ee
Considering the variation of the action at face value, there is no way to cancel the term in the third line of \eqref{gvS} against another term; thus an additional requirement would be that
\be\label{c2}
3\,\eta^{MN}\,\hat T_{M[JK}\,\hat
T_{I]LN}+3\,\rho^{M}{}_{[I}\,\partial_{\underline{M}}\hat T_{KJ]L}+
\rho^{M}{}_{L}\,\partial_M\hat T_{IJK}=0~.
\ee
Since this is a differential condition for the fluxes, it is naturally
interpreted as the implementation of the Bianchi identities in the DFT
membrane sigma-model. This is confirmed by 
noting that the first term is in fact antisymmetric in all four indices $(IJKL)$, while the second and the third term combine to a single term antisymmetric in these indices; in other words, we rewrite the equation as 
\be\label{c2b}
3\,\eta^{MN}\,\hat T_{M[JK}\,\hat
T_{IL]N}+4\,\rho^{M}{}_{[I}\,\partial_{\underline{M}}\hat T_{KJL]}=0~.
\ee
This is indeed the correct formula for the Bianchi identities in DFT, see e.g. \cite{dftflux3}, \emph{after imposing the strong constraint}. 
 Substitution into \eqref{c2b} of the explicit expressions for the DFT fluxes together with the anchor components from \eqref{rhoplus} leads to its expanded form
 \bea 
{\cal{D}}_{[i}H_{jkl]} &=& \sfrac 32\, H_{m[ij}\,f_{kl]}{}^m~,\nn\\[4pt]
 {\cal D}_{[i}f_{jk]}{}^l - \sfrac 13\,\widetilde{\cal D}^l H_{ijk} 
 &=&Q_{[i}{}^{lm}\,H_{jk]m} -f_{[ij}{}^m\,f_{k]m}{}^l~,\nn\\[4pt]
 {\cal D}_{[i}Q_{j]}{}^{kl} + \widetilde{\cal D}^{[k}f_{ij}{}^{l]}
 &=& \sfrac 12\, f_{ij}{}^m\, Q_m{}^{kl} +\sfrac 12\, H_{ijm}\,R^{mkl} - 2\,Q_{[i}{}^{m[k}\,f_{j]m}{}^{l]}~,\label{obi}\\[4pt]
\widetilde{\cal D}^{[i}Q_l{}^{jk]} -\sfrac 13\, {\cal D}_l R^{ijk} &=& f_{lm}{}^{[i}\,R^{jk]m}-Q_m{}^{[ij}\,Q_l{}^{k]m}~,\nn\\[4pt]
\widetilde{\cal D}^{[i}R^{jkl]} &=& \sfrac{3}{2}\,R^{m[ij}\, Q_m{}^{kl]}~,\nn
 \eea
 where 
 \be{\cal D}_i=\partial_i+B_{ji}\,\tilde{\partial}^j \qquad \mbox{and}
 \qquad \tilde{D}^i=\tilde{\partial}^i+\beta^{ji}\,{\cal D}_j~,
\ee
 and we used the identifications \eqref{eq:hatTIJKHfQR}.
 Recall that these are expressions in a holonomic frame; the corresponding expressions for a non-holonomic frame may be found using similar methods.

However, there is a delicate issue here. The second term in the first line of \eqref{gvS} cannot be cancelled and thus it would give rise to a gauge anomaly. How can this be? 
In order to avoid this, one may impose the following constraint 
\be\label{anomaly} 
\rho_{KL}\,\partial^I\rho^K{}_M\,\epsilon^M\,F_I\w A^L=
\rho_{KL}\,\partial_i\rho^K{}_M\,\epsilon^M\,F^i\w
A^L+\rho_{KL}\,\tilde{\partial}^i\rho^K{}_M\,\epsilon^M\,F_i\w A^L
=0~, 
\ee   
where we opened up only the index contracted among the derivative and $F$.
We discuss this point and its relation to the strong constraint of DFT systematically in Section \ref{sec5}. The very presence of this term also explains why the Bianchi identities above are only valid when a constraint is used. In accord with \cite{dftflux3}, we could just impose 
\be \label{biz}
3\,\eta^{MN}\,\hat T_{M[JK}\,\hat T_{IL]N}+4\,\rho^{M}{}_{[I}\,\partial_{\underline{M}}\hat T_{KJL]}={\cal Z}_{IJKL}~,
\ee
where ${\cal Z}$ is a 4-form. As we will discuss in Section \ref{sec5}
in terms of a modified Jacobi identity, after solving the strong
constraint \eqref{eq:strongconstraint} this 4-form can be consistently set to zero and the Bianchi identities are recovered as above. However, there is a way to relax this. We can introduce the 4-form ${\cal Z}$ as a Wess-Zumino term 
on an extension of the membrane worldvolume to four dimensions, as in~\cite{Hansen:2009zd}. Thus we take a four-dimensional worldvolume $\S_4$ such that $\partial\S_4=\S_3$ and the action{\footnote{If $\S_3$ is a manifold with boundary, as we have assumed before, then $\S_4$ must be a manifold with corners of codimension two in order to support this Wess-Zumino term, analogously to the situation discussed in~\cite{Mylonas:2012pg}. If the boundary $\S_3=\partial\S_4$ consists of two faces $\S_3^\pm$, i.e. $\S_3=\S_3^+\cup\S_3^-$ and $\partial\S_3=\S_3^+\cap\S_3^-$, then different boundary conditions have to be implemented on $\S_3^+$ and $\S_3^-$ in order to reproduce the fields of the pertinent worldsheet sigma-model on their intersection.}} 
\be
 \hat{S}[\X,A,F]=S+\int_{\S_4}\,\sfrac 1{4!}\,{\cal T}_{IJKL}\,\dd \X^I\w\dd\X^J\w\dd\X^K\w\dd\X^L~.
 \ee
If the 4-form ${\cal T}$ is closed, 
 \be 
 \dd {\cal T}=0~,
 \ee
then the gauge variation of this action vanishes provided that \eqref{biz} holds with
 \be 
 {\cal Z}_{IJKL}=\sfrac 12\,\rho^M{}_I\,\rho^N{}_J\,\rho^P{}_K\,\rho^Q{}_L\,{\cal T}_{MNPQ}~,
 \ee
  and the variation of the auxiliary 2-form $F_I$ is modified to 
 \bea 
 \d_{\epsilon}F_K&=&-\epsilon^J\,\big(\partial_K\rho^{I}{}_{J}\,F_I-(\partial_K \hat T_{ILJ}+\sfrac 16\,\rho^M{}_J\,\rho^N{}_L\,\rho^P{}_I\,{\cal T}_{MNPK})\,A^I\w A^L \nn\\
 && \qquad +\,\sfrac 16\,\rho^M{}_J\,{\cal T}_{MNPK}\,\dd \X^N\w\dd \X^P+\sfrac 16 \,\rho^M{}_J\,\rho^P{}_L\,{\cal T}_{MNPK}\,\dd\X^N\w A^L\big)~.
 \eea
In this way, even after the strong constraint is solved, the underlying geometric structure is not precisely a Courant algebroid, but a Courant algebroid twisted by this closed 4-form~${\cal T}$.

\section{The DFT algebroid structure}
\label{sec5}

In Section \ref{sec2} we mentioned that the geometric structure of DFT lies between the two Courant algebroid structures over $T^{\ast}M$ and $M$ respectively. Let us call the first one the \emph{large} Courant algebroid and the second the \emph{canonical} Courant algebroid. 
Here we would like to understand better what the intermediate
structure is. First, we know what it is not; it cannot be a Courant
algebroid. The quickest way to see this is to note that the canonical
Courant algebroid is associated with the fields $(X^i,A^I,F_i)$ of the Courant sigma-model, while
the large Courant algebroid is associated with the fields
$(\X^I,\A^{\hat{I}},\F_I)$ of the large Courant sigma-model respectively. Recalling that $i=1,\ldots,d,\;I=1,\ldots, 2d,\;\hat I=1,\ldots,4d$, we see that in both cases the number of  1-forms $A$ is double the number of fields corresponding to the target manifold coordinates $X$ or auxiliary fields $F$. This is true in any Courant algebroid.
However, in the DFT case the relevant data comprise the
fields $(\X^I,A^I,F_I)$ and the number of all fields is the same, since they all carry the same index. Another, maybe more intuitive way to understand this is the following:  the canonical Courant algebroid  defined over a $d$-dimensional target has an $O(d,d)$-invariant metric on its vector bundle, the large Courant algebroid defined over a $2d$-dimensional target has an $O(2d,2d)$-invariant metric on its vector bundle, while in DFT case we  have a $2d$-dimensional target (as in the large Courant algebroid) but an $O(d,d)$-invariant metric (as in the canonical Courant algebroid). The goal of this section is to establish a more precise
criterion for this statement, and to properly define the new geometric
structure. 

\subsection{The role of the strong constraint}

A Courant algebroid comes with a set of axioms (see Appendix~\ref{seca}). In local coordinates these axioms lead to three equations, one algebraic and two differential, given in \eqref{ca1}--\eqref{ca3}. These are obviously valid in both the canonical and the large Courant algebroids. In the canonical case, the algebraic equation stems from one of the properties of a Courant algebroid $E$ over $M$, 
\be \label{capro4}
\langle {\cal D}f,{\cal D}g\rangle_E=0~,
\ee
which in local coordinate form reads as 
\be \label{aaa}
\rho^i{}_{I}\,\eta^{IJ}\,\rho^j{}_J\,\partial_if\,\partial_jg=0~,
\ee
for all functions $f,g\in C^{\infty}(M)$.

On the other hand, in DFT the situation differs. As explained in the previous sections, instead of the map $\rho=(\rho^{i}{}_{J}): E\to TM$, the role of the anchor in DFT is played by $\rho_+=(\rho^{I}{}_{J}):L_+\to T(T^{\ast}M)$. In a general parametrization, the components of $\rho_+$ are given in \eqref{rhoplus}. At this stage, using \eqref{rhoplus}, it is useful to compute
\be \label{pre1}
\rho^K{}_I\,\eta^{IJ}\,\rho^{L}{}_J=\eta^{KL}~.
\ee   
This directly implies that
\be \label{aaa2}
\rho^K{}_{I}\,\eta^{IJ}\,\rho^L{}_J\,\partial_Kf\,\partial_Lg=\eta^{KL}\,\partial_Kf\,\partial_Lg~,
\ee
for $f,g\in C^\infty(T^\ast M)$, and the right-hand side is  in
general  non-vanishing. Thus, one immediately sees the failure of the Courant algebroid structure for general $\rho_+$. 
Had $\rho_+$ been an anchor map in a Courant algebroid, the right-hand side of \eqref{aaa2} would have been zero, as in \eqref{aaa}.
 Its vanishing is precisely the strong constraint \eqref{eq:strongconstraint}. In other words, before imposing the strong constraint the relevant structure cannot be a Courant algebroid, but it can become such when the strong constraint is imposed. The expression \eqref{aaa2} can be written without reference to a local coordinate system as
\be \label{scdfta}
\langle {\cal D}_+f,{\cal D}_+g\rangle_{L_+}=\sfrac 14\, \langle \dd f,\dd g\rangle_{L_+}~,
\ee
where 
\be 
\langle {\cal D}_+f,A\rangle_{L_+}=\sfrac 12\, \rho_+(A)f~,
\ee
or, in local coordinates,
\be 
{\cal D}_+f=\sfrac 12 \,\rho^K{}_L\,\partial_Kf\,\eta^{LJ}\,e_{J}^+~.
\ee
Thus \eqref{scdfta} should be one of the properties of the DFT geometric structure before imposing the strong constraint.

From a different point of view, the local coordinate form \eqref{aaa} 
of the Courant algebroid  property \eqref{capro4} may be obtained directly from the classical master equation (see Appendix~\ref{seca}). As explained in Appendix~\ref{seca},  the Courant algebroid data can be  recovered  from a differential graded manifold ${\cal M}$ equipped with a degree-2 symplectic form $\omega$ and a degree-3 Hamiltonian function $\Theta$. In particular, ${\cal M}$ is equipped with local Darboux coordinates $(x^i, A^I, F_i)$ of degree 0,1 and 2 respectively, while the symplectic form $\omega=\dd x^i\w \dd F_i+\sfrac 12\eta_{IJ}\dd A^I\w\dd A^J$  is utilized to construct the graded Poisson bracket. 
With the most general Hamiltonian function \eqref{Th}, the classical master equation $\{\Theta,\Theta\}=0$ yields three conditions, the first of which reads as
\be 
  \big(\rho^k{}_I\,\eta^{IJ}\,\rho^{l}{}_J\big)\,F_k\,F_l=0~.
  \ee
  From the point of view of the membrane sigma-model, the $F_i$  correspond to  the auxiliary worldvolume 2-forms introduced in \eqref{scasm}.
  In this spirit, in the case of DFT, the classical master equation leads instead to  
  \be \label{pre1b}
  \big(\rho^K{}_I\,\eta^{IJ}\,\rho^{L}{}_J\big)\,F_K\,F_L=\eta^{KL}\,F_K\,F_L =:F^K\,F_K~,
  \ee
  and the right-hand side is in general non-vanishing.
This was also derived in \cite{Heller:2016abk}. One immediately observes that this can be zero when, for instance, $F^i=0$. Recalling that $F^i$ is the conjugate variable to $\widetilde X_i$, we conclude that when nothing depends on the dual coordinates this obstruction to the Courant algebroid structure is eliminated. In other words, the solution of the strong constraint reduces the DFT structure to a Courant algebroid structure. Note that different solutions of the strong constraint are naturally implemented in this discussion. For example, in the opposite case of eliminating all target space coordinates $X^i$, the conjugate variable $F_i$ is eliminated and \eqref{pre1b} gives again zero.
According to this discussion, it is now clear how the gauge anomaly encountered in Section~\ref{sec42} is accounted for. The relevant term appears in \eqref{anomaly}.
Now solving the strong constraint as 
$\tilde{\partial}^i=0$ and $F^i=0$, i.e. eliminating dual coordinates,
renders this expression zero. The same is true for the alternative
choice $\partial_i=0$ and $F_i=0$, or any other mixed choice that
solves the strong constraint and eliminates half of the coordinates; the different choices are related by $O(d,d)$ transformations, and both \eqref{pre1b} and \eqref{anomaly} are $O(d,d)$-invariant. 

However, this is not the only relation we should examine, since there are two additional ones. In our case, these are given by the two conditions \eqref{c1} and \eqref{biz}. For clarity, we summarize all relevant data in Table~\ref{table1}. The local coordinate expressions appearing in the third column clarify in which sense the DFT structure lies between the two Courant algebroids. The first equation was already discussed above. The second equation in the DFT case also exhibits a non-trivial right-hand side, which is zero in the case of Courant algebroids. It appears in the gauge anomaly and it is zero when the strong constraint is imposed.  

\begin{center}
\begin{table}[t]
\begin{center}
	\begin{tabular}{|c|c|c|}
		\hline {Algebroid structure} & {Fields} & {Axioms in local coordinates} 
		\\[4pt] \hhline{|=|=|=|}
		Large Courant & $(\X^I,\A^{\hat I},\F_I)$ & 
		\begin{tabular}{c}\\
			$\eta^{\hat{I}\hat{J}}\,\rho^K{}_{\hat{I}}\,\rho^L{}_{\hat{J}}=0$\\[2pt] $2\,\rho^L{}_{[\hat{I}}\,\partial_{{L}}\rho^K{}_{\hat{J}]}
			-\eta^{\hat{M}\hat{N}}\,\rho^K{}_{\hat{M}}\,T_{\hat{N}\hat{I}\hat{J}}=0 
			$\\[2pt] $4\,\rho^{{M}}{}_{[\hat{L}}\,\partial_{{M}}T_{\hat{I}\hat{J}\hat{K}]}+
			3\,\eta^{\hat{M}\hat{N}}\,
			T_{\hat{M}[\hat{I}\hat{J}}\,T_{\hat{K}\hat{L}]\hat{N}}=0$\\[4pt]\end{tabular} \\[4pt] \hline 
		DFT & $(\X^I,A^{I},F_I)$ & 
		\begin{tabular}{c}
			\\$\eta^{{I}{J}}\,\rho^K{}_{{I}}\,\rho^L{}_{{J}}=\eta^{KL}$\\[2pt] 
			$2\,\rho^{L}{}_{[I}\,\partial_{\underline{L}}\rho^{K}{}_{J]}-\eta^{MN}\,\rho^{K}{}_{M}\,\hat T_{NIJ}= \rho_{L[I}\,\partial^K\rho^L{}_{J]}$\\[2pt] $4\,\rho^{M}{}_{[L}\,\partial_{\underline{M}}\,\hat T_{IJK]}+3\,\eta^{MN}\,\hat T_{M[IJ}\,\hat T_{KL]N}={\cal Z}_{IJKL}$\\[4pt]
		\end{tabular} 
		\\[4pt] \hline 
		Canonical Courant &$(X^i,A^{I},F_i)$ &
		\begin{tabular}{c}\\
			$\eta^{IJ}\,\rho^k{}_{I}\,\rho^l{}_J=0$\\[2pt] $2\,\rho^l{}_{[I}\,\partial_l\rho^k{}_{J]}-\eta^{MN}\,\rho^k{}_M\,T_{NIJ}=0 
			$ \\[2pt] 
			$4\,\rho^m{}_{[L}\,\partial_mT_{IJK]}+3\,\eta^{MN}\,
			T_{M[IJ}\,T_{KL]N}=0$\\[4pt]\end{tabular}\\\hline
	\end{tabular}
\end{center}
\caption{\small The fields and local coordinate expressions for the axioms of the three different geometric structures encountered. With reference to the classical master equation, the three sub-rows in the last column of each row are the 0-form coefficients in front of the 4-forms $\F_K\,\F_L/F_K\, F_L/F_k\, F_l$ for the first sub-row, $\F_K\,\A^{\hat{I}}\,\A^{\hat{J}}/F_K\, A^I\, A^J/F_k\, A^I\, A^J$ for the second sub-row, and $\A^{\hat{I}}\, \A^{\hat{J}}\,\A^{\hat{K}}\,\A^{\hat{L}}/A^I\, A^J\, A^K\, A^L/A^I\, A^J\, A^K\, A^L$ for the third sub-row, respectively. Indices run as $i=1,\dots,d$, $I=1,\dots,2d$ and $\hat I=1,\dots,4d$. }\label{table1}
\end{table}
\end{center}

\subsection{Global formulation and Courant algebroids}

Now our goal is to express these relations without reference to a
local coordinate system, thereby obtaining a set of axioms and
properties that the DFT structure should satisfy in general, similarly
to Definition~\ref{ca1def} in the case of a Courant algebroid. For
this, we will examine properties 1--5 of Definition~\ref{ca1def} by
replacing the Courant bracket with the C-bracket \eqref{cbragen}, the
fiber metric $\langle\, \cdot \,,\, \cdot \,\rangle_E$ with
$\langle\, \cdot \,,\, \cdot \,\rangle_{L_+}$, and the anchor $\rho$ with
$\rho_+$, and examine the resulting geometric structure, which is not known a priori since this is not a Courant algebroid structure. We do not impose the strong constraint in this process.

First, for the Jacobi-like identity, one obtains
\bea \label{preJ}
\cbral\cbral A,B\cbrar_{L_+},C\cbrar_{L_+}+\text{cyclic}={\cal D}_+
{\cal N}_+(A,B,C) + {\cal Z}(A,B,C)+{\sf SC}_{\rm Jac}(A,B,C)~,
\eea
where ${\cal N}_+$ is the analog of the Nijenhuis operator for the C-bracket,
\be 
{\cal N}_+(A,B,C)=\sfrac 13\, \langle \cbral A,B\cbrar_{L_+},C\rangle_{L_+}+\text{cyclic}~,
\ee
and ${\cal Z}$ is a 4-form with components as given in Table
\ref{table1}. The DFT $(3,1)$-tensor ${\sf SC}_{\rm Jac}$ vanishes upon imposing the strong constraint and its explicit local form is given by
\bea \label{sc1}
{\sf SC}_{\rm Jac}(A,B,C)^L&=&-\sfrac 12\, \big(A^I\,\partial_J B_I\,\partial^JC^L-B^I\,\partial_J A_I\,\partial^JC^L\big)
\nn\\
&& \qquad -\, \rho_{I[J}\,\partial_M
\rho^I{}_{N]}\,\big(A^J\,B^N\,\partial^M C^L-\sfrac 12\,
C^J\,A^K\,\partial^M B_K\,\eta^{NL}\nn\\ && \qquad \qquad \qquad \qquad
\qquad +\,\sfrac 12\, C^J\,B^K\,\partial^M A_K\,\eta^{NL}\big) +\text{cyclic} ~.
\eea
We observe that the C-bracket does not satisfy the very first of the axioms in Definition~\ref{ca1def}, which confirms once more the claim that the structure is not a Courant algebroid. At this point one might suspect that the relevant structure is that of a pre-Courant algebroid, which fails to be a Courant algebroid precisely due to the violation of property 1 in Definition~\ref{ca1def}. However, we can already infer that this is not the case, since for a pre-Courant algebroid property 4 in Definition~\ref{ca1def} continues to hold, while here we have already seen that it is in general violated in \eqref{scdfta}. 

For the Leibniz rule (property 3 in Definition~\ref{ca1def}), a straightforward calculation reveals that 
\be 
\cbral A,f\,B\cbrar_{L_+}=f\,\cbral A,B\cbrar_{L_+}+\big(\rho_+(A)f\big)\,B-\langle A,B\rangle_{L_+}\,{\cal D}_+f~,
\ee
for all functions $f\in C^{\infty}(T^*M)$.
In other words, the Leibniz rule is not modified with respect to the (pre-)Courant algebroid structure.

Next we move on to the analog of the compatibility condition expressed as property 5 in Definition~\ref{ca1def}. We find
\be 
\langle \cbral C,A\cbrar_{L_+}+{\cal D}_+\langle C,A\rangle_{L_+},B\rangle_{L_+}+\langle A,\cbral C,B\cbrar_{L_+}+{\cal D}_+\langle C,B\rangle_{L_+}\rangle_{L_+}=\rho_+(C)\langle A,B\rangle_{L_+}~.
\ee
Thus we also find an unmodified compatibility condition for the DFT structure. 

Finally, we examine the homomorphism property for $\rho_+$. A direct computation leads to 
\be \label{prehomo}
\rho_+\cbral A,B\cbrar_{L_+}=[\rho_+(A),\rho_+(B)]+{\sf SC}_\rho(A,B)~,
\ee
where ${\sf SC}_\rho$ vanishes upon imposing the strong constraint and in local coordinates it reads as 
\be \label{sc2}
{\sf SC}_\rho(A,B)=\big(\rho_{L[I}\,\partial^K\rho^L{}_{J]}\,A^I\,B^J+\sfrac 12\, (A^I\,\partial^KB_I-B^I\,\partial^KA_I)\big)\,\partial_K~.
\ee 
Thus $\rho_+$ is not a homomorphism of bundles, but rather a ``quasi-homomorphism'' whose failure to preserve the brackets on $\G(L_+)$ and $\G(T(T^*M))$ is controlled by the strong constraint of DFT.

We can collect our discussion above into the following precise definition.
  \begin{defn}\label{dftalg1}
Let $M$ be a $d$-dimensional manifold. A \underline{DFT algebroid} on
$T^*M$ is a quadruple $(L_+,\cbral\, \cdot \,,\, \cdot \,\cbrar_{L_+},\langle
\, \cdot \,,\, \cdot \,\rangle_{L_+},\rho_+)$, where $L_+$ is vector bundle of
rank $2d$ over $T^{\ast}M$ equiped with a skew-symmetric bracket
$\cbral\, \cdot \,,\, \cdot \,\cbrar_{L_+}:\G(L_+)\otimes\G(L_+)\to\G(L_+)$, a
non-degenerate symmetric form $\langle
\, \cdot \,,\, \cdot \,\rangle_{L_+}:\G(L_+)\otimes\G(L_+)\to C^\infty(T^*M)$, and
a smooth bundle map $\rho_+:L_+\to T(T^\ast M)$, which satisfy
  	\begin{enumerate}
  	\item $\langle {\cal D}_+f,{\cal D}_+g\rangle_{L_+}=\sfrac 14\, \langle \dd f,\dd g\rangle_{L_+}~;$
  		\item \ $\cbral A,f\,B\cbrar_{L_+}=f\,\cbral A,B\cbrar_{L_+}+\big(\rho_+(A)f\big)\,B-\langle A,B\rangle_{L_+}\,{\cal D}_+f~;$ 
  		\item \ $\langle \cbral C,A\cbrar_{L_+}+{\cal D}_+\langle C,A\rangle_{L_+},B\rangle_{L_+}+\langle A,\cbral C,B\cbrar_{L_+}+{\cal D}_+\langle C,B\rangle_{L_+}\rangle_{L_+}=\rho_+(C)\langle A,B\rangle_{L_+}~;$
  		\end{enumerate}
for all $A,B,C\in \G(L_+)$ and $f,g\in C^{\infty}(T^*M)$, where ${\cal D}_+:C^\infty(T^*M)\to\G(L_+)$ is the derivative defined through $\langle {\cal D}_+f,A\rangle_{L_+}=\sfrac 12\, \rho_+(A)f$.
\end{defn}
\begin{rmk}
	A DFT algebroid as defined above is a special case of a more general structure where properties 1, 2 and 4 of Definition~\ref{ca1def} are relaxed. In Appendix~\ref{seca} we discuss this pre-DFT algebroid structure, whose supermanifold description corresponds to a symplectic nearly Lie 2-algebroid \cite{preca2}. Note that 
	although the DFT algebroid is an example of pre-DFT algebroid
        by construction, there exist pre-DFT algebroids which are not
        DFT algebroids; we spell out an explicit example in Appendix~\ref{seca4}. This outcome is reasonable in view of the fact that we reverse-engineered a definition from a set of local expressions; the general structure thus encompasses more cases than the particular case that motivated it.  
\end{rmk}
\begin{rmk}
A more constructive definition, along the lines in which we have
explicitly obtained it, would be to define a DFT algebroid as a
projection of a Courant algebroid
$(E,[\, \cdot \,,\, \cdot \,]_E,\langle\, \cdot \,,\, \cdot \,\rangle_E,\rho)$ over
$T^{\ast}M$, in the sense that there exists a surjective bundle map
$\sfp_+:E\to L_+$ which induces a bracket on $L_+$-sections $\cbral
\, \cdot \,,\, \cdot \,\cbrar_{L_+} :=\sfp_+([\sfp_+^{-1} \, \cdot \,,\sfp_+^{-1}
\, \cdot \,]_E)$, a non-degenerate bilinear form
$\langle\, \cdot \,,\, \cdot \,\rangle_{L_+}:=\langle\sfp_+^{-1} \, \cdot \,,\sfp_+^{-1}
\, \cdot \,\rangle_E$, and a bundle map $\rho_+:=\rho\circ\sfp_+^{-1} :
L_+\to T(T^\ast M)$, such that properties 1--3 of Definition
\ref{dftalg1} hold.
\end{rmk}

Note that in Definition~\ref{dftalg1} we do not require that $\rho_+$ is a homomorphism of bundles, and based on our discussion above and in Appendix~\ref{seca} we have
\begin{prop}\label{prop:DFTalgstrong}
Let $L_+$ be a DFT algebroid on $T^*M$. If the strong
constraint of DFT is imposed, then the map $\rho_+$ becomes a bundle homomorphism and $L_+$ reduces to a Courant algebroid over~$T^\ast M$.
\end{prop}
\begin{rmk}
	Although a DFT algebroid reduces to a Courant algebroid on the
        strong constraint, this is not true for the more general
        structure of a pre-DFT algebroid.{\footnote{This is a metric
        		algebroid in the terminology of~\cite{Vaisman:2012ke} and is used there to describe the
        		C-bracket and reductions to Courant algebroids in a similar way to our
        		treatment.}} In this case one encounters
        intermediate structures. Indeed, imposing $\langle {\cal
          D}f,{\cal D}g\rangle_E=0$ on a pre-DFT algebroid $E$ leads to an
        ante-Courant algebroid (see Appendix~\ref{seca3}), where
        $\rho$ is still only a quasi-homomorphism. Imposing that $\rho$ is
        a homomorphism reduces an ante-Courant algebroid to a pre-Courant
        algebroid, which only becomes a true Courant algebroid when
        the Jacobi identity is satisfied. This naturally suggests a
        weakening of the strong constraint: The strong
        constraint of DFT is sufficient to guarantee reduction of a DFT
        algebroid $L_+$ to a Courant algebroid on $T^*M$, whereas the
        weaker notion of a pre-DFT algebroid can be more generally
        reduced, in a coordinate-independent way, to a Courant
        algebroid via weaker constraints that do not necessarily imply
        the strong constraint.
	\end{rmk}
		
Having established in Proposition~\ref{prop:DFTalgstrong} what becomes of the
DFT algebroid structure when the strong constraint is imposed, let us now
examine what happens on an explicit solution of the strong
constraints. Following~\cite{Aschieri:2015roa,Freidel:2017yuv}, solving the strong constraints amounts to choosing a
polarization, which is a foliation of $T^*M$ over a $d$-dimensional submanifold $M_\cP$ which decomposes the tangent bundle as $T(T^*M)=L\oplus\widetilde L$, where the integrable distribution $L=TM_\cP$ is the tangent bundle on the leaves of the foliation and $\widetilde L$ is its dual bundle with respect to the orthogonal complement in the $O(d,d)$ metric \eqref{eta}. The strong constraint then restricts the set of admissible fields to foliated tensor fields $T_\cP$ with respect to the distribution $\widetilde L$: $\iota_{\widetilde A}T_\cP=\cL_{\widetilde A}T_\cP=0$ for all sections $\widetilde A\in\Gamma(\,\widetilde L\,)$. A polarization may be defined by introducing a projection $\cP:T(T^*M)\to L$ mapping a local frame $e_I$ of the tangent bundle $T(T^*M)$ onto the vector fields
\be 
e_i = \cP_i{}^J\, e_J \ ,
\ee
which span a $d$-dimensional subspace of the $2d$-dimensional tangent space, that is maximally isotropic with respect to the metric \eqref{eta}; in other words
\be 
\cP_i{}^K\, \eta_{KL}\, \cP_j{}^L = 0 \ .
\ee

We can also define a polarization of local coordinates,\footnote{Recalling that $M$ is assumed to be
  contractible, in the present discussion we work mostly in affine
  coordinates and assume $M=\R^d$ throughout.} which is specified by a
constant projector $\cP:T^*M\to T^*M$, $\cP^2=\cP$, of rank $d$ whose image carves out a $d$-dimensional submanifold $M_\cP\hookrightarrow T^*M$ with coordinates
\be
Z^i = \cP^i{}_J\,\X^J \ .
\ee
For example, the supergravity frame with $M_\cP=M$ is reached with
$\cP^i{}_J=(\delta^i{}_j,0)$, while the winding frame with $\widetilde
X_i=\widetilde \cP_{iJ}\,\X^J$ corresponds to the complementary
projector $\widetilde\cP = 1 - \cP$. We require the subspace $M_\cP$ to be maximally isotropic with respect to the $O(d,d)$ metric \eqref{eta}, in the sense that
\be\label{eq:maxisotropic}
\cP^i{}_K\, \eta^{KL}\, \cP^j{}_L = 0 \ .
\ee
Different choices of polarization are all related by $O(d,d)$ transformations: Acting with $\cO\in O(d,d)$ changes the polarization as
\be\label{eq:polarizationchange}
\big(\,{}^\cP_{\widetilde\cP}\,\big) \longmapsto \big(\,{}^{\cP'}_{\widetilde\cP{}'}\,\big) = \big(\,{}^\cP_{\widetilde\cP}\,\big) \, \cO \ .
\ee
The projection $\cP$ induces as usual a pullback $\cP^*$, which is right-inverse of the restriction to $M_\cP\subset T^*M$, and also a pushforward $\cP_*$, which is integration over the fibers of the bundle $T^*M\to M_\cP$.

Given a DFT algebroid $(L_+,\cbral\, \cdot \,,\, \cdot \,\cbrar_{L_+},\langle
\, \cdot \,,\, \cdot \,\rangle_{L_+},\rho_+)$ on the doubled space $T^*M$, the polarization selects a vector bundle $E_\cP:=L_+\big|_{M_\cP}$ of rank~$2d$ as the restriction of $L_+$ to the maximally isotropic submanifold $M_\cP\subset T^*M$.
Define a smooth bundle map $\rho_\cP:E_\cP\to TM_\cP$ by $\rho_\cP:=
\cP_*\circ\rho_+\circ\cP^*$, a skew-symmetric bracket $[\, \cdot \,,\,
\cdot \,]_\cP:\Gamma(E_\cP)\otimes\Gamma(E_\cP)\to \Gamma(E_\cP)$ by
$[\,\cdot\,,\,\cdot\,]_\cP:= \cP_*\big(\cbral \cP^*\, \cdot\, \,,\,
\cP^*\, \cdot\, \cbrar_{L_+}\big)$, and a non-degenerate symmetric
form
$\langle\,\cdot\,,\,\cdot\rangle_\cP:\Gamma(E_\cP)\otimes\Gamma(E_\cP)\to
C^\infty(M_\cP)$ by $\langle\,\cdot\,,\,\cdot\rangle_\cP:=
\cP_*\big(\langle \cP^*\, \cdot\, \,,\, \cP^*\, \cdot\,
\rangle_{L_+}\big)$. Changing polarization $\cP\to\cP'$ then clearly defines a natural bijection between the quadruples $(E_\cP,[\, \cdot \,,\, \cdot \,]_{E_\cP},\langle
\, \cdot \,,\, \cdot \,\rangle_{E_\cP},\rho_\cP)$ on $M_\cP$ and $(E_{\cP'},[\, \cdot \,,\, \cdot \,]_{E_{\cP'}},\langle
\, \cdot \,,\, \cdot \,\rangle_{E_{\cP'}},\rho_{\cP'})$ on $M_{\cP'}$,
as the structure maps all transform covariantly under the $O(d,d)$ transformations \eqref{eq:polarizationchange}. With these restrictions of the sections and structure maps of the DFT algebroid, it follows from \eqref{eq:maxisotropic} that the expressions \eqref{scdfta}, \eqref{sc1} and \eqref{sc2} vanish, and we have
\begin{prop}
Let $L_+$ be a DFT algebroid on $T^*M$, and let $M_\cP\subset T^*M$ be a $d$-dimensional submanifold defined by a maximally isotropic polarization $\cP$. Then the quadruple $(E_\cP,[\, \cdot \,,\, \cdot \,]_{E_\cP},\langle
\, \cdot \,,\, \cdot \,\rangle_{E_\cP},\rho_\cP)$ defined by $L_+$ and
$\cP$ is a Courant algebroid over $M_\cP$. If $M_\cP\to M_{\cP'}$ is
any $O(d,d)$ transformation of maximally isotropic submanifolds, then
the corresponding Courant algebroids on $E_\cP$ and $E_{\cP'}$ are naturally isomorphic.
\end{prop}

Let us close the present discussion by comparing our framework with the very similar constructions of~\cite{Deser:2016qkw,Heller:2016abk}, which are both rooted in the supermanifold formalism. In that language, the starting point of~\cite{Deser:2016qkw} is identical to ours, i.e. the large Courant algebroid on $E=\mathbb{T}(T^*M)$, as is their projection to $L_+$ which is described as a pre-QP-manifold; their derived bracket conditions ensuring existence of an $L_\infty$-algebra structure are a slight weakening of those corresponding to a pre-Courant algebroid (see~Appendix~\ref{seca}), and they appear to characterise our DFT algebroid and its reduction to a Courant algebroid in terms of graded geometry. On the other hand, in~\cite{Heller:2016abk} the Courant algebroid structure is relaxed from the start to regard the generalized tangent bundle on the doubled space as a pre-QP-manifold itself; their construction of the strong constraint is also a slight weakening of the derived bracket structure of a pre-Courant algebroid, but they do not appear to have a version of our DFT algebroid structure. Our DFT algebroid picture in this sense seems to be somewhat weaker than the structures discussed in~\cite{Deser:2016qkw,Heller:2016abk}.
	
\section{Sigma-models with dynamical fiber metric}
	\label{sec6}
	
In Sections~\ref{sec4} and~\ref{sec5} we saw that the DFT membrane
sigma-model is gauge-invariant provided that the constraint
\eqref{anomaly} is imposed, which is satisfied for instance when the
strong constraint of DFT holds. Motivated by the natural geometric weakenings of
the strong constraint that we encountered in Section~\ref{sec5}, in this section we would like to challenge this result and examine to what extent one can write a gauge-invariant sigma-model of the type \eqref{dftsm} \emph{without} imposing additional constraints. 

The new ingredient we introduce in this section is a dynamical metric
$\eta(\X)$. In other words we promote the  metric $\eta$, which
controls the choice of polarization in \eqref{polarise}, to a
dynamical field and examine the consequences of such an
assumption. This will take us beyond DFT, where $\eta$ is fixed to
\eqref{eta}. Previous discussions of the global geometry of DFT have
also considered such a dynamical metric, as in
e.g.~\cite{Freidel:2015pka,Freidel:2017yuv}. More notably, in \cite{Hansen:2009zd} where sigma-models were used to derive a definition of a Courant algebroid twisted by a closed 4-form, the fiber metric is also dynamical.

The first consequence of introducing an $\X$-dependent metric $\eta$ is that its projection to the DFT structure gives rise to a modified C-bracket. Indeed, recall that our strategy in deriving the DFT ingredients was to rewrite all large Courant algebroid data in terms of $\A_{\pm}$ using the expressions \eqref{plusminusinv}. Now $\widetilde{\A}_I$ is modified by the $\X$-dependence of $\eta$ and thus it will yield 
terms with derivatives acting on $\eta$ whenever a derivative operator acts on it. Taking this into account, we calculate
\be \label{etabra}
\cbral A,B\cbrar_{L_+,\eta}:=\sfp_+\big([\sfp_+(\A),\sfp_+(\B)]_E\big)=\cbral A,B\cbrar_{L_+}+S(A,B)~,
\ee	
where in local coordinate form 
\be 
S(A,B)=S^L{}_{IJ}\,A^I\,B^J\,e^+_L:=\eta^{LK}\,\rho^M{}_{[I}\, \partial_{\underline{M}}\,\eta_{J]K}\,A^I\,B^J\,e^+_L~.
\ee
Thus the twist of the C-bracket is modified to include a
$\partial\eta$-type term. At the level of the membrane sigma-model
\eqref{dftsm}, this correction is not visible because $\eta$ is
symmetric, namely
 $\langle\cbral A,A\cbrar_{L_+,\eta},A\rangle_{L_+}=\langle\cbral A,A\cbrar_{L_+},A\rangle_{L_+}$. 
However, the additional twist has the following effect in the gauge structure of the theory. Considering the transformations 
\bea\label{gt1b}
\delta_\epsilon \X^I&=&\rho^{I}{}_{J}(\X)\,\epsilon^J~,\\[4pt] \label{gt2b}
\delta_\epsilon A^I&=&\dd\epsilon^I+\big(\eta^{IJ}(\X)\,{\hat T}_{JKL}(\X)+S^I{}_{KL}(\X)\big)\,A^K\,\epsilon^L~,
\eea
the variation of the worldvolume derivative $D\X^I$ becomes
\be 
\d_{\epsilon}D\X^I=\epsilon^J\,\partial_K\rho^{I}{}_{J}\,D\X^K+\big(2\,\rho^{K}{}_{[L}\,\partial_K\rho^{I}{}_{M]}-\rho^{I}{}_{J}\,\eta^{JK}\,\hat T_{KLM}-\rho^I{}_J\,S^J{}_{LM}\big)\,A^L\,\epsilon^M~.
\ee
Then $D\X^I$ can be made exactly covariant by requiring the vanishing of the second term, which gives the relation
\be 
\rho^I{}_J\,S^J{}_{LM}=\rho_{N[L}\,\partial^I\rho^N{}_{M]}~,
\ee
or equivalently
\be 
\rho^K{}_{[I}\,\partial_K\eta_{L]J}=\rho_{J}{}^{K}\,\rho^{M}{}_{[I}\,\partial_{\underline{K}}\rho^{N}{}_{L]}\,\eta_{MN}
~.
\ee
The advantage now is that the anomaly term of the gauge variation of the action disappears and at the same time no new terms of the type $A\w\dd A$ are generated. In particular, the gauge variation of the action \eqref{dftsm} gives
\bea
\delta_\epsilon S &=& \int_{\Sigma_3}\,\bigg(\d_{\epsilon}F_K\w D\X^K+\epsilon^J\,\Big(\partial_K\rho^{I}{}_{J}\,F_I-\partial_K\eta_{JL}\,\dd A^L \nn\\ 
&& \qquad \qquad -\, \big(\partial_K \hat T_{JIL}-\partial_K\eta_{IM}\,(\eta^{MN}\,\hat{T}_{NLJ}+S^M{}_{LJ})\big)\,A^I\w A^L\Big)\w D\X^K\nn\\
&& \qquad \qquad \qquad  +\, \epsilon^L\,\big(\eta_{MQ}\,(\eta^{PQ}\,\hat T_{PJK}+S^Q{}_{JK})\,(\eta^{MN}\,\hat
T_{NIL}+S^M{}_{IL}) \nn\\
&& \qquad \qquad \qquad \qquad +\, \rho^{M}{}_{I}\,\partial_M \hat T_{KJL}+\sfrac 13\, \rho^{M}{}_{L}\,\partial_M\hat T_{IJK}\big)\,A^I\w A^J\w A^K\bigg)~.\label{gvS2}
\eea
 Thus, with an appropriate transformation rule for $F_I$, the membrane
 sigma-model action is gauge-invariant provided that the last term
 vanishes. This has the additional consequence that, when a 4-form
 Wess-Zumino term is included as explained in Section \ref{sec42}, the
 strong constraint is no longer a necessary condition for the
 gauge invariance of the extended action~$\hat S$. 

One also needs to check the closure of the algebra of
gauge transformations. Assuming that the gauge parameters do not
change under gauge variation, i.e. they do not depend on $\X$ but only on the worldvolume coordinates, we calculate
\bea
(\delta_\lambda\,\delta_\epsilon
-\delta_\epsilon\,\delta_\lambda)\X^I=2\,\rho^{K}{}_{[L}\, \partial_{\underline
  K}\rho^{I}{}_{J]}\,\lambda^L\,\epsilon^J \ .
\eea
Using the expression for the DFT fluxes \eqref{c1} we have 
\bea
  (\delta_\lambda\,\delta_\epsilon -\delta_\epsilon\,\delta_\lambda)\X^I & =& \big(\rho^{I}{}_{N}\,\eta^{NS}\,\hat T_{SLJ}+\rho_{N[L}\,\partial^I\rho^{N}{}_{J]}\big)\,\lambda^L\,\epsilon^J
\nn\\[4pt] &=& \rho^{I}{}_{N}\, \big(\eta^{NS}\,{\hat
  T}_{SLJ}+S^N{}_{LJ}\big)\, \lambda^L\,\epsilon^J  ~.
\eea
For gauge parameters which are independent of $\X$ we have
\be\label{defxi}
\cbral \lambda,\epsilon\cbrar_{L_+,\eta}=\lambda^L\,\epsilon^J\,\cbral
e^+_L,e^+_J\cbrar_{L_+,\eta} =\lambda^L\,\epsilon^J\,\big(\eta^{NS}\,{\hat
    T}_{SLJ}+S^N{}_{LJ}\big)\, e^+_N~,
\ee
so we can define a new gauge parameter
\be
\xi=\xi^N\,e^+_N:=\cbral \l,\epsilon\cbrar_{L_+,\eta}~,\ee
such that 
\be
(\delta_\lambda\,\delta_\epsilon
-\delta_\epsilon\,\delta_\lambda)\X^I=\rho^{I}{}_{N}\,\xi^N =\delta_{\xi}\X^I~,
\ee
namely the algebra of gauge transformations closes on $\X$.
The gauge variation of $A^I$ gives 
\bea
(\delta_\lambda\,\delta_\epsilon -\delta_\epsilon\,\delta_\lambda)A^I
&=& \dd\xi^I+C^I{}_{KL}\,A^K\,\xi^L
-\partial_NC^I{}_{JK}\,\l^J\,\epsilon^K\, D\X^N \nn\\ && -\,
\big(3\,\rho^{K}{}_{[N}\,\partial_{\underline K}C^I{}_{LM]}-3\,C^I{}_{K[L}\,C^K{}_{MN]}\big)\,A^M\,\l^N\,\epsilon^L~,
\eea
where we used the shorthand notation $C^I{}_{JK}(\X):=\eta^{IL}\,{\hat
  T}_{LJK}(\X)+S^I{}_{JK}(\X)$. The first two terms combine to the
expected result and the third term vanishes on the 
equations of motion for $F_I$. The last term should vanish as a consequence of the Jacobi
identity for the bracket \eqref{etabra},  and indeed for the case of
constant $\eta$ this term vanishes under application of the anchor map and using the
strong constraint \eqref{eq:strongconstraint}.  
So it would seem that one needs the strong constraint for closure of
the algebra of gauge transformations. 

However, let us check what happens with the gauge transformations of $\rho_+(A)=\rho^I{}_J\,A^J\, \partial_I$.  From the gauge variations \eqref{gt1b} and \eqref{gt2b} we obtain 
\bea
\delta_\epsilon \big(\rho^I{}_J\,
A^J\big)=\dd\big(\rho^I{}_J\,\epsilon^J\big)-\partial_K\rho^I{}_J\,
\epsilon^J\,D\X^K \ ,
\eea
again using \eqref{c1}.  Now we check the closure of gauge
transformations on $V^I=\rho^I{}_J\, A^J$, still assuming that
$\delta_\epsilon\lambda=0$, and we find
\bea
 (\delta_\lambda\,\delta_\epsilon
 -\delta_\epsilon\,\delta_\lambda)V^I=\dd\big(\rho^I{}_J\,\xi^J\big){-}\big(2\,
\rho^M{}_{[N}\,\partial_{\underline M}\partial_{\underline K}\rho^I{}_{J]} +2\,\partial_M\rho^I{}_{[J}\,\partial_{\underline K}\rho^M{}_{N]}\big)\,D\X^K\,\lambda^N\,\epsilon^J~.
  \eea
Therefore the algebra of gauge transformations of $V:=\rho_+(A)$
closes on the equations of motion for the auxiliary field $F_I$. In the
correspondence with the flux formulation of DFT discussed in Section~\ref{sec4}, one can regard
$\rho^I{}_J$ as a duality twist matrix, and $V^I$ as the
physical fields obtained after gauge-fixing and reduction. Thus, by
using a dynamical fiber metric $\eta(\X)$, the algebra of gauge
transformation closes on the physical fields $\X^I,V^I$ without use of
the strong constraint \eqref{eq:strongconstraint}. 

\paragraph{Acknowledgments.}
We would like to thank Olaf Lechtenfeld for discussions and
participation in the initial stages of this project, and Olaf Hohm, Branislav Jur\v{c}o, Felix Rudolph,  Peter Schupp, David Svoboda and Satoshi
Watamura for helpful discussions and correspondence. The results of this paper were presented at the Workshop on ``String Dualities and Geometry'' in San Carlos de Bariloche, Argentina, from January~15--19, 2018; A.Ch. and L.J. would like to thank the organizers. We acknowledge support by COST (European Cooperation in Science and Technology) in the framework of the Action MP1405 QSPACE. The
work of A.Ch.,  L.J. and F.S.K. was supported by the Croatian Science
Foundation under the Project IP-2014-09-3258 and by the H2020 Twinning
Project No. 692194 ``RBI-T-WINNING''.  The work of R.J.S. was
supported by the Consolidated Grant ST/P000363/1 
from the UK Science and Technology Facilities Council.

\appendix 

\section{From Courant algebroids to DFT algebroids}
\label{seca}

In this appendix we provide a brief account of Courant algebroids and
some of their natural generalizations. We begin with the two equivalent
definitions of Courant algebroid given in~\cite{liu} and \cite{dee1},
stating the axioms and properties of the geometric structure. 
We further provide the local coordinate expressions of these axioms,
and discuss them in the spirit of the main text of this paper. Then we
present the notions of a pre-Courant algebroid \cite{preca} and of a
4-form twisted Courant algebroid \cite{Hansen:2009zd},  
whose equivalence is discussed in\cite{prehequiv}. Finally, we introduce the notions of an ante-Courant algebroid and a pre-DFT algebroid as natural generalizations of the pre-Courant algebroid structure, and further discuss their relation to the metric algebroid of~\cite{Vaisman:2012ke} and their description in terms of graded geometry. We provide examples for all structures in Appendix~\ref{seca4}.

\subsection{Courant algebroids}

The notion of a Courant algebroid, essentially introduced in \cite{courant}, was systematically defined in~\cite{liu}.
\begin{defn}\label{ca1def}
	Let $M$ be a $d$-dimensional manifold. A \underline{Courant
          algebroid} on $M$ is a quadruple
        $(E,[\, \cdot \,,\, \cdot \,],\langle\, \cdot \,,\, \cdot \,\rangle,\rho)$ consisting of a vector bundle $E \rightarrow M$, a skew-symmetric bracket on its sections, a non-degenerate symmetric bilinear form on $E$, and a smooth bundle map $\rho: E\to TM$, satisfying
	\begin{enumerate}
		\item \ $[[A,B],C]+\text{cyclic}= {\cal D}{\cal N}(A,B,C)~;$
		\item \ $\rho [A,B]=[\rho(A),\rho(B)]~;$
		\item \ $[A,f\,B]=f\,[A,B]+\big(\rho(A)f\big)\,B-\langle A,B\rangle\, {\cal D}f$~;
		\item \ $\rho \circ {\cal D}=0 \quad \Longleftrightarrow \quad \langle {\cal D}f, {\cal D}g \rangle=0~;$
		\item \ $\rho(C)\langle A,B\rangle = \langle [C,A]+{\cal D}\langle C,A\rangle,B\rangle + \langle A,[C,B]+{\cal D}\langle C,B\rangle\rangle ~;$
	\end{enumerate}
	where 
	\be 
	{\cal N}(A,B,C)=\sfrac 13\, \langle [A,B],C\rangle +\text{cyclic}~,
	\ee
	and the differential operator ${\cal D}: C^{\infty}(M)\to \G(E)$ is defined by
	\be 
	\langle {\cal D}f, A\rangle=\sfrac 12\,\rho(A)f~,
	\ee
for any 
	$A,B,C\in\G(E)$ and $f\in C^{\infty}(M)$.
\end{defn}

In applications to generalized geometry and DFT one is interested in \emph{exact Courant algebroids},
whose underlying vector bundles fit into the short exact sequence
\be
0\longrightarrow T^*M \xrightarrow{ \ \rho^* \ } E \xrightarrow{ \
  \rho \ } TM \longrightarrow 0 \ ,
\ee
where $\rho^*:T^*M\to E$ denotes the transpose map of $\rho$. If there
is $H$-flux on $M$, then a choice of $B$-field defines a Lagrangian
splitting $\lambda:TM\to E$, and locally $E$ is the Whitney sum of the
tangent and cotangent bundles of $M$ (see e.g.~\cite{Saemann:2012ab}). This defines the
\emph{generalized tangent bundle}. As we are interested in local
considerations in the present paper (equivalently $M$ is
contractible), we assume $E=TM\oplus T^*M$ throughout.

Properties 1--5 in Definition~\ref{ca1def} are not meant to be a
minimal set of axioms defining the structure, since some of them imply
the others~\cite{Uchino2002}. Minimally one would only have to assume
properties 1 and 5, together with any one of properties 2, 3 or 4.
Let us discuss the meaning of these properties and also write
them in a local coordinate form. For this, we introduce a local basis
$e^I$, $I=1,\dots,2d$, of sections of $E$, which we expand as
$A=A_I\,e^I$. The map $\rho$ is called the \emph{anchor} and it has
components $(\rho^{i}{}_J)=(\rho^i{}_j,\rho^{ij})$, where $i=1,\dots,d$. In this basis we write the local coordinate form of the relevant operations as
\bea 
[e^I,e^J]&=&\eta^{IK}\,\eta^{JL}\,T_{KLM}\,e^M~,\label{lc1}\\[4pt]
\langle e^I,e^J\rangle&=&\sfrac 12\,\eta^{IJ}~,\label{lc2}\\[4pt]
\rho(e^I)f&=&\eta^{IJ}\,\rho^i{}_{J}\,\partial_if~,\label{lc3} \\[4pt]
{\cal D}f&=&{\cal D}_If\,e^I=\rho^i{}_I\,\partial_if\, e^I~,\label{lc4}
\eea 
where the bundle metric $\eta$ on $E$ has split signature $(d,d)$, and ${\cal D}$ is the pullback of the
exterior derivative $\dd$ by the transpose map $\rho^*$.

Property 1 is the modified Jacobi identity; it states that the bracket
of the Courant algebroid is not a Lie bracket due to a ${\cal
  D}$-exact form obstruction characterized in terms of the
\emph{Nijenhuis operator} $\mc N$. Property 3 is simply the Leibniz rule for the bracket on $E$. In local coordinates, after a computation using the expressions \eqref{lc1}--\eqref{lc4} and the Leibniz rule, property 1 is equivalent to the three equations
\bea 
\label{ca1} \eta^{IJ}\,\rho^i{}_{I}\,\rho^j{}_J&=&0~,\\[4pt]
\label{ca2} \rho^i{}_{I}\,\partial_i\rho^j{}_J-\rho^i{}_J\,\partial_i\rho^j{}_I-\eta^{KL}\,\rho^j{}_K\,T_{LIJ}&=&0~,\\[4pt]
\label{ca3} 4\,\rho^i{}_{[L}\,\partial_iT_{IJK]}+3\,\eta^{MN}\,
T_{M[IJ}\,T_{KL]N}&=&0~.
\eea
Property 2 states that the map $\rho$ is a homomorphism of bundles,
i.e. it is compatible with the bracket on $\G(E)$ and the usual Lie
bracket of vector fields on $\G(TM)$; its local expression is
identical to \eqref{ca2}, thus it follows from properties 1 and
3. Property 5 is a compatibility condition and it is satisfied
identically when the local expressions are used. Finally, property 4, $\langle {\cal D}f, {\cal D}g \rangle=0$, is written in local coordinates as 
\be 
\eta^{IJ}\,\rho^i{}_{I}\,\rho^j{}_J\,\partial_if\,\partial_jg=0~.
\ee
Thus we observe that it is identically satisfied due to \eqref{ca1},
and it also follows from the previous properties. It is interesting to note that this property involves the product of two derivatives acting on functions on $M$. As such it is reminiscent of the strong constraint of DFT. Indeed, as we show in the main text, it is precisely the violation of \eqref{ca1} that leads to the strong constraint. However, at the level of the Courant algebroid there is clearly no such additional assumption. 

The local coordinate expression for the skew-symmetric bracket, called
the \emph{Courant bracket}, may be
obtained by using the Leibniz rule and the expressions \eqref{lc1}--\eqref{lc4}. A direct calculation leads to 
\bea
[A,B]&=&\big(\rho^l{}_J\,(A^J\,\partial_lB_K-B^J\,\partial_lA_K)-\sfrac
12\, \rho^l{}_K\,(A^J\,\partial_lB_J-B^J\,\partial_lA_J)\big)\,e^K
\nn\\ && +\,A^L\,B^M\,T_{LMK}\,e^K~, \label{courantgeneral}
\eea
where indices are raised with the inverse metric $\eta^{-1}$. For the
special case of the \emph{standard Courant algebroid}, where the
anchor $\rho:E\to TM$ is the projection to the tangent bundle, the
metric is induced by the natural pairing between $TM$ and $T^*M$, and
the map ${\cal D}:C^\infty(M)\to \G(E)$ is given by ${\cal
  D}f=\dd f$, one has $\rho^{i}{}_{J}=(\d^i{}_j,0)$ and
writing $e^I=(\partial_i,\dd x^i)$ the
formula \eqref{courantgeneral} reads
\bea
[A,B]_{\rm s}&=&
\big(A^l\,\partial_lB^k-B^l\,\partial_lA^k\big)\,\partial_k+\big(A^l\,\partial_lB_k-B^l\,\partial_lA_k-\sfrac
12\, A^l\,\partial_kB_l+\sfrac 12\,B^l\,\partial_kA_l \nn\\
&&-\sfrac
12\, A_l\,\partial_kB^l+\sfrac 12\,B_l\,\partial_kA^l 
+A^l\,B^m\,H_{lmk}\big)\,\dd x^k~,
\eea
which is the local coordinate expression for the standard $H$-twisted Courant bracket  
\bea 
[A,B]_{\rm s}=[A_V,B_V]+{\cal L}_{A_V}B_F-{\cal L}_{B_V}A_F-\sfrac 12\, \dd (\iota_{A_V}B_F-\iota_{B_V}A_F)+H(A_V,B_V)~,
\eea  
where $A=A_V+A_F\in \G(E)$ with $A_V\in\G(TM)$ and
$A_F\in\G(T^{\ast}M)$. However, the expression \eqref{courantgeneral}
is evidently more general and may in fact be written in intrinsic geometric terms as 
\bea 
[A,B]&=&[A_V,B_V]+{\cal L}_{A_F}B_V-{\cal L}_{B_F}A_V+\sfrac 12\, \dd_{\ast} (\iota_{A_V}B_F-\iota_{B_V}A_F)
\\
&+&[A_F,B_F]+{\cal L}_{A_V}B_F-{\cal L}_{B_V}A_F-\sfrac 12\, \dd
(\iota_{A_V}B_F-\iota_{B_V}A_F)+T(A,B)~, \nn
\eea
as in \cite{liu}. (Here $\dd$ and $\dd_{\ast}$ are exterior
differential operators on the tangent and cotangent bundles of $M$,
respectively, see e.g. \cite{Chatzistavrakidis:2015vka} for details.) 
Whenever one deals with a Courant algebroid other than the standard
one, this more general bracket should be used (see e.g.~\cite{Deser:2014mxa,Chatzistavrakidis:2015vka}.) 

An alternative definition of a Courant algebroid, appearing in
\cite{dee3} (see also \cite{Severa:2017oew}), uses instead a binary
operation which is often called the \emph{Dorfman bracket}, although
it is not skew-symmetric. It is defined by
\be
A\circ B:= [A_V,B_V]+{\cal L}_{A_V}B_F-\iota_{B_V}\dd A_F \ ,
\ee
and it is related to the Courant bracket by skew-symmetrization
\be 
[A,B]=A\circ B-B\circ A~.
\ee 
\begin{defn}\label{ca2def}
Let $M$ be a $d$-dimensional manifold. A \underline{Courant algebroid}
on $M$ is a quadruple
$(E,\, \cdot \,\circ\, \cdot \,,\langle\, \cdot \,,\, \cdot \,\rangle,\rho)$ consisting of a vector bundle $E \rightarrow M$, a binary operation on its sections, a non-degenerate symmetric bilinear form on $E$, and a smooth bundle map $\rho: E\to TM$, satisfying:
	\begin{enumerate}
		\item \ $A\circ (B\circ C)=(A\circ B)\circ C+B\circ(A\circ C)~;$
		\item \ $\rho (A\circ B)=[\rho(A),\rho(B)]~;$
		\item \ $A\circ (f\,B)=f\,(A\circ B)+\big(\rho(A)f\big)\,B$~;
		\item \ $A\circ A={\cal D}\langle A,A\rangle~;$
		\item \ $\rho(C)\langle A,B\rangle = \langle C\circ A,B\rangle + \langle A,C\circ B\rangle ~;$
	\end{enumerate}
for any 
	$A,B,C\in\G(E)$ and $f\in C^{\infty}(M)$.
	\end{defn}
Definitions~\ref{ca1def} and~\ref{ca2def} are completely equivalent, as
proven in \cite{dee3}, with the binary operation given by 
\be \label{dorfmandef}
A\circ B = [A,B]+{\cal D}\langle A,B\rangle~.
\ee The convenience of the latter definition is that (a) unlike the
Courant bracket, the Dorfman bracket satisfies a Jacobi-like identity, and (b) in the Leibniz rule and the compatibility condition the additional ``anomaly'' terms of Definition~\ref{ca1def} are now absent.

Finally, let us briefly discuss the relation to differential graded (dg-)manifolds. 
First recall that a \emph{QP$n$-manifold} is a triple $(\mc
M,\omega,Q)$ consisting 
of an $n$-graded manifold $\mc M$, a degree~$n$ symplectic structure
$\omega$, and a degree~1 vector field $Q$ which is nilpotent, $Q^2=0$, called a 
\emph{homological vector field}, satisfying the compatibility condition
\be\label{eq:LQomega}
{\mc L}_Q\,\omega=0~.
\ee
Because of \eqref{eq:LQomega}, the homological vector field $Q$ gives rise to a degree~$n+1$
\emph{Hamiltonian function} $\Theta\in C^\infty(\mc M)$ as 
\be
Q=\{\Theta,\, \cdot \,\}~,
\ee
where the bracket is the graded Poisson bracket defined from $\omega$. The nilpotency of $Q$ implies the \emph{classical master equation}
\be \label{cme}
\{\Theta,\Theta\}=0~,
\ee
which in the AKSZ construction essentially guarantees the gauge invariance of the corresponding BV action and the closure of the 
gauge algebra. QP$n$-manifolds are sometimes also refered to as
\emph{symplectic Lie $n$-algebroids}~\cite{Severa2001}, which
arise from $n$-graded vector bundles $\mc M$ over their degree~0 body $M:=\mc M_0$.

For our purposes, we are interested in the case $n=2$. We can
introduce local Darboux coordinates $(x^i,A^I,F_i)$ on $\mc M$ of
degree~0, 1 and 2, respectively, such that
\be
\omega = \dd x^i\w\dd F_i+\sfrac 12\,\eta_{IJ}\, \dd A^I\w\dd A^J \ .
\ee
Then the most general Hamiltonian function $\Theta$ is
given in these coordinates
by~\cite{dee2}
\be \label{Th}
\Theta=\rho^i{}_I(x)\, F_i\,A^I-\sfrac 1{3!}\,T_{IJK}(x)\,A^I\,A^J\,A^K~,
\ee 
and the classical master equation \eqref{cme} gives precisely the three conditions
\eqref{ca1}--\eqref{ca3}, see e.g.~\cite{Ikeda:2012pv}.
In other words, QP2-manifolds, or symplectic Lie 2-algebroids, are in
a one-to-one correspondence with Courant algebroids, which is the celebrated 
Roytenberg theorem~\cite{dee1}; in this correspondence, functions of
degree~1 on $\mc M$ are identified with sections $\G(E)$ of a vector
bundle $E\to M$ whose structure maps are given by the derived bracket construction. In particular, exact Courant
algebroids on a manifold $M$ can be recovered from QP2-manifolds with underlying 2-graded
manifold $\mc M=T^*[2]T[1]M$.\footnote{The notation $[n]$ indicates grade-shift of the
fiber degree by $n$.}

\subsection{Pre-Courant algebroids}

The structure of a Courant algebroid may be generalized in the direction of relaxing the Jacobi identity in its definition. This was considered in \cite{preca}.
\begin{defn}\label{preca1def}
With the same conventions as above, a \underline{pre-Courant
  algebroid} on $M$ is a quadruple $(E,[\, \cdot \,,\, \cdot
\,],\langle\, \cdot \,,\, \cdot \,\rangle,\rho)$ which satisfies only properties 
2--5 of Definition~\ref{ca1def}.
\end{defn}
The crucial difference here is that property 1 is no longer necessarily satisfied. Furthermore, one can define the corresponding generalization of Definition~\ref{ca2def}, by relaxing its property~1. Schematically:
\be \label{scheme1}
\small
\text{Pre-Courant algebroid} \ \xleftarrow{ \
    \cancel{\scriptsize\circled{1}} \ } \ \text{Courant algebroid}
\normalsize
\ee

In a similar fashion, the violation of the Jacobi identity may be
expressed in terms of a 4-form, defined as in~\cite{Hansen:2009zd}.
\begin{defn}\label{hcadef}
With the same conventions as above, let ${\cal T}$ be a closed 4-form on
$M$. A \underline{${\cal T}$-twisted Courant algebroid} on $M$ is a
quadruple $(E,\, \cdot \,\circ\, \cdot \,,\langle\, \cdot \,,\, \cdot \,\rangle,\rho)$
satisfying properties 2--4 of Definition~\ref{ca2def} together with
\be 
A\circ (B\circ C)=(A\circ B)\circ C+B\circ(A\circ C)+\rho^{\ast}\,{\cal T}\big(\rho(A),\rho(B),\rho(C)\big)~.
\ee
\end{defn}
This definition shows that the violation of the Jacobi identity
is controlled by a 4-form{\footnote{In \cite{Hansen:2009zd} this
    4-form is denoted by $H$. Here we use a different notation in
    order to avoid confusion with the NS--NS 3-form flux $H$.}} ${\cal T}$.
As discussed in \cite{prehequiv}, the two definitions are essentially equivalent. 

Furthermore, following~\cite{dee1}, in~\cite{preca2} the structure
corresponding to a pre-Courant algebroid in the supermanifold
framework is defined as a \emph{symplectic almost Lie 2-algebroid}. In
this case, the classical master equation \eqref{cme} is no longer
satisfied, but is weakened to 
\be
\{\{\Theta,\Theta\},f\}=0~,
\ee
for any function $f\in C^\infty(M)$, where in local Darboux coordinates
the Hamiltonian function $\Theta\in C^\infty({\cal M})$ is given as in \eqref{Th}. 

\subsection{Ante-Courant algebroids and pre-DFT algebroids}
\label{seca3}

Further generalization of the Courant algebroid structure can be
achieved by relaxing the homomorphism property of the anchor map
$\rho$ and property 4 of Definition~\ref{ca1def}. As we have seen in Section \ref{sec5}, this is the appropriate setting for DFT before the strong constraint is imposed. However, in general properties 2 and 4 may be relaxed independently. This becomes clear with the following definitions and the examples discussed in Appendix \ref{seca4}. 
 \begin{defn}\label{prepreca1def}
Let $(E,[\, \cdot \,,\, \cdot \,],\langle\, \cdot \,,\, \cdot
\,\rangle,\rho)$ be a quadruple with the same conventions as
above. With reference to Definition~\ref{ca1def}, we
call it an \underline{ante-Courant algebroid} on $M$
if it only satisfies properties 3, 4 and 5, and a \underline{pre-DFT algebroid} on $M$ 
if it only satisfies properties 3 and 5.
\end{defn}
Schematically, this enhances the picture \eqref{scheme1} to 
\be \label{scheme2}
\small
\begin{matrix}\text{Pre-DFT} \\ \text{algebroid}\end{matrix} \ \xleftarrow{ \ \cancel{\scriptsize\circled{4}} \ } \ 
\begin{matrix}\text{Ante-Courant} \\ \text{algebroid}\end{matrix} \ \xleftarrow{ \ \cancel{\scriptsize\circled{2}} \
} \ \begin{matrix}\text{Pre-Courant} \\ \text{algebroid}\end{matrix} \ \xleftarrow{ \ \cancel{\scriptsize\circled{1}}
  \ } \ 
\begin{matrix}\text{Courant} \\ \text{algebroid}\end{matrix}
\normalsize
\ee
What we have shown in the main text is that a DFT algebroid is a
special case of a pre-DFT algebroid, such that the properties 1, 2 and 4 are violated in a \emph{dependent} way. In other words, imposing property 4, namely the strong constraint, on the DFT algebroid leads directly to a Courant algebroid without stopping at the intermediate structures. Schematically:
\be 
\small
\text{Large Courant algebroid} \ \xrightarrow{ \ \mathsf{p}_+ \ } \ 
\text{DFT algebroid} \ \xrightarrow{ \ {\scriptsize\circled{4}} \ } \ \text{Courant algebroid}
\normalsize
\ee

\begin{rmk}
	In~\cite{Vaisman:2012ke}, a \underline{metric algebroid} is defined as a quadruple
	$(E,\, \cdot \,\circ\, \cdot \,,\langle\, \cdot \,,\, \cdot \,\rangle,\rho)$ satisfying properties 3, 4 and 5 of Definition \ref{ca2def}. Although it looks like this structure corresponds to an ante-Courant algebroid, 
	the situation is more subtle. When $\rho$ is not a homomorphism, properties 4 of Definitions 
	\ref{ca1def} and \ref{ca2def} do not directly follow from each other. Therefore, when an antisymmetric 
	bracket is introduced in \cite{Vaisman:2012ke}, a metric algebroid does not necessarily satisfy property 4 
	of Definition \ref{ca1def}. Thus an ante-Courant algebroid is always a metric algebroid but not conversely. On the other 
	hand,  assuming \eqref{dorfmandef} we conclude that a metric algebroid is equivalent with a pre-DFT algebroid. 
	\end{rmk}

In the supermanifold description, the structure corresponding to a pre-DFT algebroid was
identified as a \emph{symplectic nearly Lie 2-algebroid}
in~\cite{preca2}, which consists of a 2-graded superbundle $\mc M$ over a
manifold $M$, a non-degenerate Poisson bracket of degree $-2$, and a Grassmann
odd function $\Theta\in C^\infty(\mc M)$ of degree 3. Using these
data and the derived bracket construction, one can show that the derived (Dorfman) bracket
satisfies the Leibniz
rule and the compatibility property (properties 3 and 5 of
Definition~\ref{ca2def}).  Therefore the skew-symmetrization of the
Dorfman bracket, which is the C-bracket of DFT in our case, satisfies properties 3 and 5 of Definition~\ref{ca1def}.

Moreover, the failure of properties 1 and 2 in the definition of a
Courant algebroid is given in \cite{preca2} in terms of third order higher
derived brackets generated by $\{\Theta,\Theta\}$ as
$\{\{\{\{\Theta,\Theta\},A\},B\},C\}$ and
$\{\{\{\{\Theta,\Theta\},f\},A\},B\}$ respectively, for $A,B,C\in
\Gamma(E)$ and $f\in C^\infty(M)$. Explicit calculation, using the component expressions and taking into account the appropriate skew-symmetrization, shows that these obstructions are  exactly the ones given  in \eqref{preJ} and \eqref{prehomo}.

\subsection{Examples}
\label{seca4}

In order to compare with the results obtained in the main text regarding DFT, it is instructive to 
 examine some 
characteristic cases of Courant algebroids, and their generalizations above, with twists.

\subsubsection*{The standard Courant algebroid} 

The standard Courant algebroid is the simplest case corresponding to
the choice of anchor $\rho=(\text{id},0)$, the projection to the
tangent bundle, which in components reads
\be 
\rho^{i}{}_{J}=(\d^i{}_{j},0)~.
\ee
The condition \eqref{ca1} is identically satisfied without further
restrictions. The condition \eqref{ca2} implies, after opening the Courant algebroid indices, 
that $T_{jk}{}^i=T_k{}^{ij}=T^{ijk}=0$, or, in standard notation in the context of string backgrounds with fluxes, $f=Q=R=0$. This means that only $H$-flux is permitted for this anchor, leading 
to the $H$-twisted standard Courant algebroid. 
Indeed, for a 3-form NS--NS flux $H$ satisfying the Bianchi identity $\dd H=0$, the
condition \eqref{ca3} is also automatically satisfied. 
Alternatively, one may think of \eqref{ca3} as imposing the Bianchi identity.

\subsubsection*{Non-standard Courant algebroids and their generalizations}

Let us now go beyond the choice of projection for the anchor. One possibility is to
choose $\rho=(0,\beta^{\sharp})$ for some $(0,2)$-tensor
$\beta\in\G(TM\otimes TM)$ with
corresponding bundle map $\beta^\sharp:T^*M\to TM$ induced by the
canonical dual pairing between the tangent and cotangent bundles. In components this reads
\be 
\rho^{i}{}_{J}=(0,\b^{ij})~.
\ee
Then once more the condition \eqref{ca1} is identically satisfied. The
condition \eqref{ca2} now implies that 
\bea 
\beta^{kl}\,H_{lij}=0 \ , \qquad \beta^{kl}\,f_{lj}{}^i=0 \qquad
\mbox{and} \qquad \beta^{li}\,\partial_l\beta^{jk}-\beta^{lk}\,\partial_l\beta^{ji}+\beta^{jl}\,Q_l{}^{ik}=0~.
\eea
In principle this allows for all fluxes to be non-vanishing; notably, the $R$-flux does not even appear in these conditions and thus it is 
not constrained by condition \eqref{ca2}. 

For example, if $\beta=\Pi$ is a non-degenerate Poisson bivector, in which case the Schouten bracket with itself vanishes, $[\Pi,\Pi]_{\text{S}}=0$, then a Courant algebroid is obtained 
as $H=f=0$ and $Q_i{}^{jk}=\partial_i\Pi^{jk}$. Furthermore, the condition \eqref{ca3} leads to the additional requirement 
$[\Pi,R]_{\text{S}}=0$,
or in local coordinates
\be 
\Pi^{m[l}\,\partial_m R^{ijk]}+\sfrac 32\, R^{m[ij}\,\partial_m\Pi^{kl]}=0~,
\ee
which is the Bianchi identity in this instance.
 This case was studied for example in \cite{Bessho:2015tkk}. It plays
 a role in our discussion in Section~\ref{sec34}. In addition, as
 noticed in \cite{preca2}, when no condition is assumed between the
 Poisson structure $\Pi$ and the trivector $R$, one obtains a simple
 example of pre-Courant algebroid. Finally, as also discussed in
 \cite{preca2}, if one discards the assumption that $\beta$ is a
 Poisson bivector, namely $[\beta,\beta]_{\text{S}}\ne 0$, then the
 pre-Courant algebroid structure is further relaxed, this being an
 example of a symplectic nearly Lie 2-algebroid. In our language, this
 example constitutes an ante-Courant algebroid, since although the
 Jacobi identity and the homomorphism property for $\rho$ are
 obstructed, this choice of anchor satisfies the property $\rho\circ
 {\cal D}=0$. Moreover, one may directly check that the condition
 \eqref{c1} for the DFT fluxes is not satisfied. As we showed in the
 main text, the DFT equations are compatible only with a pre-DFT
 algebroid structure.  

A combination of the above choices leads to an even larger class of
examples for Courant algebroids. Specifically, consider $ \rho=(\text{id},\beta^{\sharp}) $,
which in local coordinates reads
\be 
\rho^{i}{}_{J}=(\d^i{}_j,\b^{ij})~.
\ee
Then \eqref{ca1} implies that 
\be 
\beta^{(ij)}=0~,
\ee
thus $\beta$ has to be a bivector, though not necessarily Poisson. Additionally, \eqref{ca2} leads to the conditions
\bea 
\label{mix1} f_{ij}{}^k+\beta^{kl}\,H_{lij}&=&0~,\\[4pt]
\label{mix2} \partial_k\beta^{ij}-Q_k{}^{ji}+\beta^{jl}\,f_{lk}{}^i&=&0~,\\[4pt]
\label{mix3} \beta^{li}\,\partial_l\beta^{jk}-\beta^{lk}\,\partial_l\beta^{ji}-R^{jik}-\beta^{jl}\,Q_l^{ik}&=&0~.
\eea
We emphasize once more that these are conditions on the fluxes which
ensure that the structure consistently defines a Courant algebroid. They
yield the potential expressions for fluxes in generalized geometry, as discussed in the main text, along with the Bianchi identities that are obtained from \eqref{ca3}.   

Let us make two final noteworthy observations. First, suppose we would like to have a pure $R$-flux. Thus we set $H=f=Q=0$, which leads to $\partial_i\beta^{jk}=0$ and thus $R=0$. We conclude that even for such general anchors, there is no pure $R$-flux Courant algebroid. 
Second, the most general expressions for fluxes are obtained 
using a coordinate-dependent anchor $\rho^i{}_j=e^i{}_{j}(x)$ instead
of just the projection to the tangent bundle. Then one may associate
the resulting structure to the fluxes in a non-holonomic frame.


\begin{thebibliography}{99}
	
	\bibitem{stncg1}
	M.~R.~Douglas and C.~M.~Hull,
	``D-branes and the noncommutative torus,''
	JHEP {\bf 9802} (1998) 008
	[arXiv:hep-th/9711165].
	
	\bibitem{stncg2}
	C.-S.~Chu and P.-M.~Ho,
	``Noncommutative open string and D-brane,''
	Nucl.\ Phys.\ B {\bf 550} (1999) 151--168
	[arXiv:hep-th/9812219].
	
	\bibitem{stncg3}
	N.~Seiberg and E.~Witten,
	``String theory and noncommutative geometry,''
	JHEP {\bf 9909} (1999) 032
	[arXiv:hep-th/9908142].

\bibitem{Douglas:2001ba}
  M.~R.~Douglas and N.~A.~Nekrasov,
  ``Noncommutative field theory,''
  Rev.\ Mod.\ Phys.\  {\bf 73} (2001) 977--1029
  [arXiv:hep-th/0106048].

\bibitem{Szabo:2001kg}
  R.~J.~Szabo,
  ``Quantum field theory on noncommutative spaces,''
  Phys.\ Rept.\  {\bf 378} (2003) 207--299
  [arXiv:hep-th/0109162].
  	
	\bibitem{stnag1}
	R.~Blumenhagen and E.~Plauschinn,
	``Nonassociative gravity in string theory?,''
	J.\ Phys.\ A {\bf 44} (2011) 015401
	[arXiv:1010.1263 [hep-th]].
	
	\bibitem{stnag2}
	D.~L\"ust,
	``T-duality and closed string noncommutative (doubled) geometry,''
	JHEP {\bf 1012} (2010) 084
	[arXiv:1010.1361 [hep-th]].
	
	\bibitem{stnag3}
	R.~Blumenhagen, A.~Deser, D.~L\"ust, E.~Plauschinn and F.~Rennecke,
	``Non-geometric fluxes, asymmetric strings and nonassociative geometry,''
	J.\ Phys.\ A {\bf 44} (2011) 385401
	[arXiv:1106.0316 [hep-th]].
	
	\bibitem{Mylonas:2012pg}
	D.~Mylonas, P.~Schupp and R.~J.~Szabo,
	``Membrane sigma-models and quantization of non-geometric flux backgrounds,''
	JHEP {\bf 1209} (2012) 012
	[arXiv:1207.0926 [hep-th]].

\bibitem{Plauschinn:2012kd}
  E.~Plauschinn,
  ``Non-geometric fluxes and nonassociative geometry,''
  PoS CORFU {\bf 2011} (2011) 061
  [arXiv:1203.6203 [hep-th]].
  
	\bibitem{stnag4}
	D.~L\"ust,
	``Twisted Poisson structures and noncommutative/nonassociative closed string geometry,''
	PoS CORFU {\bf 2011} (2011) 086
	[arXiv:1205.0100 [hep-th]].

\bibitem{Mylonas:2014aga}
  D.~Mylonas, P.~Schupp and R.~J.~Szabo,
  ``Nonassociative geometry and twist deformations in non-geometric string theory,''
  PoS ICMP {\bf 2013} (2013) 007
  [arXiv:1402.7306 [hep-th]].

\bibitem{Blumenhagen:2014sba}
  R.~Blumenhagen,
  ``A course on noncommutative geometry in string theory,''
  Fortsch.\ Phys.\  {\bf 62} (2014) 709--726
  [arXiv:1403.4805 [hep-th]].
  
\bibitem{Barnes:2016cjm}
  G.~E.~Barnes, A.~Schenkel and R.~J.~Szabo,
  ``Working with nonassociative geometry and field theory,''
  PoS CORFU {\bf 2015} (2016) 081
  [arXiv:1601.07353 [hep-th]].

	\bibitem{Hull:2009sg}
	C.~M.~Hull and R.~A.~Reid-Edwards,
	``Non-geometric backgrounds, doubled geometry and generalised T-duality,''
	JHEP {\bf 0909} (2009) 014
	[arXiv:0902.4032 [hep-th]].
	
	\bibitem{doubled1}
	M.~J.~Duff,
	``Duality rotations in string theory,''
	Nucl.\ Phys.\ B {\bf 335} (1990) 610--620.
	
	\bibitem{doubled2}
	A.~A.~Tseytlin,
	``Duality symmetric formulation of string worldsheet dynamics,''
	Phys.\ Lett.\ B {\bf 242} (1990) 163--174.
	
	\bibitem{doubled3}
	W.~Siegel,
	``Two vierbein formalism for string inspired axionic gravity,''
	Phys.\ Rev.\ D {\bf 47} (1993) 5453--5459
	[arXiv:hep-th/9302036].
	
	\bibitem{Siegel:1993th}
	W.~Siegel,
	``Superspace duality in low-energy superstrings,''
	Phys.\ Rev.\ D {\bf 48} (1993) 2826--2837
	[arXiv:hep-th/9305073].
	
	\bibitem{doubled4}
	C.~M.~Hull,
	``A geometry for non-geometric string backgrounds,''
	JHEP {\bf 0510} (2005) 065
	[arXiv:hep-th/0406102].
	
	\bibitem{doubled5}
	C.~M.~Hull,
	``Doubled geometry and T-folds,''
	JHEP {\bf 0707} (2007) 080
	[arXiv:hep-th/0605149].
	
	\bibitem{dft1}
	C.~M.~Hull and B.~Zwiebach,
	``Double field theory,''
	JHEP {\bf 0909} (2009) 099
	[arXiv:0904.4664 [hep-th]].
	
	\bibitem{dft2}
	C.~M.~Hull and B.~Zwiebach,
	``The gauge algebra of double field theory and Courant brackets,''
	JHEP {\bf 0909} (2009) 090
	[arXiv:0908.1792 [hep-th]].
	
	\bibitem{dft3}
	O.~Hohm, C.~M.~Hull and B.~Zwiebach,
	``Background independent action for double field theory,''
	JHEP {\bf 1007} (2010) 016
	[arXiv:1003.5027 [hep-th]].
	
	\bibitem{dft4}
	O.~Hohm, C.~M.~Hull and B.~Zwiebach,
	``Generalized metric formulation of double field theory,''
	JHEP {\bf 1008} (2010) 008
	[arXiv:1006.4823 [hep-th]].
	
	\bibitem{Freidel:2015pka}
	L.~Freidel, R.~G.~Leigh and D.~Minic,
	``Metastring theory and modular spacetime,''
	JHEP {\bf 1506} (2015) 006
	[arXiv:1502.08005 [hep-th]].
	
	\bibitem{dftrev1}
	G.~Aldazabal, D.~Marqu\'es and C.~N\'u\~nez,
	``Double field theory: A pedagogical review,''
	Class.\ Quant.\ Grav.\  {\bf 30} (2013) 163001
	[arXiv:1305.1907 [hep-th]].
	
	\bibitem{dftrev2}
	D.~S.~Berman and D.~C.~Thompson,
	``Duality symmetric string and M-theory,''
	Phys.\ Rept.\  {\bf 566} (2014) 1--60
	[arXiv:1306.2643 [hep-th]].
	
	\bibitem{dftrev3}
	O.~Hohm, D.~L\"ust and B.~Zwiebach,
	``The spacetime of double field theory: Review, remarks, and outlook,''
	Fortsch.\ Phys.\  {\bf 61} (2013) 926--966
	[arXiv:1309.2977 [hep-th]].
	
		\bibitem{courant} T.~J.~Courant, 
	``Dirac manifolds,''
	Trans. Amer. Math. Soc. {\bf 319} (1990) 631--661.
	
	\bibitem{liu}
	Z.-J.~Liu, A.~Weinstein and P.~Xu,
	``Manin triples for Lie bialgebroids,''
	J.\ Diff.\ Geom.\  {\bf 45} (1997) 547--574
	[arXiv:dg-ga/9508013].
	
		\bibitem{dee3}
	D.~Roytenberg, 
	``Courant algebroids, derived brackets and even symplectic
	supermanifolds,''
	PhD Thesis, University of California at Berkeley
	[arXiv:math.DG/9910078].
	
	\bibitem{Severa:2017oew}
	P.~\v{S}evera,
	``Letters to Alan Weinstein about Courant algebroids,''
	arXiv:1707.00265 [math.DG].
	
	\bibitem{Hitchin:2004ut}
	N.~Hitchin,
	``Generalized Calabi-Yau manifolds,''
	Quart.\ J.\ Math.\  {\bf 54} (2003) 281--308
	[arXiv:math-dg/0209099].
	
	\bibitem{Gualtieri:2003dx}
	M.~Gualtieri,
	``Generalized complex geometry,''
	PhD Thesis, University of Oxford
	[arXiv:math-dg/0401221].
	
\bibitem{cg}
  G.~R.~Cavalcanti and M.~Gualtieri,
  ``Generalized complex geometry and T-duality,''
 CRM Proc. Lect. Notes {\bf 50} (2010) 341--366
  [arXiv:1106.1747 [math.DG]].

\bibitem{Bouwknegt:2003zg}
  P.~Bouwknegt, K.~Hannabuss and V.~Mathai,
  ``T-duality for principal torus bundles,''
  JHEP {\bf 0403} (2004) 018
  [arXiv:hep-th/0312284].
 
	\bibitem{Freidel:2017yuv}
	L.~Freidel, F.~J.~Rudolph and D.~Svoboda,
	``Generalised kinematics for double field theory,''
	JHEP {\bf 1711} (2017) 175
	[arXiv:1706.07089 [hep-th]].
	
	\bibitem{Vaisman:2012ke}
	I.~Vaisman,
	``On the geometry of double field theory,''
	J.\ Math.\ Phys.\  {\bf 53} (2012) 033509
	[arXiv:1203.0836 [math.DG]].

\bibitem{Deser:2014mxa}
A.~Deser and J.~Stasheff,
``Even symplectic supermanifolds and double field theory,''
Commun.\ Math.\ Phys.\  {\bf 339} (2015) 1003--1020
[arXiv:1406.3601 [math-ph]].

\bibitem{Deser:2016qkw}
A.~Deser and C.~Saemann,
``Extended Riemannian geometry I: Local double field theory,''
arXiv:1611.02772 [hep-th].

	\bibitem{Heller:2016abk}
	M.~A.~Heller, N.~Ikeda and S.~Watamura,
	``Unified picture of non-geometric fluxes and T-duality in double field theory via graded symplectic manifolds,''
	JHEP {\bf 1702} (2017) 078
	[arXiv:1611.08346 [hep-th]].
	
	\bibitem{dee1}
	D.~Roytenberg, 
	``On the structure of graded symplectic supermanifolds and
        Courant algebroids,''
        Contemp. Math. {\bf 315} (2002) 169--186
	[arXiv:math.SG/0203110].
	
	\bibitem{Alexandrov:1995kv}
	M.~Alexandrov, M.~Kontsevich, A.~Schwarz and O.~Zaboronsky,
``The geometry of the master equation and topological quantum field theory,'' Int.~J.~Mod.~Phys.~A {\bf 12} (1997) 1405--1429
[arXiv:hep-th/9502010].

	\bibitem{dee2}
	D.~Roytenberg, 
	``AKSZ--BV formalism and Courant algebroid-induced topological
        field theories,''
        Lett. Math. Phys. {\bf 79} (2007) 143--159
	[arXiv:hep-th/0608150].
	
	\bibitem{Park:2000au}
	J.-S.~Park,
	``Topological open $p$-branes,''
	in: {\em Symplectic Geometry and Mirror Symmetry}, eds. K.~Fukaya, Y.-G.~Oh, K.~Ono and G.~Tian (World Scientific, 2001), pp.~311--384
	[arXiv:hep-th/0012141].
	
	\bibitem{Ikeda:2002wh}
	N.~Ikeda,
	``Chern-Simons gauge theory coupled with BF-theory,''
	Int.\ J.\ Mod.\ Phys.\ A {\bf 18} (2003) 2689--2702
	[arXiv:hep-th/0203043].
	
	\bibitem{Hofman:2002jz}
	C.~Hofman and J.-S.~Park,
	``BV quantization of topological open membranes,''
	Commun.\ Math.\ Phys.\  {\bf 249} (2004) 249--271
	[arXiv:hep-th/0209214].
	
\bibitem{Aschieri:2015roa}
P.~Aschieri and R.~J.~Szabo,
``Triproducts, nonassociative star products and geometry of $R$-flux string compactifications,''
J.\ Phys.\ Conf.\ Ser.\  {\bf 634} (2015)  012004
[arXiv:1504.03915 [hep-th]].

	\bibitem{Chatzistavrakidis:2015vka}
	A.~Chatzistavrakidis, L.~Jonke and O.~Lechtenfeld,
	``Sigma-models for genuinely non-geometric backgrounds,''
	JHEP {\bf 1511} (2015) 182
	[arXiv:1505.05457 [hep-th]].
	
	\bibitem{Bessho:2015tkk}
	T.~Bessho, M.~A.~Heller, N.~Ikeda and S.~Watamura,
	``Topological membranes, current algebras and $H$-flux--$R$-flux duality based on Courant algebroids,''
	JHEP {\bf 1604} (2016) 170
	[arXiv:1511.03425 [hep-th]].
	
	\bibitem{Coimbra:2011nw}
	A.~Coimbra, C.~Strickland-Constable and D.~Waldram,
	``Supergravity as generalised geometry I: Type II theories,''
	JHEP {\bf 1111} (2011) 091
	[arXiv:1107.1733 [hep-th]].
	
	\bibitem{Kontsevich:1997vb}
	M.~Kontsevich,
	``Deformation quantization of Poisson manifolds,''
	Lett.\ Math.\ Phys.\  {\bf 66} (2003) 157--216
	[arXiv:q-alg/9709040].
	
	\bibitem{Cattaneo:1999fm}
	A.~S.~Cattaneo and G.~Felder,
	``A path integral approach to the Kontsevich quantization formula,''
	Commun.\ Math.\ Phys.\  {\bf 212} (2000) 591--611
	[arXiv:math.QA/9902090].
	
	\bibitem{Shelton:2005cf}
	J.~Shelton, W.~Taylor and B.~Wecht,
	``Non-geometric flux compactifications,''
	JHEP {\bf 0510} (2005) 085
	[arXiv:hep-th/0508133].
	
	\bibitem{dftflux2}
	G.~Aldazabal, W.~Baron, D.~Marqu\'es and C.~N\'u\~nez,
	``The effective action of double field theory,''
	JHEP {\bf 1111} (2011) 052
	[Erratum: JHEP {\bf 1111} (2011) 109]
	[arXiv:1109.0290 [hep-th]].
	
	\bibitem{dftflux1}
	D.~Geissb\"uhler,
	``Double field theory and $\mathcal{N}=4$ gauged supergravity,''
	JHEP {\bf 1111} (2011) 116
	[arXiv:1109.4280 [hep-th]].
	
	\bibitem{dftflux3}
	D.~Geissb\"uhler, D.~Marqu\'es, C.~N\'u\~nez and V.~Penas,
	``Exploring double field theory,''
	JHEP {\bf 1306} (2013) 101
	[arXiv:1304.1472 [hep-th]].
	
	\bibitem{dftflux4}
	R.~Blumenhagen, X.~Gao, D.~Herschmann and P.~Shukla,
	``Dimensional oxidation of non-geometric fluxes in type II orientifolds,''
	JHEP {\bf 1310} (2013) 201
	[arXiv:1306.2761 [hep-th]].
	
	\bibitem{Halmagyi}
	N.~Halmagyi,
	``Non-geometric backgrounds and the first order string sigma-model,''
	arXiv:0906.2891 [hep-th].

\bibitem{Bakas:2016nxt}
  I.~Bakas, D.~L\"ust and E.~Plauschinn,
  ``Towards a worldsheet description of doubled geometry in string theory,''
  Fortsch.\ Phys.\  {\bf 64} (2016) 730--747
  [arXiv:1602.07705 [hep-th]].
	
	\bibitem{Uchino2002}
	K.~Uchino,
	``Remarks on the definition of a Courant algebroid,''
	Lett. Math. Phys. {\bf 60} (2002) 171--175
	[arXiv:math.DG/0204010].
	
		\bibitem{preca}
	I.~Vaisman, 
	``Transitive Courant algebroids,''
	Int. J. Math. Sci. {\bf 2005} (2005) 1737--1758.
	
	\bibitem{Hansen:2009zd}
	M.~Hansen and T.~Strobl,
	``First class constrained systems and twisting of Courant algebroids by a closed 4-form,''
	in: {\em Fundamental Interactions: A Memorial Volume for Wolfgang
		Kummer}, eds. D.~Grumiller, A.~Rebhan and D.~V.~Vassilevich (World
	Scientific, 2010), pp.~115--144
	[arXiv:0904.0711 [hep-th]].
	
	 \bibitem{preca2}
	A.~J.~Bruce and J.~Grabowski,
	``Pre-Courant algebroids,''
	arXiv:1608.01585 [math-ph].
	
\bibitem{prehequiv}
Z.-J.~Liu, Y.~Sheng and X.~Xu,
``The Pontryagin class for pre-Courant algebroids,''  
J.\ Geom.\  Phys.  {\bf 104} (2016) 148--162
[arXiv:1205.5898 [math-ph]].

\bibitem{Svoboda:2018rci}
D.~Svoboda,
``Algebroid structures on para-Hermitian manifolds,''
arXiv:1802.08180 [math.DG].


\bibitem{Blumenhagen:2016vpb}
  R.~Blumenhagen and M.~Fuchs,
  ``Towards a theory of nonassociative gravity,''
  JHEP {\bf 1607} (2016) 019
  [arXiv:1604.03253 [hep-th]].
	  
\bibitem{Aschieri:2017sug}
  P.~Aschieri, M.~Dimitrijevi\'c \'Ciri\'c and R.~J.~Szabo,
  ``Nonassociative differential geometry and gravity with non-geometric fluxes,''
JHEP {\bf 1802} (2018) 036
  [arXiv:1710.11467 [hep-th]].

	\bibitem{Ikeda:2012pv}
	N.~Ikeda,
	``Lectures on AKSZ sigma-models for physicists,''
	in: {\em Noncommutative Geometry and Physics 4},
        eds. Y.~Maeda, H.~Moriyoshi, M.~Kotani and S.~Watamura (World
        Scientific, 2017), pp.~79--170
	[arXiv:1204.3714 [hep-th]].

\bibitem{Andriot:2012wx}
  D.~Andriot, O.~Hohm, M.~Larfors, D.~L\"ust and P.~Patalong,
  ``A geometric action for non-geometric fluxes,''
  Phys.\ Rev.\ Lett.\  {\bf 108} (2012) 261602
  [arXiv:1202.3060 [hep-th]].
 
\bibitem{Condeescu:2012sp}
  C.~Condeescu, I.~Florakis and D.~L\"ust,
  ``Asymmetric orbifolds, non-geometric fluxes and noncommutativity in closed string theory,''
  JHEP {\bf 1204} (2012) 121
  [arXiv:1202.6366 [hep-th]].
  
  \bibitem{Freidel:2017wst}
  L.~Freidel, R.~G.~Leigh and D.~Minic,
  ``Intrinsic non-commutativity of closed string theory,''
  JHEP {\bf 1709} (2017) 060
  [arXiv:1706.03305 [hep-th]].

\bibitem{Andriot:2012an}
  D.~Andriot, O.~Hohm, M.~Larfors, D.~L\"ust and P.~Patalong,
  ``Non-geometric fluxes in supergravity and double field theory,''
  Fortsch.\ Phys.\  {\bf 60} (2012) 1150--1186
  [arXiv:1204.1979 [hep-th]].
	
\bibitem{Grana:2008yw}
  M.~Gra\~na, R.~Minasian, M.~Petrini and D.~Waldram,
  ``T-duality, generalized geometry and non-geometric backgrounds,''
  JHEP {\bf 0904} (2009) 075
  [arXiv:0807.4527 [hep-th]].

\bibitem{Bakas:2013jwa}
  I.~Bakas and D.~L\"ust,
  ``3-cocycles, nonassociative star products and the magnetic paradigm of $R$-flux string vacua,''
  JHEP {\bf 1401} (2014) 171
  [arXiv:1309.3172 [hep-th]].

\bibitem{Mylonas:2013jha}
  D.~Mylonas, P.~Schupp and R.~J.~Szabo,
  ``Non-geometric fluxes, quasi-Hopf twist deformations and nonassociative quantum mechanics,''
  J.\ Math.\ Phys.\  {\bf 55} (2014) 122301
  [arXiv:1312.1621 [hep-th]].
  
\bibitem{Barnes:2014ksa}
  G.~E.~Barnes, A.~Schenkel and R.~J.~Szabo,
  ``Nonassociative geometry in quasi-Hopf representation categories I: Bimodules and their internal homomorphisms,''
  J.\ Geom.\ Phys.\  {\bf 89} (2015) 111--152
  [arXiv:1409.6331 [math.QA]].

\bibitem{Bakas:2015gia}
  I.~Bakas and D.~L\"ust,
  ``T-duality, quotients and currents for non-geometric closed strings,''
  Fortsch.\ Phys.\  {\bf 63} (2015) 543--570
  [arXiv:1505.04004 [hep-th]].

\bibitem{Kupriyanov:2015dda}
  V.~G.~Kupriyanov and D.~V.~Vassilevich,
  ``Nonassociative Weyl star products,''
  JHEP {\bf 1509} (2015) 103
  [arXiv:1506.02329 [hep-th]].

\bibitem{Blumenhagen:2013zpa}
R.~Blumenhagen, M.~Fuchs, F.~Ha\ss ler, D.~L\"ust and R.~Sun,
``Nonassociative deformations of geometry in double field theory,''
JHEP {\bf 1404} (2014) 141 
[arXiv:1312.0719 [hep-th]].

\bibitem{Kokenyesi2018}
Z.~K\"ok\'enyesi, A.~Sinkovics and R.~J.~Szabo,
``Double field theory for the A/B-models and topological S-duality in generalized geometry,''
  arXiv:1805.11485 [hep-th].
  
\bibitem{Chatzistavrakidis:2013wra}
  A.~Chatzistavrakidis, L.~Jonke and O.~Lechtenfeld,
  ``Dirac structures on nilmanifolds and coexistence of fluxes,''
  Nucl.\ Phys.\ B {\bf 883} (2014) 59--82
  [arXiv:1311.4878 [hep-th]].

	\bibitem{Blumenhagen1}
	R.~Blumenhagen, A.~Deser, E.~Plauschinn and F.~Rennecke,
	``Bianchi identities for non-geometric fluxes: From quasi-Poisson structures to Courant algebroids,''
	Fortsch.\ Phys.\  {\bf 60} (2012) 1217--1228
	[arXiv:1205.1522 [hep-th]].
	
\bibitem{Saemann:2012ab}
  C.~Saemann and R.~J.~Szabo,
  ``Groupoids, loop spaces and quantization of 2-plectic manifolds,''
  Rev.\ Math.\ Phys.\  {\bf 25} (2013) 1330005
  [arXiv:1211.0395 [hep-th]].

\bibitem{Severa2001}
P.~\v{S}evera,
``Some title containing the words `homotopy' and `symplectic',
e.g. this one,''
Travaux Math. {\bf 16} (2005) 121--137
[arXiv:math.SG/0105080].

\end{thebibliography}
\end{document}